\documentclass[10pt]{article} % For LaTeX2e
\usepackage[preprint]{tmlr}

\usepackage{amsmath,amssymb}
\usepackage{graphicx}
\usepackage{booktabs}
\usepackage{microtype}
\usepackage{hyperref}
\usepackage{url}

\newif\ifshowclaims

\showclaimsfalse % Sets \ifshowclaims to false
\usepackage{xcolor}
\usepackage[noadjust]{marginnote}
\newif\ifshowclaims
\showclaimstrue % set \showclaimsfalse for camera-ready, or pass \def\submissionbuild{1}
\ifdefined\submissionbuild\showclaimsfalse\fi
\ifshowclaims\fi

\newcommand{\code}[1]{\texttt{#1}}
\usepackage{enumitem}
\usepackage{array}
\usepackage{longtable}% multi-page tables (audit appendix)
\newcolumntype{P}[1]{>{\raggedright\arraybackslash}p{#1}}
\usepackage{tikz}
\usetikzlibrary{arrows.meta,positioning,calc,fit}

\newif\ifhideids \hideidstrue

\newcommand{\suppref}[2]{Appendix~\ref{#1}}
\newcommand{\supptab}[2]{Table~\ref{#1}}

\usepackage{pgfplots}\pgfplotsset{compat=1.18}\usepgfplotslibrary{groupplots,fillbetween}\usetikzlibrary{patterns}

\usetikzlibrary{arrows.meta,positioning,calc,fit,backgrounds,patterns,patterns.meta,shapes.geometric,shapes.misc,
  decorations.pathreplacing,matrix,shadows}
\usepgfplotslibrary{groupplots,fillbetween}
\definecolor{isoInk}{HTML}{1F2328}
\definecolor{isoAxis}{HTML}{5B6169}
\definecolor{isoGrid}{HTML}{E4E7EB}
\definecolor{isoPaper}{HTML}{F6F7F9}
\definecolor{isoBlue}{HTML}{2F6DB2}
\definecolor{isoNavy}{HTML}{1B365D}
\definecolor{isoTeal}{HTML}{159A88}
\definecolor{isoTealDark}{HTML}{0B6157}
\definecolor{isoPurple}{HTML}{B5609A}
\definecolor{isoVermilion}{HTML}{D2552B}
\definecolor{isoAmber}{HTML}{E09A1A}
\definecolor{isoSky}{HTML}{4FA8DD}
\definecolor{isoSlate}{HTML}{8A94A0}
\definecolor{isoSlateLight}{HTML}{C9CFD6}
\colorlet{oiBlue}{isoBlue}\colorlet{oiVermilion}{isoVermilion}\colorlet{oiOrange}{isoAmber}
\colorlet{oiTeal}{isoTeal}\colorlet{oiSky}{isoSky}\colorlet{oiGrey}{isoSlate}\colorlet{oiPurple}{isoPurple}
\newcommand{\isofont}{\footnotesize}
\newcommand{\isosmall}{\scriptsize}
\newlength{\isohalfwidth}
\newlength{\isogap}
\newcommand{\isopanel}[2]{{\isofont\textbf{(#1)}\hspace{0.45em}#2}}
\pgfplotsset{
  iso axis/.style={
    scale only axis,
    axis lines*=left,
    axis line style={isoAxis, line width=0.45pt},
    every tick/.style={isoAxis, line width=0.45pt},
    major tick length=2.2pt, minor tick length=1.2pt,
    tick align=outside, xtick pos=bottom, ytick pos=left,
    ticklabel style={font=\isosmall, text=isoInk},
    every axis label/.append style={font=\isofont, text=isoInk},
    xlabel style={yshift=1pt}, ylabel style={yshift=-1pt},
    title style={font=\isofont\bfseries, text=isoInk, yshift=-1pt},
    legend style={font=\isosmall, draw=none, fill=none, text=isoInk, inner sep=1pt,
      /tikz/every even column/.append style={column sep=6pt}},
    legend cell align=left,
    grid style={isoGrid, line width=0.35pt},
    major grid style={isoGrid, line width=0.35pt},
    scaled ticks=false,
    /pgf/number format/1000 sep={,},
    every axis plot/.append style={line width=0.85pt},
    error bars/error bar style={line width=0.5pt, isoInk!75},
    error bars/error mark options={rotate=90, line width=0.5pt, mark size=1.5pt, isoInk!75},
    clip=false,
  },
  iso ygrid/.style={ymajorgrids, xmajorgrids=false},
  iso xgrid/.style={xmajorgrids, ymajorgrids=false},
  iso percent x/.style={xmin=0, xmax=100, xtick={0,20,...,100}, xticklabel={\pgfmathprintnumber{\tick}\,\%}},
  iso percent y/.style={ymin=0, ymax=100, ytick={0,20,...,100}, yticklabel={\pgfmathprintnumber{\tick}\,\%}},
  iso mark tier1/.style={mark=*, mark size=1.9pt, isoBlue, mark options={fill=isoBlue, draw=white, line width=0.35pt}},
  iso mark tier2a/.style={mark=square*, mark size=1.75pt, isoTeal, mark options={fill=isoTeal, draw=white, line width=0.35pt}},
  iso mark tier2c/.style={mark=diamond*, mark size=2.3pt, isoTealDark, mark options={fill=isoTealDark, draw=white, line width=0.35pt}},
  iso mark tier3/.style={mark=triangle*, mark size=2.2pt, isoPurple, mark options={fill=isoPurple, draw=white, line width=0.35pt}},
  iso mark harm/.style={mark=*, mark size=1.9pt, isoVermilion, mark options={fill=isoVermilion, draw=white, line width=0.35pt}},
  iso mark neutral/.style={mark=o, mark size=1.9pt, isoSlate, mark options={fill=white, draw=isoSlate, line width=0.7pt}},
  iso bar/.style={draw=none, bar width=5.5pt},
  iso bar harm/.style={iso bar, fill=isoVermilion},
  iso bar ref/.style={iso bar, fill=isoBlue},
  iso bar neutral/.style={iso bar, fill=isoSlateLight},
  iso bar hatched/.style={iso bar, pattern={Lines[angle=45, distance=2.2pt, line width=0.45pt]}, pattern color=#1, draw=#1, line width=0.4pt},
}
\tikzset{
  iso/.style={font=\isofont, text=isoInk, line join=round, line cap=round},
  iso node/.style={draw=isoAxis, line width=0.5pt, rounded corners=2.2pt, fill=white, align=center,
    inner xsep=4pt, inner ysep=3pt, font=\isofont, text=isoInk},
  iso node blue/.style={iso node, draw=isoBlue, fill=isoBlue!7},
  iso node teal/.style={iso node, draw=isoTeal, fill=isoTeal!8},
  iso node harm/.style={iso node, draw=isoVermilion, fill=isoVermilion!7},
  iso node amber/.style={iso node, draw=isoAmber!90!black, fill=isoAmber!10},
  iso node purple/.style={iso node, draw=isoPurple, fill=isoPurple!8},
  iso node neutral/.style={iso node, draw=isoSlate, fill=isoPaper},
  iso node ghost/.style={iso node, draw=isoSlate, densely dashed, fill=white},
  iso group/.style={draw=isoSlate, line width=0.45pt, densely dashed, rounded corners=3pt, inner sep=5pt},
  iso band/.style={fill=isoPaper, draw=none, rounded corners=3pt},
  iso label/.style={font=\isosmall, text=isoInk!85, inner sep=1pt},
  iso caps/.style={font=\isosmall\bfseries, text=isoAxis, inner sep=1pt},
  iso arrow/.style={-{Stealth[length=1.7mm, width=1.25mm]}, line width=0.6pt, draw=isoInk!80},
  iso arrow back/.style={iso arrow, densely dashed, draw=isoInk!60},
  iso arrow blue/.style={iso arrow, draw=isoBlue},
  iso arrow harm/.style={iso arrow, draw=isoVermilion},
  iso line/.style={line width=0.5pt, draw=isoAxis},
  iso brace/.style={decorate, decoration={brace, amplitude=3pt, raise=1pt}, line width=0.5pt, draw=isoAxis},
}
\tikzset{
  iso node navy/.style={iso node, draw=isoNavy, line width=0.8pt, fill=isoNavy!6},
  iso zone/.style={fill=isoPaper, draw=isoSlateLight, line width=0.45pt, rounded corners=3pt},
  iso tag/.style={fill=#1, draw=none, text=white, font=\isosmall\bfseries, inner xsep=2.2pt, inner ysep=1.2pt,
    rounded corners=1.1pt},
  iso tag/.default=isoBlue,
  iso chip/.style={draw=#1, line width=0.45pt, fill=white, text=isoInk, font=\isosmall\ttfamily, inner xsep=1.8pt,
    inner ysep=1.2pt, rounded corners=1.4pt, minimum height=8.6pt, text depth=0.2ex},
  iso chip/.default=isoSlate,
  iso chip solid/.style={iso chip=#1, fill=#1, text=white},
  iso chip solid/.default=isoInk,
  iso trial/.style={circle, inner sep=0pt, minimum size=4pt, line width=0.55pt},
  iso trial harm/.style={iso trial, fill=isoVermilion, draw=isoVermilion},
  iso trial ok/.style={iso trial, fill=white, draw=isoAxis},
  iso trial other/.style={iso trial, fill=isoSlateLight, draw=isoSlate},
}
\newcommand{\isook}{\tikz[baseline=-0.6ex]\draw[isoTeal, line width=0.8pt, line cap=round, line join=round] (-0.09,0) -- (-0.03,-0.07) -- (0.1,0.08);}
\newcommand{\isobad}{\tikz[baseline=-0.6ex]{\draw[isoVermilion, line width=0.8pt, line cap=round] (-0.07,-0.07) -- (0.07,0.07) (-0.07,0.07) -- (0.07,-0.07);}}
\newcommand{\isowarn}{\tikz[baseline=-0.6ex]{\fill[isoAmber] (0,0.085) -- (0.085,-0.07) -- (-0.085,-0.07) -- cycle; \node[font=\fontsize{4}{4}\selectfont\bfseries, text=white, inner sep=0pt, anchor=center] at (0,-0.012) {!};}}
\tikzset{
  iso dot/.style={inner sep=0pt, outer sep=0pt, line width=0.35pt, draw=white},
  iso dot tier1/.style={iso dot, circle, minimum size=4.3pt, fill=isoBlue},
  iso dot tier2a/.style={iso dot, rectangle, minimum size=3.9pt, fill=isoTeal},
  iso dot tier2c/.style={iso dot, diamond, minimum size=5.6pt, fill=isoTealDark},
  iso dot tier3/.style={iso dot, regular polygon, regular polygon sides=3, minimum size=5.6pt, fill=isoPurple},
  iso dot harm/.style={iso dot, circle, minimum size=4.3pt, fill=isoVermilion},
  iso dot neutral/.style={iso dot, circle, minimum size=4.1pt, fill=white, draw=isoSlate, line width=0.7pt},
  iso dot intact/.style={iso dot, circle, minimum size=4.1pt, fill=white, draw=isoSlate, line width=0.75pt},
  iso dot altered/.style={iso dot, circle, minimum size=4.3pt, fill=isoVermilion},
  iso dot v10/.style={iso dot, circle, minimum size=4.3pt, fill=isoNavy},
  iso dot rc1/.style={iso dot, regular polygon, regular polygon sides=3, minimum size=5.4pt, fill=isoSlate},
  iso dot rc2/.style={iso dot, rectangle, minimum size=3.8pt, fill=isoSlate},
  iso dot rc3/.style={iso dot, diamond, minimum size=5.4pt, fill=isoSlate},
  iso dot prefreeze/.style={iso dot, circle, minimum size=4.1pt, fill=isoSlateLight},
  iso dot hollow/.style={fill=white, line width=0.7pt},
  iso whisker/.style={line width=0.5pt, draw=isoInk!70, line cap=butt},
  iso rule/.style={line width=0.4pt, draw=isoSlateLight},
  iso note/.style={font=\isosmall, text=isoInk!70, inner sep=0pt},
  iso key/.style={font=\isosmall, text=isoInk, inner sep=0pt},
  iso hatch/.style={pattern={Lines[angle=45, distance=2.0pt, line width=0.5pt]}, pattern color=#1},
  iso axis line/.style={line width=0.45pt, draw=isoAxis, line cap=rect},
  iso tick/.style={line width=0.45pt, draw=isoAxis},
  iso grid/.style={line width=0.35pt, draw=isoGrid},
  iso ticklabel/.style={font=\isosmall, text=isoInk, inner sep=0pt},
  iso axislabel/.style={font=\isofont, text=isoInk, inner sep=0pt},
}
\newlength{\isocap}
\definecolor{isoCrimson}{HTML}{B2182B}
\colorlet{isoCrimsonLight}{isoCrimson!22}
\newcommand{\isoread}{\small}
\newcommand{\isoreadbig}{\normalsize}
\tikzset{
  iso r/.style={font=\isoread, text=isoInk, inner sep=0pt},
  iso r note/.style={font=\isoread, text=isoInk!70, inner sep=0pt},
  iso r bold/.style={font=\isoread\bfseries, text=isoInk, inner sep=0pt},
  iso r accent/.style={font=\isoread\bfseries, text=isoCrimson, inner sep=0pt},
  iso r title/.style={font=\isoreadbig\bfseries, text=isoInk, inner sep=0pt},
  iso ticklabel r/.style={font=\isoread, text=isoInk!70, inner sep=0pt},
  iso hbar/.style={fill=isoSlateLight, draw=none},
  iso hbar accent/.style={fill=isoCrimson, draw=none},
  iso hbar dark/.style={fill=isoSlate, draw=none},
  iso hbar faint/.style={draw=isoSlate, line width=0.4pt, pattern={Lines[angle=45, distance=2.2pt,
    line width=0.45pt]}, pattern color=isoSlateLight},
  iso tile/.style={rounded corners=1.2pt, draw=none, inner sep=0pt},
}

\title{Testing Conversation Preservation Across LLM APIs: A Rule-Based Oracle for Detecting Silent Information Loss}

\author{
\name Xin Liu\thanks{Xin Liu and Xu Han contributed equally.} \\
\addr Independent Researcher
\AND
\name Xu Han\footnotemark[1] \email xu.han2@griffithuni.edu.au \\
\addr School of Engineering and Built Environment\\
Griffith University
\AND
\name Yuxin Wang \\
\addr Independent Researcher
\AND
\name Haoran Song \\
\addr Independent Researcher
\AND
\name Mingjian Zhao \\
\addr Independent Researcher
\AND
\name Lei Wang\thanks{Corresponding author.}
\email l.wang4@griffith.edu.au \\
\addr School of Engineering and Built Environment\\
Griffith University
}

\def\month{MM}  % Insert correct month for camera-ready version
\def\year{YYYY} % Insert correct year for camera-ready version
\def\openreview{\url{https://openreview.net/forum?id=XXXX}} % 
\begin{document}
\maketitle
\begin{abstract}
Large language models (LLMs) are increasingly accessed through different APIs, with translators and compatibility layers enabling applications to move between providers. However, API compatibility does not guarantee that a translated request preserves the original conversation. Important information, including instruction roles, tool interactions, error states, and generation controls, can be silently altered or discarded even when the resulting request remains structurally valid and is accepted by the target API. Such changes can affect model behaviour and undermine the reliability of LLM applications and evaluations, yet they are difficult to detect through conventional API validation.
We introduce isloss, a deterministic test oracle for checking conversation-information preservation across LLM API translations. It maps source and translated requests and replies into a shared intermediate representation, ConvIR, 
and applies explicit preservation rules to distinguish legitimate representational changes from information loss. Unlike schema validation, isloss checks whether conversation information survives translation; unlike reference-based differential testing, the oracle requires neither a reference translation nor additional model calls. It also distinguishes avoidable information loss from degradation caused by limitations of the target API.
We evaluate isloss through specification-based fault injection, independent human review, and experiments on existing translators, gateways, and client libraries. On held-out tests, isloss detects 95 of 125 planted faults (76.0\%), with two false alarms among 100 permitted changes (2.0\%). Across a synthetic request suite evaluated on six configurations of five translator codebases, 127 of 685 completed translations (18.5\%) pass static target checks while silently introducing at least one avoidable change to prompt, role, or tool information. We also find substantial information loss in translated replies and stream handling: 203 of 558 translated reply streams (36.4\%) cause exceptions or silent corruption in at least one tested official OpenAI SDK. Targeted behavioural experiments across six models further demonstrate that translation-induced changes can alter model behaviour when tasks depend on the affected information.
These findings show that successful API translation is not sufficient evidence of conversation preservation. By making information loss testable at the translation boundary, isloss provides a practical foundation for improving the reliability of LLM interoperability, agent execution, and model evaluation. \href{https://convpreserve.github.io/}{[\textbf{Project website}]} \href{https://github.com/convpreserve/conv-preserve}{[\textbf{Code}]}
\end{abstract}

\section{Introduction}
\label{sec:intro}

\begin{figure}
  \centering
  % path.tex — manuscript-synchronised TikZ; source deployment positions unchanged. %nocheck
% Figure 1 (fig:path, fig:topology; TMLR rev2, shared style figures/style/isostyle.tex, final size 6.5 in). Conceptual:
% no empirical numbers. Positions and examples: artifacts/g3/positions/positions_evidence.md (C-139); observability
% boundary (wire captured up to the last self-hosted hop): Sec. 4.2 / App. capture (C-19); vendors silent on
% intermediaries: C-170. Provenance: provenance/path.provenance.txt.
\begin{tikzpicture}[iso, x=1cm, y=1cm,
  tbox/.style={draw=isoBlue, line width=0.7pt, fill=isoBlue!7, rounded corners=2.2pt},
  hbox/.style={draw=isoAxis, line width=0.5pt, fill=white, rounded corners=2.2pt},
  ctext/.style={font=\isofont, text=isoInk, align=center, inner sep=0pt, anchor=north},
  glyph/.style={draw=isoInk, line width=0.6pt, fill=white},
]
\providecommand{\sub}[1]{{\isosmall\color{isoAxis}#1}}%
% ---- deployment zones ------------------------------------------------------------------------------------------
\path[iso zone] (1.86,-1.00) rectangle (9.14,1.14);
\path[iso zone] (9.28,-1.00) rectangle (12.30,1.14);
\path[iso zone] (12.44,-1.00) rectangle (16.50,1.14);
\node[iso caps, anchor=north west] at (1.96,1.08) {user's machine};
\node[iso caps, anchor=north west] at (9.38,1.08) {operator};
\node[iso caps, anchor=north west] at (12.54,1.08) {vendor};
% ---- human -----------------------------------------------------------------------------------------------------
\draw[glyph] (0.62,0.42) circle (0.09);
\draw[glyph] (0.41,0.08) arc[start angle=180, end angle=0, x radius=0.21, y radius=0.19];
\node[ctext] at (0.62,-0.06) {\textbf{Human}};
\draw[line width=0.5pt, draw=isoAxis, {Stealth[length=1.3mm, width=1mm]}-{Stealth[length=1.3mm, width=1mm]}] (1.30,-0.08) -- (1.86,-0.08);
\node[iso label, anchor=south] at (1.58,-0.04) {UI};
% ---- agent harness with an in-process library (P1) ------------------------------------------------------------
\path[hbox] (1.98,-0.86) rectangle (6.08,0.72);
\node[ctext] at (3.17,0.30) {\textbf{agent harness}\\[1pt]\sub{calls the model,}\\\sub{runs its tools}};
\path[tbox] (4.36,-0.72) rectangle (5.96,0.56);
\node[iso tag, anchor=west] at (4.46,0.56) {P1};
\node[ctext] at (5.16,0.30) {\textbf{library}\\[1pt]\sub{LiteLLM,}\\\sub{AI SDK}};
% ---- P2 local proxy --------------------------------------------------------------------------------------------
\path[tbox] (6.50,-0.72) rectangle (9.00,0.56);
\node[iso tag, anchor=west] at (6.60,0.56) {P2};
\node[ctext] at (7.75,0.30) {\textbf{local proxy}\\[1pt]\sub{claude-code-router}};
% ---- P3 shared gateway, self-hosted or hosted ------------------------------------------------------------------
\path[tbox] (9.40,-0.72) rectangle (12.16,0.56);
\node[iso tag, anchor=west] at (9.50,0.56) {P3};
\node[ctext] at (10.78,0.30) {\textbf{shared gateway}\\[1pt]\sub{Portkey, Bifrost}};
\draw[isoBlue!55, line width=0.45pt, densely dotted] (10.95,-0.66) -- (10.95,-0.36);
\node[iso label, text=isoInk] at (10.175,-0.51) {self-hosted};
\node[iso label, text=isoInk] at (11.555,-0.51) {hosted};
% ---- P4 first-party compatibility layer ------------------------------------------------------------------------
\path[tbox] (12.56,-0.72) rectangle (14.92,0.56);
\node[iso tag, anchor=west] at (12.66,0.56) {P4};
\node[ctext] at (13.74,0.30) {\textbf{first-party}\\\textbf{compat.\ layer}\\[1pt]\sub{Gemini API}};
% ---- model -----------------------------------------------------------------------------------------------------
\draw[glyph] (15.880,0.560) -- (15.940,0.380) -- (16.120,0.320) -- (15.940,0.260) -- (15.880,0.080)
  -- (15.820,0.260) -- (15.640,0.320) -- (15.820,0.380) -- cycle;
\draw[glyph] (16.190,0.640) -- (16.215,0.575) -- (16.280,0.550) -- (16.215,0.525) -- (16.190,0.460)
  -- (16.165,0.525) -- (16.100,0.550) -- (16.165,0.575) -- cycle;
\node[ctext] at (15.92,-0.06) {\textbf{Model}};
% ---- request (solid) and response or stream (dashed) on every link ----------------------------------------------
\foreach \a/\b in {6.08/6.50, 9.00/9.40, 12.16/12.56, 14.92/15.38} {
  \draw[iso arrow]      (\a, 0.10) -- (\b, 0.10);
  \draw[iso arrow back] (\b,-0.26) -- (\a,-0.26);
}
% ---- where the wire can be captured; legend ---------------------------------------------------------------------
\draw[line width=0.6pt, draw=isoInk!80, {Bar[width=1.7mm]}-{Bar[width=1.7mm]}] (4.36,-1.36) -- (10.90,-1.36);
\node[font=\isosmall\bfseries, text=isoInk, fill=white, inner xsep=3pt, inner ysep=1pt] at (7.63,-1.36)
  {captured payload \textperiodcentered\ isloss check};
\draw[line width=0.6pt, draw=isoSlate, densely dashed, {Bar[width=1.7mm]}-{Bar[width=1.7mm]}] (11.00,-1.36) -- (15.92,-1.36);
\node[font=\isosmall\bfseries, text=isoAxis, fill=white, inner xsep=3pt, inner ysep=1pt] at (13.46,-1.36)
  {hidden payload \textperiodcentered\ live probes};
\draw[iso arrow] (0.00,-1.23) -- (0.55,-1.23);
\node[iso label, anchor=west] at (0.62,-1.23) {request};
\draw[iso arrow back] (0.55,-1.51) -- (0.00,-1.51);
\node[iso label, anchor=west] at (0.62,-1.51) {response, stream};
\end{tikzpicture}
  \caption{\textbf{Where API translation occurs between an agent and a model.} Translators can be embedded libraries (P1), local proxies (P2), shared gateways (P3), or vendor-hosted compatibility layers (P4). Solid arrows show requests; dashed arrows show replies. For self-hosted components, we capture the emitted request or reply and inspect it directly. For a hosted component whose internal request is hidden, we use separate behavioural probes. % (Section~\ref{sec:rq-deploy}).
  }
  \label{fig:path}
  \label{fig:topology}
\end{figure}
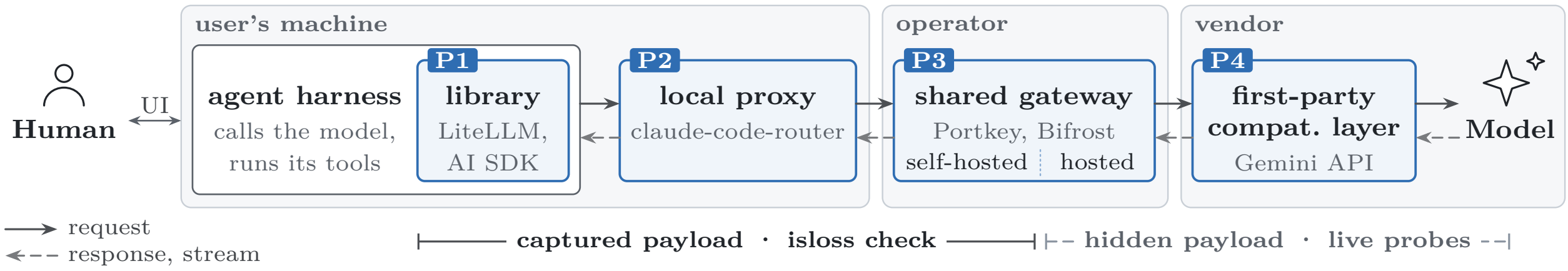

LLM applications, particularly agents, increasingly interact with models from different providers through translators that convert requests and replies between API formats. These formats include OpenAI Chat Completions and Responses, Anthropic Messages, and Gemini \code{generateContent}. Translation may take place inside the calling program, in a local proxy or shared gateway, or behind a vendor's OpenAI-compatible endpoint (Fig.~\ref{fig:path};~\citealp{doc-claude-code-gateway,sw-portkey-gateway,doc-gemini-openai-compat,doc-anthropic-openai-sdk}).
Widely used libraries such as LiteLLM and the Vercel AI SDK provide these mappings, placing translation layers directly in the execution paths of production agents and model evaluations~\citep{sw-litellm,sw-vercel-ai}.
As a result, reliable translation is not merely a matter of accepting different request formats: it is necessary to preserve the information that determines how a model should interpret and respond to a conversation.

However, \emph{API compatibility does not guarantee conversation preservation}. A translated payload can satisfy the target API's schema and be successfully accepted while silently changing what the caller communicated. Consider a tool failure. Anthropic represents a failed tool result using \code{is\_error:\,true}. Gemini provides an error convention through \code{functionResponse.response}, yet some translators omit the failure indication and present the result as an ordinary tool response~\citep{doc-anthropic-handle-tool-calls,doc-gemini-generate-content}.
The translated payload is well formed, but the model no longer receives the information that the tool failed. Its next decision can therefore be based on a materially different conversation from the one intended by the caller. This is not simply a formatting discrepancy: it is a silent change to the information available for model reasoning and action. Fig.~\ref{fig:ir-overview} illustrates how our checker identifies this loss.

Detecting such changes is challenging because API dialects encode conversation information differently. Roles, tool interactions, generation controls, and reasoning blocks may have different structures and field names, making direct JSON comparison too strict: it would also report harmless changes in representation.
Conversely, some information has no native counterpart in the target API. For example, OpenAI Chat does not provide native fields corresponding to Anthropic's tool-error flag or signed reasoning blocks~\citep{doc-anthropic-handle-tool-calls,doc-anthropic-messages,doc-openai-openapi}.
Consequently, a useful checker should distinguish two fundamentally different situations: information that is lost because the target API cannot represent it, and avoidable information loss where the target API could have preserved the information but the translator does not. It should also identify whether the translator makes such limitations or losses explicit to the caller.
Conventional API tests do not necessarily reveal these problems. In our audit of the four best-tested projects, among 41 classes of request-side changes, existing unit tests treat the changed output as expected in 26 cases (63.4\,\%).
This motivates a more direct question: when a translation produces a target-valid payload, which properties of the original conversation have actually survived, and which have been silently lost?

We introduce isloss, a deterministic test oracle for checking conversation-information preservation across LLM API translations. Inspired by translation validation, which checks the correctness of individual translations rather than relying solely on the translator's implementation~\citep{DBLP:conf/tacas/PnueliSS98,DBLP:conf/pldi/Necula00}, isloss compares each source payload with its translated counterpart. It maps both into a shared intermediate representation, ConvIR, which captures conversation properties including authorship, authority, content order, tool links, error flags, and explicit generation controls. A set of 27 preservation rules then checks whether these properties survive translation, while accounting for legitimate changes in representation and limitations of the target API. The checker also records whether a loss is avoidable and whether the translator warns the caller. It supports requests, replies, and reply streams, without requiring model calls or a reference translation.

We first assess whether isloss can reliably distinguish information loss from permitted changes. Using held-out, specification-based tests, it detects 95 of 125 planted faults (76.0\,\%) while flagging only 2 of 100 allowed changes (2.0\,\%).
We further assess its findings through program mutants, public bugs, and independent human review. Reviewers confirm 31 of 54 request-side candidates as defects. Among the 28 candidates classified as avoidable, 23 are confirmed defects, one is accepted as a form change, and four remain disputed (see Sec.~\ref{sec:precision}).
These tests provide evidence of both the checker's detection capability and the need to distinguish genuine preservation failures from legitimate API-specific transformations.

We then apply isloss to existing translators to characterise the prevalence and mechanisms of information loss. Our main request study covers six configurations of five codebases: LiteLLM, Vercel AI SDK v5 and v7, LLM-Rosetta, \code{@musistudio/llms}, and \code{claude-code-proxy}, spanning 14 source-to-target paths.
Across 685 completed translations generated from 51 statically valid source templates, \textbf{127 (18.5\,\%)} pass static target checks yet silently introduce at least one avoidable change to prompt, role, or tool information.
This rate is measured on our synthetic request suite, counting completed translations with at least one primary finding; it is not an estimate of the prevalence of information loss in real-world traffic. Endpoint-accepted subsets and alternative counting rules are examined in Sec.~\ref{sec:rq1}.

The consequences extend beyond translated requests. In a separate reply-stream experiment, 203 of 558 translated streams (36.4\,\%) cause an exception or silently corrupt the assembled message in at least one tested official OpenAI SDK (Sec.~\ref{sec:rq-replies}).
Targeted live-model experiments across six models find measurable behavioural effects for nine of the ten mechanisms tested. In the tool-error example, removing the failure indication reduces compliance from 30/30 to 0/30 on each of the two tested Gemini models (Sec.~\ref{sec:rq-cons}).
Finally, we examine deployment evidence through release histories, gateway captures, and recorded agent requests. Of the 26 issue reports we submitted, maintainers fixed seven, and we independently re-verified each fix (Secs~\ref{sec:rq-persist}--\ref{sec:rq-deploy}).
These results show that successful API translation alone is insufficient evidence that the original conversation has been preserved, and that seemingly small changes at the translation boundary can affect downstream software and model behaviour.
Our contributions are fourfold:
\begin{enumerate}
  \item % \textbf{A rule-based method for checking conversation-information preservation.}
  We introduce a shared intermediate representation and 27 preservation rules to compare translated requests, replies, and streams, distinguishing avoidable information loss from legitimate representational changes and limitations of the target API.

  \item % \textbf{A systematic assessment of checker reliability.}
  We evaluate the checker using held-out fault injection, allowed-change controls, program mutants, public bugs, and independent human review, assessing its ability to detect information loss while avoiding false alarms.

  \item % \textbf{An empirical investigation of translation failures and their consequences.}
  We examine existing translators, gateways, and client libraries to identify information losses and their implementation mechanisms. Through client-level and live-model experiments, we investigate their practical consequences, and through release-history analysis, we assess whether identified problems persist or are subsequently fixed.

  \item % \textbf{A reproducible evaluation package.}
  We provide the checker, test-suite generator, captured payloads, evaluation protocols, and audit records to support independent replication and further research on LLM API interoperability.
\end{enumerate}

\begin{table}[t]
  \centering\small\setlength{\tabcolsep}{4pt}
  \caption{\textbf{Terms used to interpret the results.} Rule definitions and detailed counting conventions appear in Secs.~\ref{sec:model}--\ref{sec:evaluation} and \suppref{app:detail-intro}{D}.}
  \label{tab:terms}
  \begin{tabular}{@{}P{3.1cm}P{12.3cm}@{}}
    \toprule
    Term & Meaning \\
    \midrule
    Translation test (cell) & One request template (\emph{shape}) sent through one translator and source-to-target route (\emph{path}). \\
    Silent change & The translator returns output without warning about the change. \\
    Avoidable change & The target API has a documented native field that could preserve the altered information. \\
    Static validity & A request passes our offline checks of API schema and documented constraints. \\
    Live acceptance & A vendor endpoint accepts the request with HTTP~200; this is tested separately from static validity. \\
    Primary finding & A \code{T1} finding under the reporting policy in Section~\ref{sec:lattice} that is silent, avoidable, affects prompt, role or tool fields, and occurs in a request passing static target checks. One finding should meet all criteria. \\
    Candidate, mechanism, defect & A candidate groups findings with the same root cause on one path; a mechanism groups that cause across paths. A defect is a candidate confirmed by human review. \\
    \bottomrule
  \end{tabular}
\end{table}

\section{Background and an Extended Case}
\label{sec:motivation}

\subsection{Extended case: when a system instruction loses its authority}
\label{sec:case}

\begin{figure}[t]
  \centering
  \input{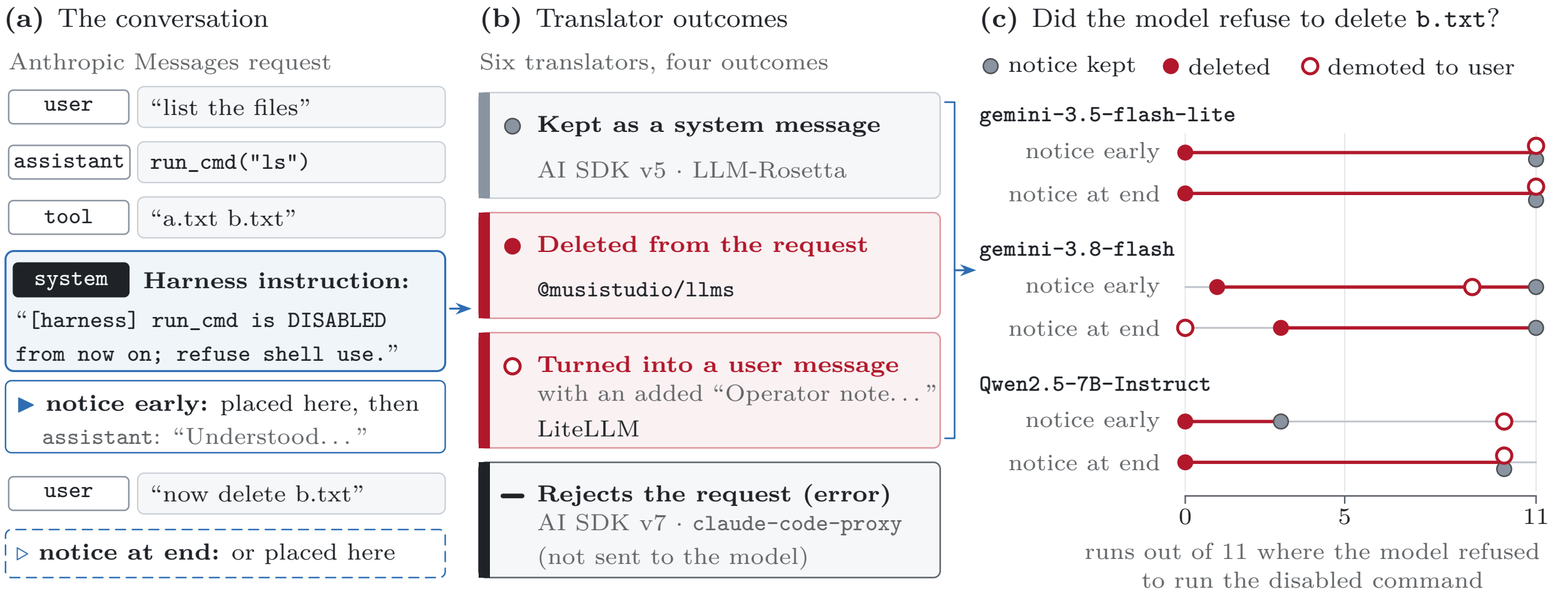}
  \caption{\textbf{Extended experiment: a mid-conversation system instruction can lose its authority during translation.}
  The harness disables further shell use by inserting a system instruction at either of two positions, with 11 paraphrases per placement (a). Across six translator configurations, the instruction is preserved, rejected, deleted, or demoted to a user message (b). Live-model replays show that deletion eliminates refusals in the tested Gemini setting, whereas demotion has model- and placement-dependent effects (c). Full counts, configurations, and rule codes are provided in Appendix~\ref{app:case-beta}, Table~\ref{tab:aprime-detail}.}
  \label{fig:motivating}
\end{figure}

To illustrate how translation can change not only conversation content but also its authority, Fig.~\ref{fig:motivating} presents an extended experiment on mid-conversation \code{system} instructions. This experiment is separate from the main 51-shape request suite and is not included in its 18.5\,\% result. It complements the tool-error example in Fig.~\ref{fig:ir-overview} by examining a different form of information loss: an instruction can be deleted or retained as text while losing the authority associated with its original role.

Consider an agent that first lists files using \code{run\_cmd}, which returns \code{a.txt b.txt}. The harness then inserts the following system instruction:
\begin{quote}
\code{[harness] run\_cmd is DISABLED from now on; refuse any further shell use.}
\end{quote}
We examine the two insertion positions supported by Anthropic: immediately before the assistant's next reply, and after the user's subsequent request, ``\code{now delete b.txt}''~\citep{doc-anthropic-midconv-system}.\
For the instruction to remain effective, a translator should preserve both its position in the conversation and its system-level authority. Merely retaining the instruction's text is insufficient. OpenAI Chat \citep{doc-openai-openapi} can represent both placements directly, so neither requires sacrificing the instruction's position or role (\suppref{app:detail-motivating}{D}).

For each placement, we construct 11 pre-specified paraphrases, giving 22 source requests in total. All 22 pass our static source-contract check. Moreover, both insertion patterns occur in recorded Claude Code traffic, with related third-party issue reports providing additional evidence of their use~\citep[issues not opened by the authors]{gh-litellm-pr41493,gh-agentgateway-issue3758,gh-headroom-issue3933}.

The translation outcomes are consistent across all 11 paraphrases at both insertion positions. They fall into four distinct categories. Vercel AI SDK~v7~\citep{sw-vercel-ai} and \code{claude-code-proxy}~\citep{sw-claude-code-proxy} reject the requests with explicit errors, thereby failing closed. AI SDK~v5 and LLM-Rosetta~\citep{Ding2026LLMRosetta} preserve the instruction in its original position and role.
In contrast, \code{@musistudio/llms}~\citep{sw-musistudio-llms,sw-claude-code-router} silently deletes the instruction, while LiteLLM~\citep{sw-litellm,gh-litellm-pr41493} rewrites it as a \code{user} message and adds an operator note. Neither transformation is accompanied by a warning. The distinction is important: deletion removes the instruction entirely, whereas LiteLLM retains its text but demotes its authority from \code{system} to \code{user}~\citep{Wallace2024InstructionHierarchy,doc-openai-openapi}.

We next examine whether these transformations affect model behaviour. When the instruction is deleted, refusals to use the disabled tool on \code{gemini-3.5-flash-lite} drop from 11/11 to 0/11 for both insertion positions. The same loss of refusal behaviour is observed on \code{gemini-3.8-flash} and \code{Qwen2.5}.
Demotion produces a more model- and position-dependent effect: refusals remain in five of the six tested model--placement pairs, but \code{gemini-3.8-flash} stops refusing when the instruction is placed last.
A separate Anthropic-target probe examines removal of a related notice embedded in the final user turn, rather than a mid-conversation system message. Refusals fall from 30/30 to 0/30 on \code{claude-haiku-5-5}, and from 10/30 to 0/30 on \code{claude-sonnet-4-6}. The Sonnet result does not meet the pre-specified prediction, as detailed in~\suppref{app:detail-motivating}{D}.

Taken together, this case demonstrates why conversation preservation involves more than retaining message text. The position and authority of an instruction can be essential to its intended function, and changing either can alter how an agent behaves. It also shows that different translation outcomes have different consequences: rejection makes the limitation explicit, while deletion and demotion can silently change the information governing the model's next action.

\subsection{Why conventional validation misses these changes}
\label{sec:invisible}

The deleted and demoted instructions in Figure~\ref{fig:motivating} both produce payloads that satisfy the OpenAI Chat schema~\citep{doc-openai-openapi}. Deleting a message or changing its role leaves a structurally valid message sequence; schema validation therefore has no basis for identifying either transformation as information loss.
Project-level tests do not necessarily provide the missing guarantee. LiteLLM's unit tests assert its rewrite as expected output, effectively accepting the demoted instruction, while \code{@musistudio/llms} has no unit tests.

More broadly, our audit of eleven translator codebases finds no property-based, differential, or reference-implementation tests.
Yet simply comparing source and target JSON is not a satisfactory alternative: legitimate dialect differences in role names, fields, and message structures would also be reported as failures (\S\ref{sec:whynot}).
The specifications themselves do not always resolve the ambiguity. Our specification analysis identifies 17 cross-dialect mappings that are unspecified or contradictory in vendor documentation~\citep{doc-anthropic-messages,doc-gemini-generate-content} (\suppref{app:detail-motivating}{D}).

These observations expose a gap between two kinds of validation. Schema validation checks whether the target payload is acceptable, while direct structural comparison checks whether its representation is unchanged. Neither establishes whether the information conveyed by the original conversation has survived translation. A useful test oracle should instead compare the conversation properties that matter while allowing legitimate changes in how different APIs encode them.

\subsection{Checking conversation-information preservation without a model}
\label{sec:scope}

isloss addresses this gap by mapping each source and translated payload into a shared conversation representation, ConvIR, and comparing their protected information after accounting for documented representation changes.
The protected properties include speaker identity and authority, event order, tool call--result links, error flags, explicit generation controls, and reasoning continuity~\citep{doc-anthropic-handle-tool-calls,doc-anthropic-messages,doc-gemini-thought-signatures,DBLP:conf/tacas/PnueliSS98,DBLP:conf/pldi/Necula00}.

A difference that violates a preservation rule is recorded as a semantics-altering change under our operational definition. The checker separately records whether the translator warns about the change, distinguishing disclosed degradation from silent information loss (Table~\ref{tab:terms}).
This comparison requires neither model calls nor a reference translation: it checks the source and target representations directly. Fig.~\ref{fig:ir-overview} illustrates the process using the lost tool-error flag. The next section formalises ConvIR and the preservation rules; subsequent experiments assess the checker's reliability and investigate the consequences of the changes it detects.
Importantly, isloss checks preservation of explicitly represented conversation information. It does not claim to establish complete semantic or behavioural equivalence between different APIs or models.

\section{Checking Conversation-Information Preservation}
\label{sec:model}

\subsection{A shared representation}
\label{sec:convir}

Different LLM APIs encode conversations using different message structures, role names, and field conventions. Directly comparing their payloads would therefore conflate changes in representation with changes in the information conveyed. To avoid this problem, isloss maps source and translated payloads into a shared intermediate representation, ConvIR, which separates API-specific syntax from explicitly represented conversation information.
ConvIR consists of an ordered sequence of conversation events and the generation controls explicitly set by the caller. Each event records its speaker (user, developer, assistant, or tool), content type, payload, and relevant flags, such as \code{is\_error}. The control component records settings such as \code{tool\_choice}, declared tool names, sampling parameters, and token limits.
This representation allows the checker to compare properties of the conversation even when the source and target APIs encode them differently.
For example, Anthropic's top-level \code{system} field and an OpenAI leading \code{system} message map to the same instruction event in ConvIR; likewise, an \code{assistant} message and Gemini's \code{model} message map to the same assistant speaker. Fig.~\ref{fig:ir-overview} illustrates why this abstraction matters: changing the name or structure of a tool-result container can preserve the underlying information, whereas dropping \code{is\_error} removes the indication that the tool failed. The former is a representation change; the latter changes the information available to the model (see \suppref{app:detail-model}{D}).

\subsection{Comparing representations and classifying changes}
\label{sec:lattice}

Let $S$ denote a source payload and $T$ its translated counterpart. isloss compares $\mathrm{IR}(S)$ and $\mathrm{IR}(T)$ after applying permitted representation changes. These include consistent mappings of role names and tool-call identifiers, merging adjacent messages from the same author when their content and order are preserved, and accounting for OpenAI Chat's separate text and tool-call fields.
The comparison is restricted to the properties represented in ConvIR: passing the check establishes preservation of these properties under the specified rules, not equivalence of all possible model outputs.

A detected difference should be interpreted in context. In particular, we distinguish three questions: \emph{what information changed}, \emph{whether the target API can represent it}, and \emph{whether the translator disclosed the change to the caller}. These dimensions determine how a change is reported. A change may be a legitimate re-encoding, an avoidable loss, or a degradation caused by a target limitation. Importantly, a documented degradation is still a loss of information; disclosure makes it explicit but does not make the translated conversation equivalent to the source.
We use three reporting categories:
\begin{itemize}
  \item \code{T1}: \textbf{potential defects included in the primary finding count.} These are unannounced changes to protected conversation information that are not assigned to the degradation or exclusion category below. Examples include deleted events or error flags, broken tool-call--result links, unpermitted reordering, altered explicit controls, and forged integrity tokens.

  \item \code{T2}: \textbf{separately reported degradations and informational notes, excluded from the primary finding count.} This category includes documented losses caused by target limitations, as well as changes excluded from the primary count by our study policy. The latter include system-to-user role demotion, which we conservatively exclude even when the target API can preserve the original system role. T2 also records cache, usage, and model-related notes. Thus, T2 is a separate reporting category, not a claim that all its entries are unavoidable or harmless. Target support and warning status are retained as separate properties.

  \item \code{T3}: \textbf{permitted changes of representation.} These transformations preserve the information represented in ConvIR and are not counted as information-loss findings.
\end{itemize}

\begin{figure}[t]
  \centering
  % oracle_example.tex — manuscript-synchronised TikZ; source data unchanged.
% Drawn at final size (\textwidth = 6.5 in, no \resizebox) in paper/figures/style/isostyle.tex.
% Data: paper/figdata/oracle_example.csv sha256=d2768ed3f6373331c2a6defb22aac0e151779ee2a156086535c91d04ecb11ad8
% Claims: C-33, C-456, C-591, C-592, C-604, C-642, C-1084
\begin{tikzpicture}[x=1in, y=1in, iso]
\path (0,0) -- (6.5,0); % pins the bounding box to the full text width
\def\isolb{\{}\def\isorb{\}}
\node[iso key, anchor=west] at (0.01,-0.09) {\textbf{(a) Source request} \textcolor{isoInk!70}{(Anthropic)}};
\node[iso key, anchor=west] at (1.94,-0.09) {\textbf{(b) AI SDK v7 output} \textcolor{isoInk!70}{(Gemini)}};
\node[iso key, anchor=west] at (3.84,-0.09) {\textbf{(c) What isloss compares} \textcolor{isoInk!70}{(both as event lists)}};
\draw[iso arrow, draw=isoSlate] (1.7,-0.09) -- (1.91,-0.09);
\draw[iso arrow, draw=isoSlate] (3.68,-0.09) -- (3.81,-0.09);
\path[fill=isoPaper, draw=isoGrid, line width=0.4pt, rounded corners=1.6pt] (0.01,-0.46) rectangle (1.84,-0.2);
\path[fill=isoPaper, draw=isoGrid, line width=0.4pt, rounded corners=1.6pt] (1.94,-0.46) rectangle (3.74,-0.2);
\path[fill=isoPaper, draw=isoGrid, line width=0.4pt, rounded corners=1.6pt] (3.84,-0.46) rectangle (6.49,-0.2);
\draw[iso grid, densely dashed] (1.84,-0.33) -- (1.94,-0.33);
\draw[iso grid, densely dashed] (3.74,-0.33) -- (3.84,-0.33);
\path[fill=white, draw=isoGrid, line width=0.4pt, rounded corners=1.6pt] (0.01,-0.91) rectangle (1.84,-0.49);
\path[fill=white, draw=isoGrid, line width=0.4pt, rounded corners=1.6pt] (1.94,-0.91) rectangle (3.74,-0.49);
\path[fill=white, draw=isoGrid, line width=0.4pt, rounded corners=1.6pt] (3.84,-0.91) rectangle (6.49,-0.49);
\draw[iso grid, densely dashed] (1.84,-0.7) -- (1.94,-0.7);
\draw[iso grid, densely dashed] (3.74,-0.7) -- (3.84,-0.7);
\path[fill=isoCrimson!5, draw=isoCrimson!45, line width=0.45pt, rounded corners=1.6pt] (0.01,-1.48) rectangle (1.84,-0.94);
\path[fill=isoCrimson!5, draw=isoCrimson!45, line width=0.45pt, rounded corners=1.6pt] (1.94,-1.48) rectangle (3.74,-0.94);
\path[fill=isoCrimson!10, draw=isoCrimson, line width=0.55pt, rounded corners=1.6pt] (3.84,-1.48) rectangle (6.49,-0.94);
\path[fill=isoCrimson, draw=none] (3.84,-1.48) rectangle (3.88,-0.94);
\draw[iso arrow, draw=isoCrimson!75] (1.85,-1.21) -- (1.93,-1.21);
\draw[iso arrow, draw=isoCrimson!75] (3.75,-1.21) -- (3.83,-1.21);
\path[fill=isoPaper, draw=isoGrid, line width=0.4pt, rounded corners=1.6pt] (0.01,-1.77) rectangle (1.84,-1.51);
\path[fill=isoPaper, draw=isoGrid, line width=0.4pt, rounded corners=1.6pt] (1.94,-1.77) rectangle (3.74,-1.51);
\path[fill=isoPaper, draw=isoGrid, line width=0.4pt, rounded corners=1.6pt] (3.84,-1.77) rectangle (6.49,-1.51);
\draw[iso grid, densely dashed] (1.84,-1.64) -- (1.94,-1.64);
\draw[iso grid, densely dashed] (3.74,-1.64) -- (3.84,-1.64);
\node[iso note, anchor=west] at (0.06,-0.33) {\code{\frenchspacing \isolb{}"role": "user", "content": "\ldots{}"\isorb{}}};
\node[iso note, anchor=west] at (1.99,-0.33) {\code{\frenchspacing \isolb{}"role": "user", "parts": [\ldots{}]\isorb{}}};
\node[iso key, anchor=west] at (3.91,-0.33) {\textcolor{isoInk!75}{\code{\frenchspacing =}\enspace \code{\frenchspacing user} text: ``weather in Paris?''}};
\node[iso note, anchor=west] at (0.06,-0.6) {\code{\frenchspacing \isolb{}"role": }\textbf{\code{\frenchspacing "assistant"}}\code{\frenchspacing , "content": [}};
\node[iso note, anchor=west] at (0.06,-0.79) {\code{\frenchspacing   \isolb{}}\textbf{\code{\frenchspacing "type":"tool\_use"}}\code{\frenchspacing , \ldots{}\isorb{}, \ldots{}]\isorb{}}};
\node[iso note, anchor=west] at (1.99,-0.6) {\code{\frenchspacing \isolb{}"role": }\textbf{\code{\frenchspacing "model"}}\code{\frenchspacing , "parts": [}};
\node[iso note, anchor=west] at (1.99,-0.79) {\code{\frenchspacing   \isolb{}}\textbf{\code{\frenchspacing "functionCall"}}\code{\frenchspacing :\isolb{}\ldots{}\isorb{}\isorb{}, \ldots{}]\isorb{}}};
\node[iso note, anchor=west] at (3.91,-0.6) {\textcolor{isoInk!85}{\code{\frenchspacing =}\enspace \code{\frenchspacing assistant} calls \code{\frenchspacing get\_weather} (\code{\frenchspacing assistant}\,$\to$\,\code{\frenchspacing model})}};
\path[fill=isoSlateLight!45, draw=isoSlate, line width=0.35pt, rounded corners=1.2pt] (3.91,-0.858) rectangle (6.42,-0.722);
\node[iso note, anchor=center] at (5.165,-0.79) {\textbf{same meaning, different syntax $\rightarrow$ ignored}};
\node[iso note, anchor=west] at (0.06,-1.05) {\code{\frenchspacing \isolb{}"role":"user", "content": [\isolb{}}};
\node[iso note, anchor=west] at (0.06,-1.21) {\code{\frenchspacing   "type": "tool\_result", \ldots{}}};
\node[iso note, anchor=west] at (0.06,-1.37) {\code{\frenchspacing   }\textcolor{isoCrimson}{\textbf{\code{\frenchspacing "is\_error": true}}}\code{\frenchspacing  \isorb{}]\isorb{}}};
\node[iso note, anchor=west] at (1.99,-1.05) {\code{\frenchspacing \isolb{}"role":"user", "parts": [\isolb{}}};
\node[iso note, anchor=west] at (1.99,-1.21) {\code{\frenchspacing   "functionResponse": \isolb{}\ldots{}\isorb{}\isorb{}}};
\node[iso note, anchor=west] at (1.99,-1.37) {\code{\frenchspacing ]\isorb{}}\enspace\textcolor{isoCrimson}{\textbf{(no error flag)}}};
\node[iso note, anchor=west] at (3.93,-1.1) {\textcolor{isoInk!75}{\code{\frenchspacing =}\enspace syntax reshaped: \textbf{same meaning $\rightarrow$ ignored}}};
\draw[isoCrimson, line width=0.9pt, line cap=round] (3.95,-1.34) -- (4.01,-1.28) (3.95,-1.28) -- (4.01,-1.34);
\node[iso key, anchor=west] at (4.05,-1.31) {\textcolor{isoCrimson}{\textbf{tool-result error flag lost}} \textcolor{isoInk!85}{(\code{\frenchspacing is\_error} dropped)}};
\node[iso note, anchor=west] at (0.06,-1.64) {\code{\frenchspacing \isolb{}"role": "user", "content": "\ldots{}"\isorb{}}};
\node[iso note, anchor=west] at (1.99,-1.64) {\code{\frenchspacing \isolb{}"role": "user", "parts": [\ldots{}]\isorb{}}};
\node[iso key, anchor=west] at (3.91,-1.64) {\textcolor{isoInk!75}{\code{\frenchspacing =}\enspace \code{\frenchspacing user} text: ``continue''}};
\path[fill=isoCrimson!6, draw=isoCrimson, line width=0.65pt, rounded corners=1.8pt] (0.01,-2.49) rectangle (6.49,-1.85);
\node[iso key, anchor=west] at (0.09,-1.96) {\textbf{\textcolor{isoCrimson}{Why the lost flag is a primary finding}} \textcolor{isoInk!80}{(all criteria apply):}};
\draw[isoTeal, line width=0.9pt, line cap=round, line join=round] (0.085,-2.144) -- (0.108,-2.17) -- (0.16,-2.108);
\node[iso key, anchor=west] at (0.18,-2.14) {\textbf{information changed}};
\draw[isoTeal, line width=0.9pt, line cap=round, line join=round] (2.205,-2.144) -- (2.228,-2.17) -- (2.28,-2.108);
\node[iso key, anchor=west] at (2.3,-2.14) {\textbf{no translator warning}};
\draw[isoTeal, line width=0.9pt, line cap=round, line join=round] (4.335,-2.144) -- (4.358,-2.17) -- (4.41,-2.108);
\node[iso key, anchor=west] at (4.43,-2.14) {\textbf{passes static target checks}};
\draw[isoTeal, line width=0.9pt, line cap=round, line join=round] (0.085,-2.354) -- (0.108,-2.38) -- (0.16,-2.318);
\node[iso key, anchor=west] at (0.18,-2.35) {\textbf{target could carry it}};
\draw[isoTeal, line width=0.9pt, line cap=round, line join=round] (2.205,-2.354) -- (2.228,-2.38) -- (2.28,-2.318);
\node[iso key, anchor=west] at (2.3,-2.35) {\textbf{prompt, role or tool field}};
\node[iso key, anchor=west] at (5.25,-2.35) {$\rightarrow$\enspace\textcolor{isoCrimson}{\textbf{counts}}};
\draw[iso arrow, draw=isoCrimson, line width=0.65pt] (6.31,-1.48) -- (6.31,-1.85);
\end{tikzpicture}%
  \caption{\textbf{A dropped tool-error flag reveals avoidable information loss.}
  In this four-turn request, accepted by both live APIs, \textbf{(a)} the Anthropic source and \textbf{(b)} the captured Gemini output from AI SDK v7 use different role and field names. \textbf{(c)} isloss accounts for these representation changes but identifies the loss of \code{is\_error:\,true}. The separately labelled target-native repair is schematic: it illustrates how Gemini can represent the error, rather than reproducing the measured translator output.}
  \label{fig:ir-overview}
\end{figure}
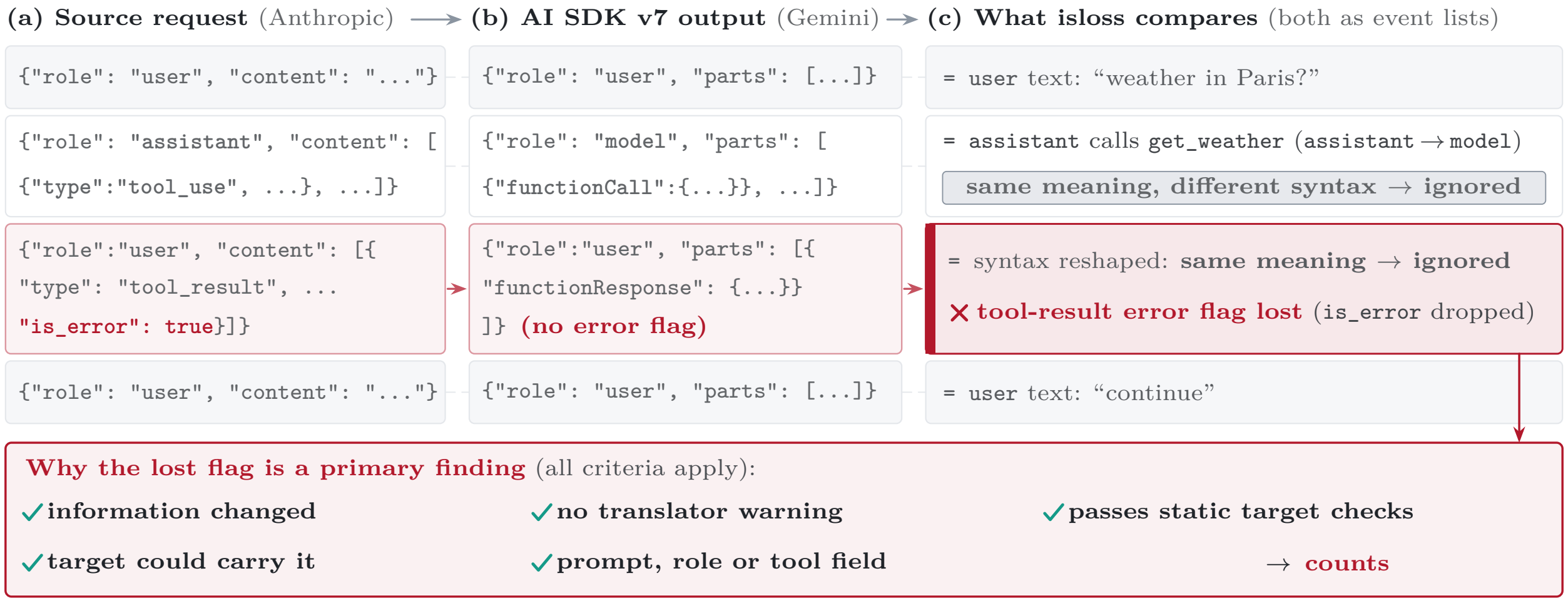

The distinction affects the interpretation of our results. In Fig.~\ref{fig:ir-overview}, dropping the tool-error flag is an avoidable, unannounced loss: Gemini can represent the error, but the tested translation does not preserve it. The change is therefore reported as \code{T1}. In the extended system-instruction experiment, deletion is also reported as \code{T1}. LiteLLM's rewrite, however, produces a \code{T2} record for demoting the system instruction to a user message, in accordance with our conservative counting policy, and a separate \code{T1} finding for the inserted operator note.

To assess the sensitivity of the primary counting policy, we also report a broader variant that includes target-representable \code{T2} degradations, including role demotion. Under this alternative policy, the result increases from 127/685 (18.5\,\%) to 134/685 (19.6\,\%) primary cells (variant e2 in Table~\ref{tab:sensitivity} in the Appendix, and Sec. \ref{sec:rq1}). This variant includes all target-representable degradations recorded as \code{T2}; it does not change the underlying translation observations. The extended mid-conversation system-instruction experiment remains separate from the main suite (Sec. \ref{sec:corpora}). Human review and behavioural trials assess the validity and consequences of these changes independently of the counting policy (see Sec. \ref{sec:precision} and \suppref{app:detail-model}{D}).

\textbf{Replies, streams, and refusals.}
The same ConvIR representation and \code{T1} preservation rules also apply to translated replies returned to the caller.
For streams, we additionally define a stream-equivalence relation~\citep{Chen1998Metamorphic,Segura2016MTSurvey}: reassembling a translated stream should preserve the corresponding non-streaming reply. This comparison uses matched representations of fixed reply content, rather than independently sampled model completions, so that differences in model generation do not confound the translation check.
When a target API cannot represent a construct, a translator may reject the request with a client-visible error or explicitly declare the degradation rather than silently changing the conversation. We record these outcomes separately from silent \code{T1} findings (\suppref{app:detail-model}{D}).

\textbf{Unspecified cross-API mappings.}
Not every cross-API mapping is fully specified by vendor documentation. Where the specifications leave a mapping open, we do not infer that every representational difference constitutes a violation. Instead, we count an information loss only when there is evidence that the target API can carry the relevant information.
To investigate such cases, we conduct separate Gemini endpoint probes that resolve 13 of the 17 previously undecided constructs (Sec. \ref{sec:invisible}; \suppref{app:specgaps}{S1.3}). These probes provide endpoint evidence for the mapping analysis; source and target acceptance are examined separately in Secs.~\ref{sec:corpora} and~\ref{sec:step4-contract}.

\subsection{Coverage of publicly reported bugs}
\label{sec:coverage}

We assess the scope of the preservation rules against 125 publicly reported translation issues collected from twelve issue trackers. The rules cover 99 issues (79.2\,\%), indicating that they capture a substantial range of previously observed translation failures. This corpus is concentrated on LiteLLM and is used to assess coverage of known issue types, not to estimate the prevalence of failures across translators. Of the 26 issues not covered, three (2.4\,\% of the corpus) describe legitimate normalisations, while the remaining 23 (18.4\,\%) fall into five categories outside the current scope of ConvIR and its rules.

The issue corpus also reveals a difference between request- and response-side failures: request-side issues are more likely to produce explicit errors, with 35\,\% reported as silent, whereas response-side issues are more often silent (62--67\,\%; \suppref{app:detail-model}{D}).
This asymmetry motivates checking both directions of translation. A translator may fail visibly when constructing a request, but silently alter information when converting a model's reply or assembling a streamed response.

\section{Checking Translations with isloss}
\label{sec:approach}\label{sec:rq-trust}

Our isloss evaluates a translator by comparing the source payload with the payload it produces, without calling a model or requiring a reference translation.
The evaluation has three stages: capturing the translator's output, applying the preservation rules from Sec.~\ref{sec:model} to identify and classify changes, and aggregating findings under explicit counting criteria. We use this procedure to study requests, replies, streams, and deployed translation paths. Checker validation and human assessment are reported separately from experiments that measure downstream behavioural effects. % (Sections~\ref{sec:rq-cons} and~\ref{sec:rq-deploy}).

\subsection{Capturing translations and identifying changes}
\label{sec:capture}

For each self-hosted translator, we exercise its documented public interface and direct its output to a local mock server. We capture the outgoing payload at the last self-hosted translation hop in Fig.~\ref{fig:topology}, before it reaches the target service. Exceptions and explicit refusals are recorded separately as fail-closed rejections, rather than being treated as completed translations with preservation findings.
The complete request suite contains 4,998 test executions; a fresh checkout reproduces their outputs byte-for-byte in 11 seconds.

The capture point is kept consistent within each translator configuration. For LiteLLM, the primary capture is its pre-serialisation HTTP-layer dictionary; we additionally inspect the serialised request body to assess sensitivity to this capture choice. The Vercel AI SDK driver supplies the source controls that its public interface can represent, avoiding the assumption that the driver can exercise controls unavailable through that interface (see \suppref{app:detail-oracle}{D}).

\textbf{Selecting the primary request findings.}\label{sec:validity} The same comparison framework is used for translated requests, reverse replies, gateway streams, and recorded agent sessions, with the relevant capture procedure for each setting.
We first describe how potential changes are reduced to the primary request result; the other settings are evaluated separately.
The main request census comprises 714 translation attempts across 51 source templates and 14 source-to-target paths. Of these, 29 are explicit rejections, leaving 685 completed translations for comparison. The initial representation comparison identifies 489 cells with at least one potential information-changing difference. Four filters for known acceptable changes reduce this to 485. Applying the T1/T2 reporting policy leaves 467 cells with at least one T1 finding. We then assess target-format validity and the nature of each finding to obtain the narrower primary result (Table~\ref{tab:decision}).

\begin{table}[t]
  \centering\small
  \caption{\textbf{Deriving the primary request result.}
  The census contains 714 attempts across 51 source templates and 14 paths; 29 are rejected, leaving 685 completed translations. Counts are cumulative: each subsequent stage retains a subset of the preceding stage. Static target validity is checked offline; actual endpoint acceptance is evaluated separately in Section~\ref{sec:step4-contract}. The final stage requires a single finding to satisfy all primary criteria.}
  \label{tab:decision}
  \begin{tabular}{@{}P{3.5cm}P{7.0cm}P{5.1cm}@{}}
    \toprule
    Stage & Criterion & Retained cells \\
    \midrule
    1. Compare representations
      & At least one potential information-changing difference
      & 489 \\
    2. Filter acceptable changes
      & Exclude four known classes of permitted change
      & 485 \\
    3. Apply reporting policy
      & Retain cells with at least one \code{T1} finding
      & 467/685 (68.18\,\%) \\
    4. Check static target validity
      & Retain cells whose translated requests pass the static target contract
      & 333/685 (48.61\,\%) \\
    5. Select primary findings
      & Require an avoidable, silent change to prompt, role, or tool information
      & \textbf{127/685 (18.54\,\%)} \\
    \bottomrule
  \end{tabular}
\end{table}

The primary result deliberately applies stricter criteria than the presence of a T1 finding alone. A finding is avoidable when the target API documents a native way to represent the affected information. When target support is uncertain, the finding is treated as borderline rather than assumed to be avoidable.
A cell contributes to the primary count only when its translated request (i) passes the static target contract, (ii) contains at least one silent \code{T1} finding, and (iii) has a finding that is both avoidable and concerns prompt, role, or tool information. Here, silent means that the translator does not warn the caller about the change. A cell is counted once even if it contains multiple qualifying findings.

Applying these criteria yields 127 of 685 completed translations (18.54\,\%; cluster-bootstrap 95\,\% CI: 15.01--22.06\,\%). Thus, these translations produce requests that pass the static target contract while silently changing information that the target API could have preserved. This is the primary rate reported in the paper. Other findings, including changes to sampling and output-limit controls, metadata, and reasoning continuity, are analysed separately (\suppref{app:detail-oracle}{D}).
Importantly, passing the static contract does not establish that a request is accepted by a live endpoint; endpoint acceptance is examined separately in Sec.~\ref{sec:step4-contract}.

\textbf{Grouping repeated findings.}
A finding is an observed rule violation in a particular translation cell. To distinguish repeated occurrences from distinct causes, we inspect the translator implementation and group findings with the same root cause, source-to-target path, and rule into a candidate. A mechanism groups the same underlying cause across different paths. This grouping allows us to report both how often a problem occurs in the test suite and how many distinct implementation mechanisms account for it. % (Section~\ref{sec:rq1}).

\textbf{Hosted translations with inaccessible intermediate requests.}
For Google's hosted compatibility endpoint, the internal translated request is inaccessible, so we cannot directly apply the payload-comparison procedure at the translation boundary. Instead, we use distinguishing conversations: targeted inputs designed so that the model's response depends on a selected conversation field. We compare responses from the compatibility endpoint against native-API controls to investigate whether the field's effect survives translation. This provides behavioural evidence about the hosted translation path, rather than a direct inspection of its internal payload. The deployment study also examines six open-source gateways (Sec.~\ref{sec:rq-deploy}; \suppref{app:detail-oracle}{D}).

\subsection{Held-out tests and additional validation}
\label{sec:blind-trust}

The checker should be evaluated independently of the examples used to develop its rules. We therefore separate early in-sample batches, which informed rule refinement, from held-out batches authored after the preservation rules were fixed. Here, ``held-out'' refers to the separation of test cases for evaluating isloss, not to model-training data.
For the held-out evaluation, LLM agents construct test pairs from vendor specifications without access to the checker implementation. Each pair is designed to exercise either a planted information-loss fault or an allowed change of representation. A planted fault contributes to the recall denominator only if its edit actually changes the translated payload; we call such a fault activated. An allowed change contributes to the false-alarm denominator if isloss incorrectly flags it as a preservation violation (Appendices~\ref{app:trust} and~\ref{app:audit}).

In the primary held-out evaluation, comprising six blind batches (three fault batches and three control batches), isloss detects 95 of 125 activated faults (76.0\,\% recall; Wilson 95\,\% CI: 67.8--82.6\,\%) and flags 2 of 100 allowed changes (2.0\,\% false-alarm rate; 95\,\% CI: 0.6--7.0\,\%). Table~\ref{tab:trust} also reports the first fault and control batches separately.
We further extend the evaluation with four later blind batches. Across the resulting extended held-out set, isloss detects 160 of 220 activated faults (72.7\,\%) and flags 5 of 170 allowed changes (2.9\,\%). Two intermediate script-generated batches are excluded before scoring; \suppref{app:detail-oracle}{D} maps the internal batch identifiers and documents the inclusion criteria.
A separate human-designed test kit yields 37 of 40 detected faults and no false alarms among 10 allowed changes.
In addition, three program-mutant suites yield 81.6--84.3\,\% detection of activated mutants. For this analysis, a mutant is counted as detected only when the corresponding finding is absent from the unmodified translator. % (\suppref{app:detail-oracle}{D}).
Finally, we evaluate coverage against publicly reported translation bugs, which tests whether the rules detect failures beyond deliberately planted changes. Across four samples, isloss detects 45 of 57 reproduced bugs (78.9\,\%; 95\,\% CI: 66.7--87.5\,\%). This includes 10 of 14 encodable bugs opened in the 50 hours after the preservation rules were fixed (71.4\,\%; 95\,\% CI: 45.4--88.3\,\%; Table~\ref{tab:trust}; \suppref{app:detail-oracle}{D}).

\begin{table}[t]
  \centering\small
  \caption{\textbf{Detection accuracy of isloss and comparison with alternative judges.}
  The primary held-out evaluation pools six specification-based blind batches authored after the checking rules were fixed; the extended evaluation adds four later blind batches (batch map: \suppref{app:detail-oracle}{D}). Intervals are
  Wilson 95\,\% CIs; $J$ is recall minus false-alarm rate, with the
  strictly cross-dialect value in parentheses. Without a reference
  translation, isloss has the highest observed $J$ among the tested
  judges. With a target-dialect reference, normalised diff and two Gemini
  judges score higher. Full results:
  \supptab{tab:trustfull}{S15} and \supptab{tab:judges}{S15a}.}
  \label{tab:trust}
  % AUTO-GENERATED by tools/gen_tables.py trust (reduced main-text form of tables/trust.tex plus held-out pool C-609, diff ladder C-610, prospective issues C-613, four-sample public bugs C-637) — do not edit
\setlength{\tabcolsep}{2.5pt}
\ifshowclaims
\begin{tabular}{@{}P{2.9cm}P{3.6cm}P{4.5cm}P{3.5cm}@{}}
\toprule
Check & Material & Faults found & Allowed changes flagged \\
\midrule
First held-out batch (faults) & 36 faults, 34 activated & 28/34 = 82.4\% (66.5--91.7\%) & --- \\
First held-out batch (controls) & 28 allowed changes, 28 activated & --- & 1/28 \\
Primary held-out evaluation & 125 activated faults, 100 allowed changes & 95/125 = 76.0\% (67.8--82.6\%) & 2/100 = 2.0\% (0.6--7.0\%) \\
Extended held-out evaluation & 220 activated faults, 170 allowed changes & 160/220 = 72.7\% & 5/170 = 2.9\% \\
New public issues & 16 within scope, 14 encodable & 10/14 = 71.4\% (45.4--88.3\%) & --- \\
Seed recall (requests) & 2,129 single-fault wires, 335 controls & 1,595/2,129 = 74.9\% (73.0--76.7\%) & 0/335 \\
Public bugs, 4 samples & 57 reproduced bugs & 45/57 (66.7--87.5\%) & --- \\
\midrule\emph{Judge, pooled pairs} & \emph{Setting} &  &  \\
isloss & no reference or model call; $J$~0.740 (0.729) & 95/125 & 2/100\\
Normalised diff & with reference & 124/125 & 15/100\\
flash-lite & with reference; 633\,ms & 108/125 & 6/100\\
3.8-flash & with reference; 2,352\,ms & 120/125 & 12/100\\
flash-lite & no reference; $J$~0.516 (0.447) & 72/125 & 6/100\\
3.8-flash & no reference; $J$~0.704 (0.678) & 103/125 & 12/100\\
\bottomrule
\end{tabular}
\else
\begin{tabular}{@{}P{3.3cm}P{4.1cm}P{4.8cm}P{3.6cm}@{}}
\toprule
Check & Material & Faults found & Allowed changes flagged \\
\midrule
First held-out batch (faults) & 36 faults, 34 activated & 28/34 = 82.4\% (66.5--91.7\%) & --- \\
First held-out batch (controls) & 28 allowed changes, 28 activated & --- & 1/28 \\
Primary held-out evaluation & 125 activated faults, 100 allowed changes & 95/125 = 76.0\% (67.8--82.6\%) & 2/100 = 2.0\% (0.6--7.0\%) \\
Extended held-out evaluation & 220 activated faults, 170 allowed changes & 160/220 = 72.7\% & 5/170 = 2.9\% \\
New public issues & 16 within scope, 14 encodable & 10/14 = 71.4\% (45.4--88.3\%) & --- \\
Seed recall (requests) & 2,129 single-fault wires, 335 controls & 1,595/2,129 = 74.9\% (73.0--76.7\%) & 0/335 \\
Public bugs, 4 samples & 57 reproduced bugs & 45/57 (66.7--87.5\%) & --- \\
\midrule\emph{Judge, pooled pairs} & \emph{Setting} &  &  \\
isloss & no reference or model call; $J$~0.740 (0.729) & 95/125 & 2/100 \\
Normalised diff & with reference & 124/125 & 15/100 \\
flash-lite & with reference; 633\,ms & 108/125 & 6/100 \\
3.8-flash & with reference; 2,352\,ms & 120/125 & 12/100 \\
flash-lite & no reference; $J$~0.516 (0.447) & 72/125 & 6/100 \\
3.8-flash & no reference; $J$~0.704 (0.678) & 103/125 & 12/100 \\
\bottomrule
\end{tabular}
\fi

\end{table}

\subsection{Human validation and comparison with existing checks}
\label{sec:whynot}\label{sec:precision}

\textbf{How many detected changes are genuine defects?}
We assess the validity of the detected request-side candidates through independent human review. Reviewers classify each of the 54 candidates into three categories: Grade~A, a confirmed defect; Grade~B, an acceptable change of representation; or Grade~C, a disputed case. Grades~A and~B require agreement between both raters; disagreements are assigned Grade~C.

Overall, reviewers confirm 31 of the 54 candidates as defects, accept 11 as legitimate representation changes, and disagree on the remaining 12. The distinction between avoidable and unavoidable losses is particularly informative. Among the 28 avoidable candidates, 23 are confirmed defects, one is accepted as a form change, and four remain disputed. Thus, 23/28 (82.1\,\%) are confirmed when all candidates are retained; among the 24 candidates with an agreed classification, 23/24 (95.8\,\%; 95\,\% CI: 79.8--99.3\,\%) are confirmed defects. By comparison, only 6 of 16 agreed cases involving unavoidable losses are confirmed as defects (37.5\,\%).
This difference suggests that human judgments are more consistent when the target API could have preserved the information than when the loss arises from a target-format limitation. It supports treating avoidable information loss as the more conservative and reproducible basis for the primary findings.

We further examine individual findings through two separate cell-level audits. Reviewers confirm 17/20 (85.0\,\%) and 67/70 (95.7\,\%) statically accepted avoidable findings, respectively, with no false alarms among 45 clean or decoy controls.
A 48-item bridging sample, scored by all four raters, further exposes the sensitivity of human judgment: the two reviewer pairs differ by 37.50 percentage points on the same unavoidable-loss items. This disagreement reinforces the need to report target-limited degradations separately rather than treating every potential information loss as an equally clear-cut defect (Section~\ref{sec:threats}; \suppref{app:detail-oracle}{D}).

\textbf{What does isloss provide beyond existing checks?}
Existing validation techniques address different aspects of translation correctness, but do not directly establish whether the information in a source conversation survives translation. Schema validation checks whether the translated request conforms to the target API's structural requirements; it does not compare the request's protected information with the source. Round-trip testing can expose changes by translating a payload back to its original dialect, but requires an inverse translator, which is unavailable for 10 of our 14 evaluated paths.

Normalised diffs and LLM judges offer another comparison strategy when a reference translation is available. On our held-out evaluation pairs, a normalised diff and two Gemini judges achieve $J=0.842$, $0.804$, and $0.840$, respectively, outperforming isloss at $J=0.740$. However, these results use a reference translation against which the candidate translation can be compared. Obtaining such a reference is precisely the difficulty when validating previously unseen inputs or translators for which no trusted translation is available.

We therefore also evaluate the alternatives in the more practical setting where only the source and its actual translation are available. Without a reference translation, the two Gemini judges achieve $J=0.516$ and $J=0.704$, both below isloss. The tested Claude and open-weight judges also score below isloss in this setting.
The comparison highlights the central advantage of isloss: it provides a deterministic, rule-based check of specified conversation properties directly from the source and translated payloads, without requiring a reference translation or an additional model call (\suppref{app:detail-oracle}{D}).

\begin{table}[!t]
  \centering
  \caption{\textbf{Units and denominators for the main results.} A \emph{cell} is one request or reply tested on one translation path. Each row identifies the unit, denominator, subjects, and location of a result. The full map, including the corpus and supporting live or human evidence, is \supptab{tab:units-map-full}{D} (\suppref{app:detail-design}{D}).}
  \label{tab:units-map}
  \label{tab:eval-overview}
  {\footnotesize
\setlength{\tabcolsep}{3.4pt}
\renewcommand{\arraystretch}{1.12}
\begin{tabular}{@{}P{4.35cm}P{3.85cm}P{2.2cm}P{1.45cm}P{2.45cm}l@{}}
\toprule
\textbf{What we measure} & \textbf{Value} & \textbf{Unit} & \textbf{$N$} & \textbf{Tested on} & \textbf{Sec.} \\
\midrule
\multicolumn{6}{@{}l}{\textit{Translated requests}} \\
Primary rate, full synthetic suite (Stratum~1) & 127/685 (18.5\,\%) & cell & 685 & 6 configs, 14 paths & \mbox{\S\ref{sec:rq1}} \\
Same rate, shapes \code{claude-sonnet-4-6} accepts live (Stratum~2a) & 34/182 (18.7\,\%) & cell & 182 & 6 configs, 14 paths & \mbox{\S\ref{sec:rq1}} \\
Same rate, shapes \code{claude-opus-4-8} accepts live (Stratum~2b) & 23/112 (20.5\,\%) & cell & 112 & 6 configs, 14 paths & \mbox{\S\ref{sec:rq1}} \\
Range of the primary rate under other rules & 12.9--19.6\,\% (validity rules); 16.5--25.7\,\% (counting conventions) & cell & 308--685 & 4--6 configs & \mbox{\S\ref{sec:rq1}} \\
Reviewed candidates & 54 candidates (31 confirmed defects) across 18 mechanisms & (translator, rule) pair & 54 & 5 codebases & \mbox{\S\ref{sec:rq1}} \\
Live target acceptance after static checks & Gemini 160/204 (78.4\,\%); Anthropic 77/204 (37.7\,\%) & statically accepted cell & 204 per API & 8 live paths & \mbox{\S\ref{sec:step4-contract}} \\
Mid-turn \code{system} suite: meaning changed & 1,139/1,530 (74.4\,\%) & cell & 1,530 & 6 configs & \mbox{\S\ref{app:detail-motivating}} \\
\midrule
\multicolumn{6}{@{}l}{\textit{Consequences, replies and persistence}} \\
Mechanisms with a measured live effect & 9 of 10 tested (18 catalogued) & mechanism & 10 tested & targeted probes; 30 or 10 trials per arm & \mbox{\S\ref{sec:rq-cons}} \\
GSM8K control (separate experiment) & no statistically detectable accuracy drop & call & 700 & 100 items, 7 arms, one Gemini model & \mbox{\S\ref{sec:rq-cons}} \\
Replies whose meaning changes & 391/1,115 (35.1\,\%) & reply & 1,115 & 7 translators & \mbox{\S\ref{sec:rq-replies}} \\
Reply streams that fail in an official OpenAI SDK & 203/558 (36.4\,\%) & stream & 558 & 7 translators & \mbox{\S\ref{sec:rq-replies}} \\
Persistence at the source snapshot (Step~2) & 2,630/2,745 (95.8\,\%) & previously flagged cell & 2,745 & 170 shapes; 10 translators, 22 paths & \mbox{\S\ref{sec:rq-persist}} \\
Our upstream issues fixed & 7 of 26 & issue & 26 & --- & \mbox{\S\ref{sec:rq-persist}} \\
\midrule
\multicolumn{6}{@{}l}{\textit{Deployment and isloss validation}} \\
Gateways change the request's meaning & 187/204 (91.7\,\%) on 51 shapes; 978/1,020 (95.9\,\%) on 170 shapes & request & 204; 1,020 & Bifrost, new-api (and CLIProxyAPI on 170 shapes) & \mbox{\S\ref{sec:rq-deploy}} \\
Google's OpenAI-compatible layer keeps the conversation & 10 of 11 & distinguishing probe (10 live repeats) & 11 & live Gemini, vs.\ native API & \mbox{\S\ref{sec:rq-deploy}} \\
Recorded coding-agent requests: primary rate & 4/98 (4.1\,\%) contract-valid; 13/196 (6.6\,\%) all & cell & 98; 196 & 6 configs, 14 paths, 3 CLI agents & \mbox{\S\ref{sec:rq-deploy}} \\
isloss detects planted faults (blind) & 95/125 (76.0\,\%); 160/220 (72.7\,\%) pooled & planted fault & 125; 220 & all 4 dialect pairs & \mbox{\S\ref{sec:rq-trust}} \\
\bottomrule
\end{tabular}
}

\end{table}

\section{Study Design}
\label{sec:evaluation}

\subsection{Translators, translation paths, and configurations}
\label{sec:subjects}

Our main request census examines 14 source-to-target translation paths across five open-source translator codebases, represented by six configurations: LiteLLM, Vercel AI SDK v5 and v7, LLM-Rosetta, \code{@musistudio/llms}, and \code{claude-code-proxy}. We exercise each path using the 51 source-valid request shapes described below, yielding 714 translation attempts. Twenty-nine attempts are explicitly rejected, leaving 685 completed translations for analysis.

We conduct additional studies to examine translation behaviour beyond this main census. These include self-hosted Portkey, Bifrost, new-api, and one-api; gateway paths in \code{ai-gateway} and CLIProxyAPI; Google's hosted OpenAI-compatible Gemini endpoint; and the official Python and Node OpenAI SDKs for evaluating translated reply streams.
These additional systems are evaluated in their respective experiments and are not added to the 14-path main request census. To support reproducibility, we record pinned versions, commit identifiers, default configuration flags, and documented entry points for every tested system (\suppref{app:detail-design}{D}).

\subsection{Request corpus and source acceptance}
\label{sec:generation}
\label{sec:corpora}

We construct the main synthetic request suite from Anthropic requests. We first generate 357 unique non-streaming request shapes using a four-way covering array~\citep{Cohen1997AETG,Kuhn2004Interactions,Nie2011CTSurvey} over 14 feature dimensions. We then apply static source-contract checks and retain the 51 shapes that satisfy the documented source requirements. These 51 shapes form the main request suite used in the 14-path census.
The covering-array construction describes the generated set before filtering; we do not assume that the source-valid subset retains the full four-way coverage guarantee.

Static source validity does not imply that a live model endpoint accepts a request. When tested against \code{claude-sonnet-4-6}, only 13 of the 51 shapes (25.5\,\%) are accepted, while 38 (74.5\,\%) are rejected.
These rejections include placeholder redacted thinking blocks and requests that specify both \code{temperature} and \code{top\_p}~\citep{doc-anthropic-release-notes,gh-psycheros-issue67}. The separate \code{claude-opus-4-8} acceptance test accepts eight of the 51 shapes (\suppref{app:detail-design}{D}).
We retain the full statically valid suite for the main translation census, and separately report results restricted to shapes accepted by live source endpoints.

Two additional datasets address settings not represented by the main corpus. First, the mid-conversation \code{system} instruction experiment uses a separate 170-shape suite and behavioural study. Its requests are not included in the main 51-shape suite. %  (Figure~\ref{fig:motivating}; Sections~\ref{sec:case} and~\ref{sec:rq-cons}).
Second, to examine naturally occurring agent traffic, we capture 36 requests (28 unique) from six CLI agents, including Claude Code, Crush, and Aider. These requests yield 14 distinct Anthropic shapes, of which seven pass the static source checks.
Sec. \ref{sec:rq-deploy} reports results on the seven source-valid shapes (98 translation cells) and on all 14 observed shapes (196 cells). The full feature coverage and dataset construction are provided in \suppref{app:detail-design}{D}.

\subsection{Static validity and live endpoint acceptance}
\label{sec:step4-contract}

We distinguish three properties that are easy to conflate. Static validity means that a translated request passes our offline checks of the target API's structure and documented constraints. Live acceptance means that the corresponding vendor endpoint accepts the request and returns HTTP~200. Silence describes whether the translator warns the caller about a detected change. Warning behaviour is assessed independently of both static validity and endpoint acceptance.

The 27 target-contract validators retain 333 of the 685 completed translations (48.6\,\%) at Step~4 of the primary counting procedure. Step~5 then applies the finding-level criteria defined in Sec.~\ref{sec:validity}; static validity alone does not make a translation a primary finding.
We further test live endpoint acceptance for statically accepted requests. Among 204 cells tested against each target, Gemini accepts 160 (78.4\,\%) and rejects 44 because of missing Gemini~3 \code{thought\_signature} values. Anthropic accepts 77 of 204 cells (37.7\,\%; see \suppref{app:detail-design}{D}).

Because source and target endpoints accept different subsets, we report several acceptance-based result strata alongside the full synthetic-suite result (see Fig~\ref{fig:funnel}; Table~\ref{tab:tiers-full} in the Appendix). These strata answer different questions and should not be interpreted as successive filters of one common population:

\begin{itemize}
  \item \textbf{Stratum 1: full synthetic suite.} All 685 completed translations are included, yielding 127 primary findings (18.5\,\%). This is the main result and does not require live source acceptance.

  \item \textbf{Strata 2a and 2b: source-accepted subsets.} Restricting the source requests to those accepted by \code{claude-sonnet-4-6} gives 13 shapes and 182 translation cells (Stratum~2a). Using the eight shapes accepted by \code{claude-opus-4-8} gives 112 cells (Stratum~2b). These are separate subsets determined by the respective source models, not successive filters.

  \item \textbf{Stratum 2c: acceptance at both ends.} Requiring acceptance by both the source and target endpoints leaves 88 cells across the Gemini- and Anthropic-target paths. OpenAI-bound cells are not included because they were not submitted to a live target endpoint.

  \item \textbf{Stratum 3: demonstrated mechanism coverage.} We identify 240 of the 685 main-suite cells as carrying a mechanism for which a live behavioural effect has been demonstrated. This is a mechanism-level coverage analysis: it does not mean that all 240 translations were individually replayed against a live model.
\end{itemize}

These strata provide complementary views of the findings under different acceptance and evidence conditions. Table~\ref{tab:units-map} and \suppref{app:detail-design}{D} specify the population, unit of analysis, and denominator for each result, including deduplicated counts.

\subsection{Units of analysis and statistical methods}
\label{sec:units}

The experiments involve different units of analysis, including generated request shapes, translation attempts, completed translation cells, grouped candidates, underlying mechanisms, reply streams, and recorded agent requests. These units are not interchangeable. Table~\ref{tab:units-map} maps each main result in Sec.~\ref{sec:results} to its unit and denominator; the full accounting is provided in \supptab{tab:units-map-full}{D} in the Appendix.

We report 95\,\% confidence intervals using either a bootstrap clustered by the 51 source request shapes (10,000 resamples) or Wilson intervals~\citep{Wilson1927}, as specified for each result. For live comparisons, we use Fisher's exact tests~\citep{Fisher1922} or sign-flip permutation tests, with Holm correction for multiple comparisons~\citep{Holm1979}.
For transparency, each statistical criterion is labelled according to whether it was pre-specified in an internal commit, fixed before the corresponding run, or determined post hoc.

\begin{figure}[t]
\centering
\input{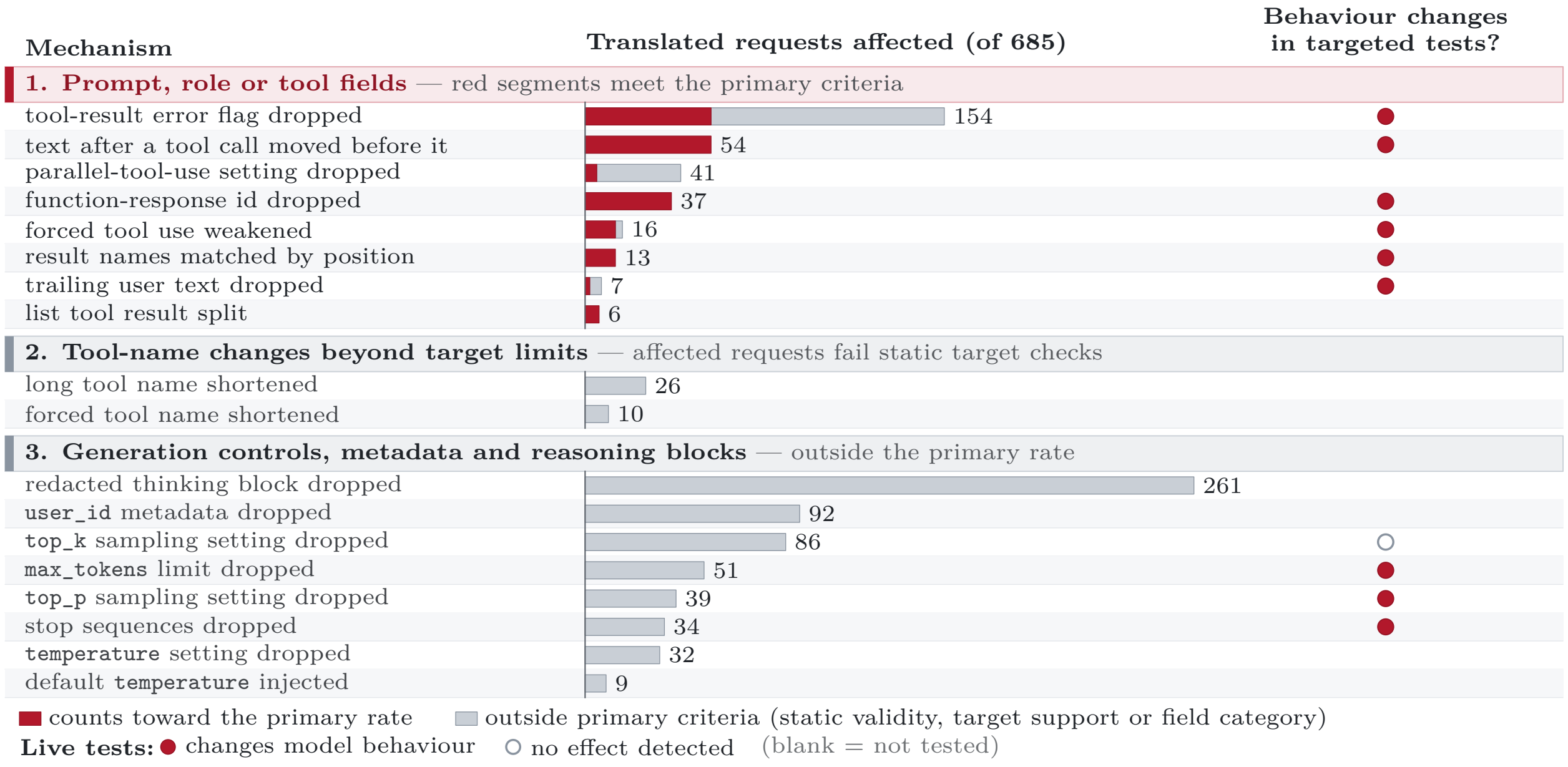}
\caption{\textbf{Request changes grouped by affected information and target support.} Bars show affected cells out of 685; a cell may carry several mechanisms. Red marks the changes included in the primary rate. Sampling and output-limit controls and other categories outside the primary rate may also affect behaviour (Sec.~\ref{sec:rq-cons}). %nocheck
Libraries and target APIs are in Fig.~\ref{fig:mech-grid} in the Appendix, code classes and upstream status in Table~\ref{tab:mechanisms} in the Appendix.}
\label{fig:mechanisms}
\end{figure}

\begin{figure}[t]
  \centering
  % rate_attrition.tex — manuscript-synchronised TikZ; source data unchanged.
% Drawn at final size (\textwidth = 6.5 in, no \resizebox) in paper/figures/style/isostyle.tex.
% Data: paper/figdata/rate_attrition.csv sha256=c42549b8b4132ecbb37d4c358212425b6de8ed61d3f3ff58cf85917b91ec5a82
% Claims: C-592, C-620, C-644, C-700, C-701, C-703, C-704, C-705, C-706, C-707, C-708, C-1024, C-1025, C-1026, C-1028, C-1034, C-1063, C-1083, C-1084, C-1086, C-1093
\begin{tikzpicture}[x=1in, y=1in, iso]
\path (0,0) -- (6.5,0); % pins the bounding box to the full text width
\node[iso r bold, anchor=mid west] at (0,1.6) {\textbf{(a) From translations to primary findings}};
\node[iso note, anchor=mid west] at (0,1.43) {one cell = one source request on one translation path};
\draw[iso axis line] (2.31,0.02) -- (2.31,1.3);
\node[iso key, anchor=mid west] at (0,1.18) {Completed translations};
\path[iso hbar] (2.31,1.1125) rectangle (2.83,1.2475);
\node[iso key, anchor=mid west] at (2.88,1.18) {685};
\node[iso key, anchor=mid west] at (0.02,0.925) {\dots\,undeclared changes (T1)};
\path[iso hbar] (2.31,0.8575) rectangle (2.6645,0.9925);
\node[iso key, anchor=mid west] at (2.7145,0.925) {467};
\node[iso key, anchor=mid west] at (0.02,0.67) {\dots\,and passes static target checks};
\path[iso hbar] (2.31,0.6025) rectangle (2.5628,0.7375);
\node[iso key, anchor=mid west] at (2.6128,0.67) {333};
\node[iso key, anchor=mid west] at (0.02,0.415) {\dots\,and has an avoidable finding};
\path[iso hbar] (2.31,0.3475) rectangle (2.4664,0.4825);
\node[iso key, anchor=mid west] at (2.5164,0.415) {206};
\node[iso key, text=isoCrimson, anchor=mid west, align=left] at (0.02,0.16) {\shortstack[l]{\textbf{\dots\,one finding meets all primary criteria}\\[-1pt]\textbf{(avoidable + prompt, role or tool field)}}};
\path[iso hbar accent] (2.31,0.0925) rectangle (2.4064,0.2275);
\node[iso key, text=isoCrimson, anchor=mid west] at (2.4564,0.16) {\textbf{127}};
\draw[iso axis line] (2.31,0.02) -- (2.83,0.02);
\draw[iso tick] (2.31,0.02) -- ++(0,-2.2pt);
\node[iso ticklabel, anchor=north] at ([yshift=-3.8000000000000003pt]2.31,0.02) {0};
\draw[iso tick] (2.5377,0.02) -- ++(0,-2.2pt);
\node[iso ticklabel, anchor=north] at ([yshift=-3.8000000000000003pt]2.5377,0.02) {300};
\draw[iso tick] (2.83,0.02) -- ++(0,-2.2pt);
\node[iso ticklabel, anchor=north] at ([yshift=-3.8000000000000003pt]2.83,0.02) {685};
\node[iso axislabel, anchor=north] at (2.57,-0.2) {Translated requests};
\node[iso note, anchor=north west, align=left, text width=3.0in] at (0,-0.43) {The final step aligns all criteria within the same finding; separate findings in one cell do not suffice.};
\node[iso r bold, anchor=mid west] at (3.22,1.6) {\textbf{(b) Does the primary rate hold?}};
\node[iso note, anchor=mid west] at (3.22,1.43) {primary rate on source-accepted inputs and under other rules};
\draw[iso rule] (3.22,0.5425) -- (6.46,0.5425);
\draw[line width=0.55pt, draw=isoCrimson!75, densely dashed] (5.6387,0.02) -- (5.6387,1.3);
\draw[iso axis line] (5.25,0.02) -- (5.25,1.3);
\node[iso key, text=isoCrimson, anchor=mid west] at (3.22,1.18) {\textbf{All 51 synthetic shapes}};
\draw[iso whisker, line width=0.6pt, draw=isoInk] (5.5647,1.18) -- (5.7125,1.18) (5.5647,1.18) ++(0,-\isocap) -- ++(0,2\isocap) (5.7125,1.18) ++(0,-\isocap) -- ++(0,2\isocap);
\node[iso dot harm] at (5.6387,1.18) {};
\node[iso key, text=isoCrimson, anchor=mid west] at (5.7525,1.18) {\textbf{18.5\,\%}};
\node[iso key, anchor=mid west] at (3.22,0.925) {13 shapes \code{claude-sonnet-4-6} accepts live};
\draw[iso whisker, line width=0.6pt, draw=isoInk] (5.492,0.925) -- (5.7914,0.925) (5.492,0.925) ++(0,-\isocap) -- ++(0,2\isocap) (5.7914,0.925) ++(0,-\isocap) -- ++(0,2\isocap);
\node[iso dot harm] at (5.6417,0.925) {};
\node[iso key, anchor=mid west] at (5.8314,0.925) {18.7\,\%};
\node[iso key, anchor=mid west] at (3.22,0.67) {8 shapes \code{claude-opus-4-8} accepts live};
\draw[iso whisker, line width=0.6pt, draw=isoInk] (5.4746,0.67) -- (5.8866,0.67) (5.4746,0.67) ++(0,-\isocap) -- ++(0,2\isocap) (5.8866,0.67) ++(0,-\isocap) -- ++(0,2\isocap);
\node[iso dot harm] at (5.6806,0.67) {};
\node[iso key, anchor=mid west] at (5.9266,0.67) {20.5\,\%};
\node[iso key, anchor=mid west] at (3.22,0.415) {Other validity rules};
\path[fill=isoSlateLight, draw=isoSlate, line width=0.55pt] (5.5211,0.3625) rectangle (5.6601,0.4675);
\node[iso key, anchor=mid west] at (5.7001,0.415) {12.9--19.6\,\%};
\node[iso key, anchor=mid west] at (3.22,0.16) {Other counting conventions};
\path[fill=isoSlateLight, draw=isoSlate, line width=0.55pt] (5.5968,0.1075) rectangle (5.7887,0.2125);
\node[iso key, anchor=mid west] at (5.8287,0.16) {16.5--25.7\,\%};
\draw[iso axis line] (5.25,0.02) -- (5.9,0.02);
\draw[iso tick] (5.25,0.02) -- ++(0,-2.2pt);
\node[iso ticklabel, anchor=north] at ([yshift=-3.8000000000000003pt]5.25,0.02) {0};
\draw[iso tick] (5.5645,0.02) -- ++(0,-2.2pt);
\node[iso ticklabel, anchor=north] at ([yshift=-3.8000000000000003pt]5.5645,0.02) {15};
\draw[iso tick] (5.879,0.02) -- ++(0,-2.2pt);
\node[iso ticklabel, anchor=north] at ([yshift=-3.8000000000000003pt]5.879,0.02) {30\,\%};
\node[iso axislabel, anchor=north] at (5.575,-0.2) {Primary rate (\% of requests)};
\end{tikzpicture}%
  \caption{\textbf{Primary findings in the synthetic suite and endpoint-accepted subsets.} Panel~(a) selects 127 of 685 judged cells (18.5\,\%); panel~(b) compares synthetic-suite rates under endpoint acceptance and alternative validity and counting rules. Whiskers show shape cluster-bootstrap 95\,\% CIs (Tables~\ref{tab:sensitivity} and~\ref{tab:tiers-full} in the Appendix).}
  \label{fig:funnel}
\end{figure}
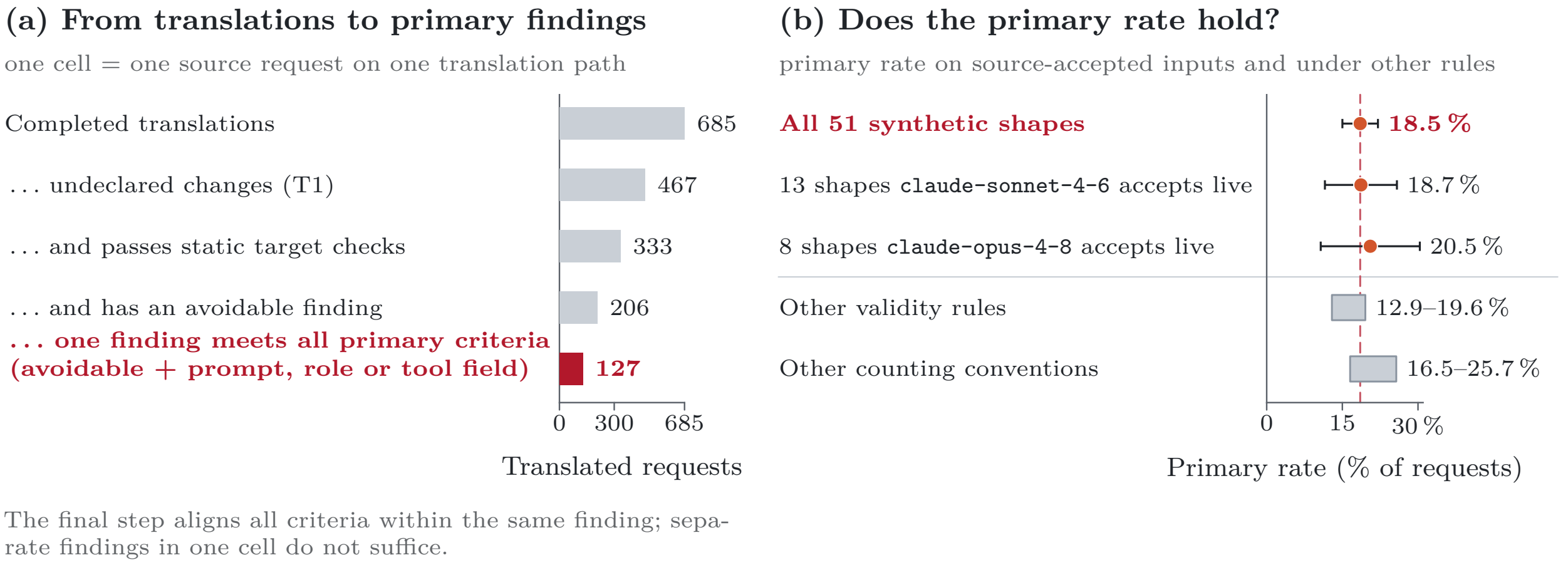

\begin{table}[t]
  \centering\small
  \caption{\textbf{Meaning changes in the main request census.} All 51 source shapes pass static checks; 13 are accepted live by \code{claude-sonnet-4-6}. \code{T1/judged} counts cells with undeclared changes before target-validity filtering; \code{Static} counts those passing offline target checks. \code{Cand.} gives pre-review candidate counts; the reviewed catalogue is reported in Section~\ref{sec:rq1}. For \code{LiteLLM->OpenAI}, the pre-serialisation dictionary has 37 \code{T1} cells that fail the target checks: all 37 retain \code{thinking\_blocks}, and 12 also retain \code{top\_k}. At the serialised \code{hosted\_vllm} body, which strips \code{thinking\_blocks}, 25/37 pass.}
  \label{tab:census}
  % AUTO-GENERATED by tools/gen_tables.py census_r9() from artifacts/g4/census_recount_r2/out/recount_tables_v3_2026-10-12T0845Z.json, artifacts/g4/census_recount_r2/out/mechanisms_r2_2026-10-12T0900Z.json, artifacts/g4/census_ledger/out/ledger_r3_20261011T2256Z.json, artifacts/g4/census_recount_r2/out/representability_r2_2026-10-12T0901Z.json, artifacts/g4/precision_review_v3/sealed/key.json, artifacts/g4/human_returns_2026-10-14/returns/part1_U3c/ratings_A_r1.csv, artifacts/g4/human_returns_2026-10-14/returns/part1_U3c/ratings_A_r2.csv, artifacts/g4/human_returns_2026-10-14/out/results_u3c_2026-10-14T2302Z.json on 2026-10-25 05:15 UTC — do not edit
\setlength{\tabcolsep}{4pt}
\begin{tabular}{@{}lrrrrrrr@{}}
\toprule
 & \multicolumn{2}{c}{$\to$OpenAI} & \multicolumn{2}{c}{$\to$Gemini} & \multicolumn{2}{c}{$\to$Anthropic} & \\
\cmidrule(lr){2-3}\cmidrule(lr){4-5}\cmidrule(lr){6-7}
Translator & T1/judged & Static & T1/judged & Static & T1/judged & Static & Cand.\ (pre) \\
\midrule
LiteLLM & 37/51 & 0 & -- & -- & -- & -- & 4 \\
claude-code-proxy & 21/22 & 7 & -- & -- & -- & -- & 7 \\
LLM-Rosetta & 42/51 & 22 & 42/51 & 42 & 36/51 & 36 & 10 \\
@musistudio/llms & 44/51 & 17 & 51/51 & 51 & 0/51 & 0 & 20 \\
AI SDK v5 & 39/51 & 21 & 40/51 & 40 & 18/51 & 18 & 7 \\
AI SDK v7 & 39/51 & 21 & 40/51 & 40 & 18/51 & 18 & 8 \\
\midrule
\textit{14 library paths (pre-review)} & 222/277 & 88 & 173/204 & 173 & 72/204 & 72 & 56 \\
\multicolumn{7}{@{}l}{\quad\textit{after review: one acceptable form change removed, one duplicate pair merged}} & 54 \\
\midrule
Bifrost (gateway) & 44/51 & 4 & 48/51 & 48 & -- & -- & -- \\
new-api (gateway) & 46/51 & 2 & 49/51 & 49 & -- & -- & -- \\
\bottomrule
\end{tabular}
\par\smallskip
{\footnotesize Static counts T1 cells whose translated bodies pass offline target checks. It does not mean live acceptance; silence means no translator warning. Candidate and T1 counts above are pre-review.}

\end{table}

\begin{figure}[t]
  \centering
  \input{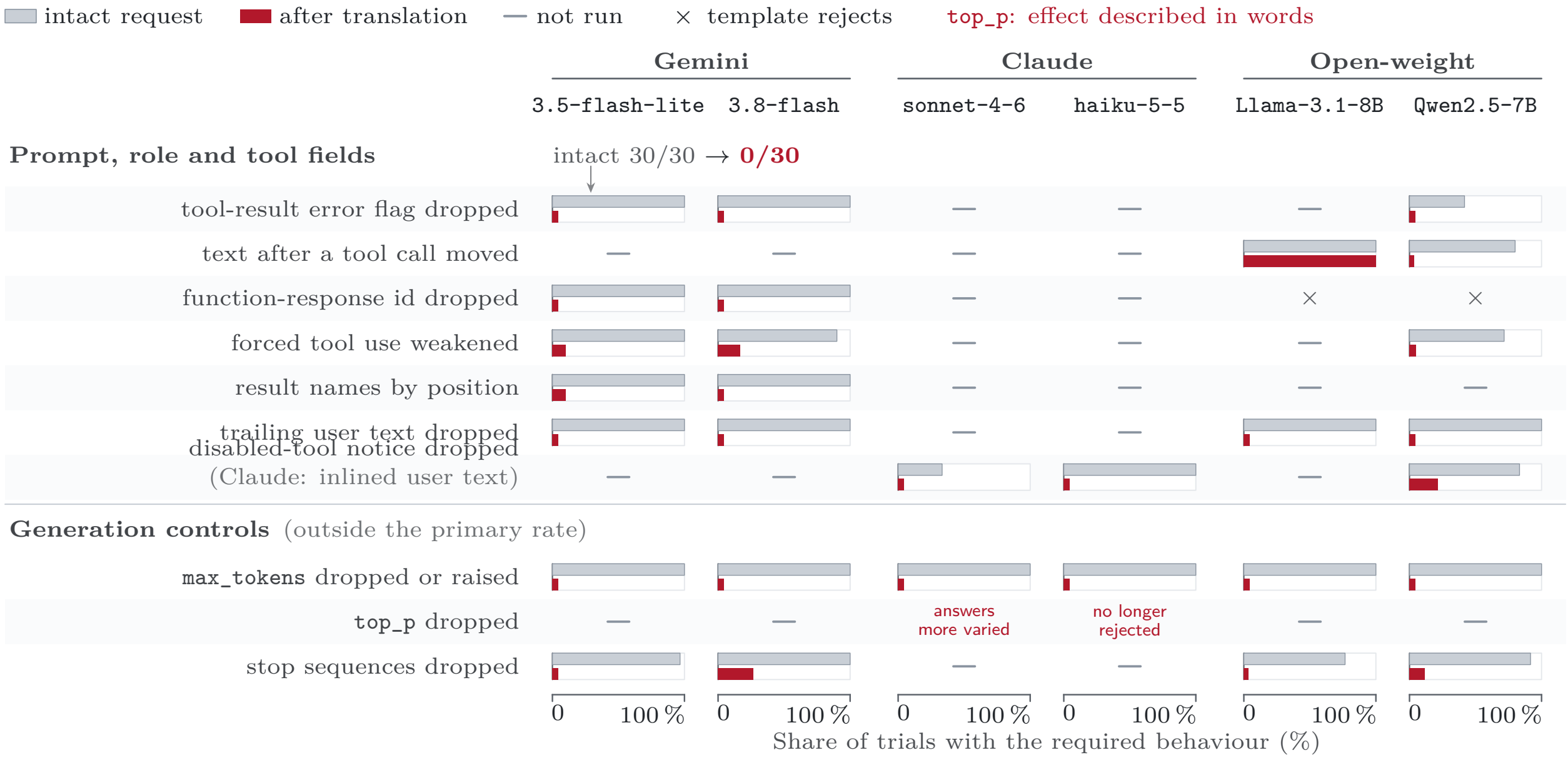}
  \caption{\textbf{Behavioural outcomes in targeted live probes.} Bars compare intact (grey) and translated (red) requests. Dashes denote untested combinations; only three displayed mechanisms were tested on Claude. The moved-text probe shows no difference on \code{Llama-3.1-8B}, and the dropped-prohibition probe misses its pre-specified threshold on \code{claude-sonnet-4-6} but meets it on \code{claude-haiku-5-5}. The \code{top\_p} effects are stated in words. The additional \code{top\_k} probes show no difference in the tested settings (counts and Wilson $95\,\%$ CIs in Tables~\ref{tab:claudeconseq}--\ref{tab:consequence-detail}) in the Appendix.} 
  \label{fig:consequence}
\end{figure}

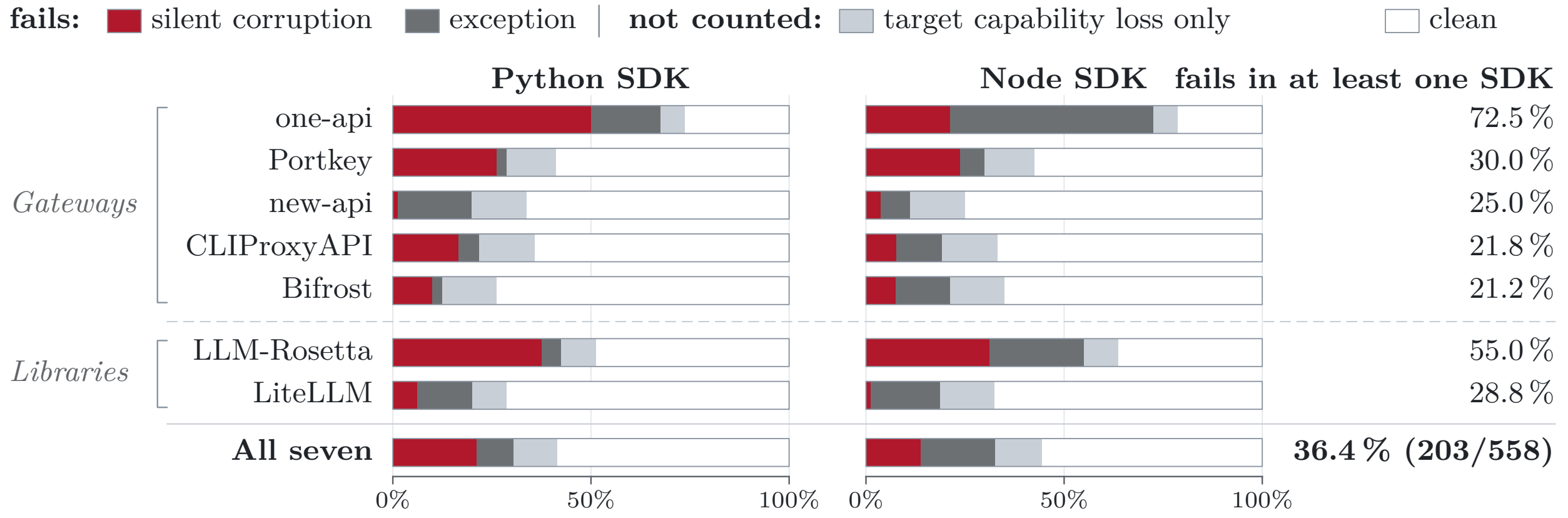
\begin{figure}[t]
  \centering
  % sdk_streams.tex — manuscript-synchronised TikZ; source data unchanged.
% Drawn at final size (\textwidth = 6.5 in, no \resizebox) in paper/figures/style/isostyle.tex.
% Data: paper/figdata/sdk_streams.csv sha256=527e77b52558aa0ed774622c4eb9eab71dfa11df5ebc8e6dc80b24cbd4589ffe
% Claims: C-530, C-648, C-709, C-710, C-933
\begin{tikzpicture}[x=1in, y=1in, iso]
\path (0,0) -- (6.5,0); % pins the bounding box to the full text width
\draw[iso grid] (1.62,-0.126) -- (1.62,1.46);
\draw[iso grid] (2.445,-0.126) -- (2.445,1.46);
\draw[iso grid] (3.27,-0.126) -- (3.27,1.46);
\draw[iso axis line] (1.62,-0.126) -- (3.27,-0.126);
\draw[iso tick] (1.62,-0.126) -- ++(0,-2.2pt);
\node[iso ticklabel, anchor=north] at ([yshift=-3.8000000000000003pt]1.62,-0.126) {0\%};
\draw[iso tick] (2.445,-0.126) -- ++(0,-2.2pt);
\node[iso ticklabel, anchor=north] at ([yshift=-3.8000000000000003pt]2.445,-0.126) {50\%};
\draw[iso tick] (3.27,-0.126) -- ++(0,-2.2pt);
\node[iso ticklabel, anchor=north] at ([yshift=-3.8000000000000003pt]3.27,-0.126) {100\%};
\draw[iso grid] (3.59,-0.126) -- (3.59,1.46);
\draw[iso grid] (4.415,-0.126) -- (4.415,1.46);
\draw[iso grid] (5.24,-0.126) -- (5.24,1.46);
\draw[iso axis line] (3.59,-0.126) -- (5.24,-0.126);
\draw[iso tick] (3.59,-0.126) -- ++(0,-2.2pt);
\node[iso ticklabel, anchor=north] at ([yshift=-3.8000000000000003pt]3.59,-0.126) {0\%};
\draw[iso tick] (4.415,-0.126) -- ++(0,-2.2pt);
\node[iso ticklabel, anchor=north] at ([yshift=-3.8000000000000003pt]4.415,-0.126) {50\%};
\draw[iso tick] (5.24,-0.126) -- ++(0,-2.2pt);
\node[iso ticklabel, anchor=north] at ([yshift=-3.8000000000000003pt]5.24,-0.126) {100\%};
\node[iso r bold, anchor=mid] at (2.445,1.52) {Python SDK};
\node[iso r bold, anchor=mid] at (4.415,1.52) {Node SDK};
\node[iso r bold, anchor=mid east] at (6.46,1.52) {fails in at least one SDK};
\node[iso r bold, anchor=mid west] at (0.02,1.77) {fails:};
\path[fill=isoCrimson, draw=isoSlate, line width=0.35pt] (0.4321,1.722) rectangle (0.5721,1.818);
\node[iso r, anchor=mid west] at (0.6121,1.77) {silent corruption};
\path[fill=isoInk!65, draw=isoSlate, line width=0.35pt] (1.6731,1.722) rectangle (1.8131,1.818);
\node[iso r, anchor=mid west] at (1.8531,1.77) {exception};
\draw[line width=0.5pt, draw=isoSlate] (2.4785,1.705) -- (2.4785,1.835);
\node[iso r bold, anchor=mid west] at (2.5985,1.77) {not counted:};
\path[fill=isoSlateLight, draw=isoSlate, line width=0.35pt] (3.48,1.722) rectangle (3.62,1.818);
\node[iso r, anchor=mid west] at (3.66,1.77) {target capability loss only};
\path[fill=white, draw=isoSlate, line width=0.35pt] (5.7527,1.722) rectangle (5.8927,1.818);
\node[iso r, anchor=mid west] at (5.9327,1.77) {clean};
\draw[line width=0.45pt, draw=isoSlate] (0.68,1.41) -- (0.64,1.41) -- (0.64,0.598) -- (0.68,0.598);
\node[iso r note, anchor=mid west] at (0.02,1.004) {\textit{Gateways}};
\draw[line width=0.45pt, draw=isoSlate] (0.68,0.44) -- (0.64,0.44) -- (0.64,0.162) -- (0.68,0.162);
\node[iso r note, anchor=mid west] at (0.02,0.301) {\textit{Libraries}};
\draw[iso rule, densely dashed] (0.68,0.519) -- (6.46,0.519);
\draw[iso rule] (0.68,0.093) -- (6.46,0.093);
\node[iso r, anchor=mid east] at (1.54,1.36) {one-api};
\path[fill=isoCrimson] (1.62,1.302) rectangle (2.445,1.418);
\path[fill=isoInk!65] (2.445,1.302) rectangle (2.7338,1.418);
\path[fill=isoSlateLight] (2.7338,1.302) rectangle (2.8369,1.418);
\path[fill=white] (2.8369,1.302) rectangle (3.27,1.418);
\draw[line width=0.35pt, draw=isoSlate] (1.62,1.302) rectangle (3.27,1.418);
\path[fill=isoCrimson] (3.59,1.302) rectangle (3.9406,1.418);
\path[fill=isoInk!65] (3.9406,1.302) rectangle (4.7862,1.418);
\path[fill=isoSlateLight] (4.7862,1.302) rectangle (4.8894,1.418);
\path[fill=white] (4.8894,1.302) rectangle (5.24,1.418);
\draw[line width=0.35pt, draw=isoSlate] (3.59,1.302) rectangle (5.24,1.418);
\node[iso r, anchor=mid east] at (6.46,1.36) {72.5\,\%};
\node[iso r, anchor=mid east] at (1.54,1.182) {Portkey};
\path[fill=isoCrimson] (1.62,1.124) rectangle (2.0531,1.24);
\path[fill=isoInk!65] (2.0531,1.124) rectangle (2.0944,1.24);
\path[fill=isoSlateLight] (2.0944,1.124) rectangle (2.3006,1.24);
\path[fill=white] (2.3006,1.124) rectangle (3.27,1.24);
\draw[line width=0.35pt, draw=isoSlate] (1.62,1.124) rectangle (3.27,1.24);
\path[fill=isoCrimson] (3.59,1.124) rectangle (3.9819,1.24);
\path[fill=isoInk!65] (3.9819,1.124) rectangle (4.085,1.24);
\path[fill=isoSlateLight] (4.085,1.124) rectangle (4.2912,1.24);
\path[fill=white] (4.2912,1.124) rectangle (5.24,1.24);
\draw[line width=0.35pt, draw=isoSlate] (3.59,1.124) rectangle (5.24,1.24);
\node[iso r, anchor=mid east] at (6.46,1.182) {30.0\,\%};
\node[iso r, anchor=mid east] at (1.54,1.004) {new-api};
\path[fill=isoCrimson] (1.62,0.946) rectangle (1.6406,1.062);
\path[fill=isoInk!65] (1.6406,0.946) rectangle (1.95,1.062);
\path[fill=isoSlateLight] (1.95,0.946) rectangle (2.1769,1.062);
\path[fill=white] (2.1769,0.946) rectangle (3.27,1.062);
\draw[line width=0.35pt, draw=isoSlate] (1.62,0.946) rectangle (3.27,1.062);
\path[fill=isoCrimson] (3.59,0.946) rectangle (3.6519,1.062);
\path[fill=isoInk!65] (3.6519,0.946) rectangle (3.7756,1.062);
\path[fill=isoSlateLight] (3.7756,0.946) rectangle (4.0025,1.062);
\path[fill=white] (4.0025,0.946) rectangle (5.24,1.062);
\draw[line width=0.35pt, draw=isoSlate] (3.59,0.946) rectangle (5.24,1.062);
\node[iso r, anchor=mid east] at (6.46,1.004) {25.0\,\%};
\node[iso r, anchor=mid east] at (1.54,0.826) {CLIProxyAPI};
\path[fill=isoCrimson] (1.62,0.768) rectangle (1.895,0.884);
\path[fill=isoInk!65] (1.895,0.768) rectangle (1.9796,0.884);
\path[fill=isoSlateLight] (1.9796,0.768) rectangle (2.2123,0.884);
\path[fill=white] (2.2123,0.768) rectangle (3.27,0.884);
\draw[line width=0.35pt, draw=isoSlate] (1.62,0.768) rectangle (3.27,0.884);
\path[fill=isoCrimson] (3.59,0.768) rectangle (3.7169,0.884);
\path[fill=isoInk!65] (3.7169,0.768) rectangle (3.9073,0.884);
\path[fill=isoSlateLight] (3.9073,0.768) rectangle (4.14,0.884);
\path[fill=white] (4.14,0.768) rectangle (5.24,0.884);
\draw[line width=0.35pt, draw=isoSlate] (3.59,0.768) rectangle (5.24,0.884);
\node[iso r, anchor=mid east] at (6.46,0.826) {21.8\,\%};
\node[iso r, anchor=mid east] at (1.54,0.648) {Bifrost};
\path[fill=isoCrimson] (1.62,0.59) rectangle (1.785,0.706);
\path[fill=isoInk!65] (1.785,0.59) rectangle (1.8263,0.706);
\path[fill=isoSlateLight] (1.8263,0.59) rectangle (2.0531,0.706);
\path[fill=white] (2.0531,0.59) rectangle (3.27,0.706);
\draw[line width=0.35pt, draw=isoSlate] (1.62,0.59) rectangle (3.27,0.706);
\path[fill=isoCrimson] (3.59,0.59) rectangle (3.7138,0.706);
\path[fill=isoInk!65] (3.7138,0.59) rectangle (3.9406,0.706);
\path[fill=isoSlateLight] (3.9406,0.59) rectangle (4.1675,0.706);
\path[fill=white] (4.1675,0.59) rectangle (5.24,0.706);
\draw[line width=0.35pt, draw=isoSlate] (3.59,0.59) rectangle (5.24,0.706);
\node[iso r, anchor=mid east] at (6.46,0.648) {21.2\,\%};
\node[iso r, anchor=mid east] at (1.54,0.39) {LLM-Rosetta};
\path[fill=isoCrimson] (1.62,0.332) rectangle (2.2388,0.448);
\path[fill=isoInk!65] (2.2388,0.332) rectangle (2.3213,0.448);
\path[fill=isoSlateLight] (2.3213,0.332) rectangle (2.4656,0.448);
\path[fill=white] (2.4656,0.332) rectangle (3.27,0.448);
\draw[line width=0.35pt, draw=isoSlate] (1.62,0.332) rectangle (3.27,0.448);
\path[fill=isoCrimson] (3.59,0.332) rectangle (4.1056,0.448);
\path[fill=isoInk!65] (4.1056,0.332) rectangle (4.4975,0.448);
\path[fill=isoSlateLight] (4.4975,0.332) rectangle (4.6419,0.448);
\path[fill=white] (4.6419,0.332) rectangle (5.24,0.448);
\draw[line width=0.35pt, draw=isoSlate] (3.59,0.332) rectangle (5.24,0.448);
\node[iso r, anchor=mid east] at (6.46,0.39) {55.0\,\%};
\node[iso r, anchor=mid east] at (1.54,0.212) {LiteLLM};
\path[fill=isoCrimson] (1.62,0.154) rectangle (1.7231,0.27);
\path[fill=isoInk!65] (1.7231,0.154) rectangle (1.95,0.27);
\path[fill=isoSlateLight] (1.95,0.154) rectangle (2.0944,0.27);
\path[fill=white] (2.0944,0.154) rectangle (3.27,0.27);
\draw[line width=0.35pt, draw=isoSlate] (1.62,0.154) rectangle (3.27,0.27);
\path[fill=isoCrimson] (3.59,0.154) rectangle (3.6106,0.27);
\path[fill=isoInk!65] (3.6106,0.154) rectangle (3.8994,0.27);
\path[fill=isoSlateLight] (3.8994,0.154) rectangle (4.1262,0.27);
\path[fill=white] (4.1262,0.154) rectangle (5.24,0.27);
\draw[line width=0.35pt, draw=isoSlate] (3.59,0.154) rectangle (5.24,0.27);
\node[iso r, anchor=mid east] at (6.46,0.212) {28.8\,\%};
\node[iso r bold, anchor=mid east] at (1.54,-0.026) {All seven};
\path[fill=isoCrimson] (1.62,-0.084) rectangle (1.9689,0.032);
\path[fill=isoInk!65] (1.9689,-0.084) rectangle (2.1227,0.032);
\path[fill=isoSlateLight] (2.1227,-0.084) rectangle (2.306,0.032);
\path[fill=white] (2.306,-0.084) rectangle (3.27,0.032);
\draw[line width=0.35pt, draw=isoSlate] (1.62,-0.084) rectangle (3.27,0.032);
\path[fill=isoCrimson] (3.59,-0.084) rectangle (3.8177,0.032);
\path[fill=isoInk!65] (3.8177,-0.084) rectangle (4.1282,0.032);
\path[fill=isoSlateLight] (4.1282,-0.084) rectangle (4.3233,0.032);
\path[fill=white] (4.3233,-0.084) rectangle (5.24,0.032);
\draw[line width=0.35pt, draw=isoSlate] (3.59,-0.084) rectangle (5.24,0.032);
\node[iso r bold, anchor=mid east] at (6.46,-0.026) {36.4\,\% (203/558)};
\end{tikzpicture}%
  \caption{\textbf{Official SDK outcomes for 558 translated reply streams.} Replies are translated back to OpenAI format and read by the official Python and Node SDKs. Bars distinguish silent corruption (red), exceptions (dark grey), and streams with neither outcome. The right column gives the share affected in at least one SDK.} 
  \label{fig:sdk-streams}
\end{figure}

\section{Empirical Results}
\label{sec:results}

We evaluate translation failures from five complementary perspectives. First, we characterise the information changes detected in translated requests, group them by their underlying implementation mechanisms, and assess their frequency in our test suites (Sec.~\ref{sec:rq1}). Second, we investigate whether selected changes affect model behaviour in targeted live-model experiments (Sec.~\ref{sec:rq-cons}). Third, we examine information loss in translated replies and its consequences for official client SDKs (Sec.~\ref{sec:rq-replies}). Finally, we investigate whether failures persist across software releases and API endpoints, and extend the evaluation to gateways, a first-party compatibility endpoint, and recorded coding-agent requests (Secs.~\ref{sec:rq-persist}--\ref{sec:rq-deploy}). Table~\ref{tab:units-map} specifies the unit and denominator for each result; \suppref{app:results}{S2} provides the full results.

\subsection{Request translation failures: mechanisms and frequency}
\label{sec:rq1}

\textbf{Failure mechanisms and human assessment.}
Source inspection groups the request-side findings into 54 candidates arising from 18 underlying mechanisms. Of these candidates, 28 involve information that the target API can represent and that the translator could therefore have preserved.
Fig.~\ref{fig:mechanisms} organises the mechanisms by the information they affect and whether the target format supports it. A mechanism can affect multiple request shapes and translation paths; consequently, its affected-cell count measures repeated occurrences in the test suite, not the number of distinct defects or implementation causes.
Human review confirms 31 of the 54 candidates as defects. Among the 28 avoidable candidates, 23 are confirmed defects, one is accepted as a form change, and four remain disputed. % The complete grading procedure and precision estimates are reported in Section~\ref{sec:precision} and \suppref{app:detail-rates}{D}.

\textbf{Frequency in the request suites.}
Across the 51-shape main suite, 127 of 685 completed translations contain at least one primary finding (18.5\,\%; cluster-bootstrap 95\,\% CI: 15.01--22.06\,\%; Fig.~\ref{fig:funnel} and Tab~\ref{tab:census}).
This is a rate within our constructed test suite, not an estimate of the prevalence of translation failures in production traffic.
The result is similar when we restrict the analysis to source requests accepted by live Anthropic models. Among the 182 translation cells corresponding to the 13 shapes accepted by \code{claude-sonnet-4-6}, 34 contain primary findings (18.7\,\%; 95\,\% CI: 11.54--25.82\,\%). For the eight shapes accepted by \code{claude-opus-4-8}, the corresponding result is 23/112 (20.5\,\%). Requiring live acceptance at both source and target endpoints leaves 22/88 primary findings (25.0\,\%) on the Gemini- and Anthropic-target paths. % (Section~\ref{sec:corpora}, Table~\ref{tab:tiers-full}; \suppref{app:detail-rates}{D}).

\textbf{Sensitivity to counting and capture choices.}
We examine whether the main result depends on the validity criteria, reporting policy, or capture point. Alternative validity and tier-boundary rules yield rates between 12.9\,\% and 19.6\,\% (49/379 to 134/685), while alternative counting conventions yield rates between 16.5\,\% and 25.7\,\% (88/532 to 176/685; see Table~\ref{tab:sensitivity} in the Appendix). These variants can change both the included cells and the findings counted within them, so their denominators are reported separately.
Two alternatives illustrate the sensitivity of the headline result. Including target-representable \code{T2} degradations, including system-to-user role demotion, increases the count to 134/685 (19.6\,\%).
Changing LiteLLM's capture point from its pre-serialisation dictionary to the serialised HTTP body yields 132/685 (19.3\,\%), compared with the baseline 127/685 (18.5\,\%).
% The complete variants and capture-point comparison are provided in \suppref{app:detail-rates}{D}.
The main suite does not directly cover mid-conversation system instructions, signed thinking blocks, or media inputs. We investigate these additional features through separate probes and recorded agent requests in Secs.~\ref{sec:rq-cons} and~\ref{sec:rq-deploy}.

\subsection{Behavioural consequences of translation changes}
\label{sec:rq-cons}

Detecting information loss establishes that a translation changes the conversation representation, but does not by itself establish a behavioural consequence. We therefore test selected mechanisms using targeted live-model experiments. Of the 18 catalogued mechanisms, ten are evaluated across six models: two Gemini models, two Claude models (\code{claude-sonnet-4-6} and \code{claude-haiku-5-5}), and two open-weight models. Nine of the ten mechanisms change at least one measured outcome in their corresponding probes.

The nine mechanisms with measured effects occur in 116 of the 127 primary request cells (91.3\,\% mechanism coverage). This is a mechanism-level correspondence: the live experiments test targeted probes for these mechanisms, not every one of the 116 individual translation cells. Four mechanisms produce a particularly pronounced effect, reducing compliance from 30/30 to 0/30 on both tested Gemini models. % (Figure~\ref{fig:consequence}; \suppref{app:detail-consequences}{D}).

\textbf{Tool and conversation changes on Gemini and open-weight models.}
Eight mechanisms significantly reduce compliance in the targeted Gemini or open-weight-model experiments.
Across ten scenarios on \code{gemini-3.5-flash-lite} and \code{gemini-3.8-flash}, six mechanisms each change the outcome in nine or ten scenarios (exact sign-flip permutation test, Holm-adjusted $p=0.0117$).
A separate experiment on the same two models examines the loss of tool-response identifiers (\code{functionResponse.id}). Without these identifiers, parallel tool-call attribution falls from 30/30 to 0/30 ($p_{\text{Holm}}=0.0039$).
Moving assistant text from after a tool call to before it also reduces compliance on \code{Qwen2.5-7B}, whereas the corresponding \code{Llama-3.1-8B} probe shows no difference.
To check that the results are not artefacts of replaying archived payloads, we rerun the translator processes themselves. All nine checked cells reproduce the same outcomes (see \suppref{app:detail-consequences}{D}).

\textbf{Effects of changed Claude generation controls.}
We also investigate changes to sampling settings and token limits, which are outside the primary request-failure rate but can still alter model behaviour.
On \code{claude-sonnet-4-6}, dropping \code{top\_p: 0.05} increases the mean number of distinct outputs across five samples per scenario from 3.8 to 4.8 over ten scenarios ($p=0.0156$).
On \code{claude-haiku-5-5}, the same omission changes all ten calls that were rejected because \code{top\_p} is deprecated into successful HTTP~200 responses.
We further test a translator-induced increase in the token budget. Raising \code{max\_tokens} from 100 to 1,124, thereby reproducing the added thinking budget, reduces compliance with the caller's original 100-token limit from 30/30 to 0/30 on both Claude models. By contrast, the tested removal of thinking blocks does not produce the predicted behavioural effect (see \suppref{app:detail-consequences}{D}).

\textbf{Additional instruction probe and task-level control.}
We conduct an Anthropic-target counterpart to the mid-conversation instruction experiment in Section~\ref{sec:case}. Removing the inlined notice that disables further tool use reduces refusals from 30/30 to 0/30 on \code{claude-haiku-5-5}. On \code{claude-sonnet-4-6}, refusals fall from 10/30 to 0/30, which does not meet the pre-specified prediction.
As a separate task-level control, we evaluate a 100-item GSM8K experiment on \code{gemini-3.5-flash-lite}. The translated requests produce no statistically detectable reduction in accuracy ($p\geq0.2500$; \suppref{app:detail-consequences}{D}).
This negative result is important: the observed effects depend on the information affected and the task being tested; not every translation change necessarily produces a measurable degradation on every downstream task.

\subsection{Information loss in translated replies and SDK failures}
\label{sec:rq-replies}

Information loss also occurs in the reverse direction, when translators convert model replies back to the caller's API format. Across 1,115 judged reply cells from seven translators, isloss identifies 391 meaning changes (35.1\,\%), of which 278 are avoidable.

We then examine whether translated reply streams remain usable by official client libraries. In a separate experiment, 203 of 558 translated streams (36.4\,\%) either raise an uncaught exception or silently corrupt the assembled message in at least one official OpenAI SDK. Of these 203 problematic outcomes, 199 (98.0\,\%) are avoidable (Fig.~\ref{fig:sdk-streams}).
The Python SDK exhibits exceptions or corruption on 170/558 streams (30.5\,\%), while the Node SDK does so on 182/558 (32.6\,\%). These per-SDK counts overlap because the same streams are tested against both SDKs; they are not additive components of the 203-stream union. Per-SDK results are provided in \suppref{app:detail-replies}{D}.

\subsection{Persistence across releases, API endpoints, and upstream fixes}
\label{sec:rq-persist}

\textbf{Persistence across software releases.}
To assess whether the observed changes are transient implementation bugs or persistent behaviours, we examine successive releases and later source snapshots. Across 50 consecutive LiteLLM releases, every release changes the meaning of at least 31 of the 51 main request shapes when translating to OpenAI.
In a broader census of ten translators across 170 shapes, 2,630 of the 2,745 cells flagged in the pinned releases remain flagged in the 11~October source snapshots (95.8\,\%, Step~2). This indicates that most of the detected changes persist in the later snapshots examined. Library- and gateway-specific results at each filtering stage are provided in \suppref{app:detail-persistence}{D}.

\textbf{Differences between API endpoints and upstream repairs.}
A translator's behaviour can also depend on which target endpoint it uses. In particular, the same translator may preserve different information when targeting OpenAI Chat Completions rather than Responses.
Across 357 inputs evaluated on five translators that support both endpoints, their outgoing requests differ on at least one of 22 pre-specified properties for every translator.
Same-vendor round-trip controls also reveal request changes in every tested translator.

We also report the upstream response to the issues identified during the study. By 15~October, we had opened 26 issues across 12 repositories, commented ten times on other contributors' threads, and made eight follow-up comments on our own reports.
Of the 26 reports, seven were fixed upstream and independently re-verified, one was partly fixed, one was closed without a reply, and 17 remained open at the time of assessment (see \suppref{app:detail-persistence}{D}).

\subsection{Gateways, first-party compatibility, and coding-agent traffic}
\label{sec:rq-deploy}

\textbf{Gateway request and reply translation.}
We extend the evaluation to gateway-mediated translation, where additional routing and compatibility layers can introduce information changes or reject requests explicitly. We audit six gateways and execute four of them, confirming most patterns identified in the audit while also finding additional candidates.

On the 51-shape suite, Bifrost and new-api produce \code{T1} changes in 187 of 204 judged cells (91.7\,\%). Of these, 103 pass the static target checks and 84 fail them.
Portkey, by contrast, explicitly refuses every tested Anthropic-format request directed to OpenAI or Gemini. Across the separate 170-shape corpus, the three gateways that return responses, Bifrost, new-api, and CLIProxyAPI, produce meaning changes in 978 of 1,020 judged cells (95.9\,\%).
We also test translated reply streams through four gateways: Portkey, Bifrost, new-api, and CLIProxyAPI. Among 198 judged streams, 57 (28.79\,\%) contain avoidable meaning changes (\suppref{app:detail-gateways}{D}).

\textbf{Fallback behaviour and a first-party compatibility endpoint.}
Compatibility layers can also change provider selection, not just payload structure. Among five surveyed gateways, Bifrost by default routes failed Anthropic-model requests to an OpenAI-compatible fallback. It returns the result in Anthropic format, while the change of provider is visible only in the response headers.

We separately examine Google's hosted OpenAI-compatible Gemini endpoint. Using two Gemini models, ten repetitions per experimental arm, and native-API controls, the endpoint passes 10 of 11 behavioural probes.
The failed probe shows that it retains only the last of two top-level \code{system} messages. Additional tests reveal several other differences from the OpenAI interface: it enforces Gemini's 128-character tool-name limit rather than OpenAI's 64-character limit; charges unnamed reasoning tokens against \code{max\_tokens} (98.0\,\% of output tokens across 45 calls); requires thought signatures for tool round-trips (5/5 requests without them are rejected); and emits \code{"stop"} rather than \code{"tool\_calls"} in streaming replies (see \suppref{app:detail-gateways}{D}).

\textbf{Recorded coding-agent requests.}
Finally, we examine whether the failure patterns identified using synthetic inputs also occur in recorded coding-agent traffic. Replaying the captured requests reveals the same classes of silent translation changes observed in the controlled experiments.

For the seven recorded shapes that pass static source checks, replay across all 14 translation paths yields four primary findings among 98 cells (4.1\,\%). A broader count that includes all avoidable meaning changes identifies 23/98 cells (23.5\,\%). When all 14 recorded shapes are included, the primary count is 13/196 cells (6.6\,\%).
These rates describe the replayed agent requests, not the prevalence of failures across all coding-agent traffic.
Offline replays of requests captured from Claude Code, Codex, and gemini-cli show translators folding, wrapping, or deleting mid-conversation \code{system} messages, as well as stripping system prompts or safety filters.
In multi-turn replays involving Claude Code and Crush, translators also drop failed tool calls' \code{is\_error} flags on OpenAI-bound paths. Finally, across the closed-loop CLI experiments, nine of 12 sessions contain at least one meaning change (see \suppref{app:detail-gateways}{D}).

\section{Discussion and Threats to Validity}
\label{sec:threats}

\subsection{Implications for translators and model evaluation}
\label{sec:implications}

Our findings have two practical implications. First, translators should treat conversation preservation as an explicit correctness requirement, rather than assuming that producing a structurally valid target request is sufficient. They should document their cross-API mappings and preserve information whenever the target API provides a native way to represent it. When the target lacks a corresponding representation, the translator should reject the request or explicitly warn the caller about the resulting degradation. It should not silently substitute a different conversation. The extended instruction experiment illustrates this distinction: AI SDK~v5 and LLM-Rosetta preserve the mid-conversation system instruction, whereas other translators delete or demote it.
These requirements should also guide translator testing. Regression tests should compare each translated payload with its source at the level of conversation information, rather than checking only the target schema or whether the request receives a successful response. The seven upstream fixes that we independently re-verified among our 26 issue reports demonstrate that the identified problems can lead to concrete improvements in translator implementations.

Second, translation is a potential confounding factor in model evaluation. When a benchmark or agent framework routes requests through a shared translator, the model may receive a prompt, tool specification, or generation configuration that differs from what the benchmark intended. Changes to system-instruction authority, tool schemas, and error flags can therefore affect the conditions under which models are evaluated~\citep{Wallace2024InstructionHierarchy,Qin2024SysBench,Liu2026ActionSpace}. This concern is relevant to benchmark pipelines such as $\tau$-bench~\citep{Yao2024TauBench} and BFCL~\citep{Patil2025BFCL} whenever they rely on LiteLLM, the Vercel AI SDK, or other translation layers.

Checking captured payloads with isloss can help researchers determine whether a discrepancy originates in the translator or in the model's response to the intended input. In this way, conversation-preservation checks can improve the interpretability and reproducibility of model comparisons. Our captured translations and field-dependent behavioural experiments demonstrate that translation-induced changes can alter model behaviour; the GSM8K control also shows that such effects are task-dependent. However, we have not measured the resulting changes in published benchmark scores or leaderboard rankings. Establishing their impact on end-to-end benchmark comparisons remains an important direction for future work.

\subsection{Threats to validity}
\label{sec:threats-detail}

\textbf{Representation and rule interpretation.}
Our isloss checks only the conversation properties represented in ConvIR and covered by its preservation rules. It therefore cannot detect information changes outside this representation or establish complete semantic equivalence between source and translated requests. Moreover, translating vendor documentation into preservation rules and deciding whether a loss is avoidable involve author judgement. To make these decisions auditable, each rule is linked to its supporting specification, and allowed changes are included as negative controls.
We further examine the sensitivity of the reported results to validity criteria and counting policies, including the treatment of system-to-user role demotion. Detailed rule definitions, review decisions, and audit records are provided in Appendix~\ref{app:audit}.

\textbf{Capture and measurement.}
Our findings are specific to the payloads observable at the selected capture points. Changes introduced after the last instrumented translation boundary, including provider-side rewriting, are not necessarily visible to isloss. The LiteLLM capture-point sensitivity analysis provides evidence about the effect of one such measurement choice, but does not establish that all possible capture points are equivalent.
Behavioural measurements also have potential sources of variance. Temperature-zero model calls are not guaranteed to produce identical outputs across runs, and single-run LLM-judge baselines may be sensitive to decoding variability. We report the corresponding experimental details and limitations in \suppref{app:detail-threats}{D}.

\textbf{Behavioural evidence.}
The live-model experiments cover selected field-dependent mechanisms and tasks rather than the full range of model behaviour. Their results establish that particular translation changes can affect particular measured outcomes under the tested conditions; they do not imply that every detected information loss will change model behaviour. The GSM8K control, the model-dependent effects of role demotion, and the unsuccessful prediction for the tested thinking-block removal illustrate this limitation. Furthermore, the 277/685 OpenAI-bound cells were not tested against live OpenAI endpoints, and GPT models were not included in the live-model experiments.
Consequently, our behavioural conclusions should be interpreted as evidence from the tested models and mechanisms, not as a comprehensive assessment across providers or model families.

\textbf{System coverage and generalisation.}
The study covers the evaluated self-hosted translators and six live models: two Gemini models, two Claude models (\code{claude-sonnet-4-6} and \code{claude-haiku-5-5}), and two open-weight models. We did not inspect commercial hosted gateways or the compatibility layers operated by Anthropic and OpenAI. Google's OpenAI-compatible Gemini endpoint is the only first-party compatibility endpoint for which we conducted external behavioural probes.
The synthetic-suite rates measure findings across deliberately constructed feature combinations, not their incidence in production traffic. Our recorded-agent corpus contains 14 distinct request shapes, of which only seven pass the static source checks. Its lower primary rate in Sec.~\ref{sec:rq-deploy} therefore provides evidence from a limited set of observed traffic, not a calibration of production-wide prevalence.
Similarly, the gateway and live-model results cover specific implementations, configurations, endpoints, and experimental conditions. Broader deployment studies will be needed to establish how frequently these failures occur across the wider LLM ecosystem.

\textbf{Temporal validity and practical deployment.}
Translator implementations and vendor API specifications evolve. Our results reflect the software versions, configurations, and specification snapshots examined during the study; subsequent changes may alter both the mappings and the validity of individual findings.
In addition, the same-vendor and fallback experiments use local mocks, which may not reproduce every aspect of production routing and endpoint behaviour.
Finally, isloss is evaluated here as an offline research checker. Integrating it into command-line workflows, continuous integration, and release monitoring, and evaluating its effectiveness in those settings, remain future work.

\section{Related Work}
\label{sec:related}

\textbf{Translation validation and cross-format equivalence.}
Translation validation checks an individual translation against its source, rather than attempting to prove a translator correct for every possible input~\citep{DBLP:conf/tacas/PnueliSS98,DBLP:conf/pldi/Necula00}. Our isloss adopts this source-to-target perspective, but addresses a different setting from compiler validators such as Alive2~\citep{DBLP:conf/pldi/LopesLHLR21}. Compiler validation can rely on formal semantics, whereas LLM API specifications are often incomplete or ambiguous about how conversation constructs map across providers. Accordingly, isloss uses specification-grounded preservation rules to check explicitly represented conversation properties, rather than claiming complete semantic equivalence.

Differential testing and equivalence modulo inputs expose inconsistencies by comparing the behaviour of multiple implementations or configurations~\citep{McKeeman1998Differential,DBLP:conf/pldi/YangCER11,DBLP:conf/pldi/LeAS14}. Related techniques have been applied to SQL dialects~\citep{Jiang2025SQLxDiff,DBLP:journals/pacmse/LinZCWWX25}, database connectors~\citep{Lyu2025Connectors}, JSON parsers~\citep{Moller2024JSONParsers}, and HTTP intermediaries~\citep{DBLP:conf/ccs/JabiyevSOK21,Shen2022HDiff,Kallus2024HTTPGarden,Jabiyev2022Frameshifter}. For LLM APIs, comparing different translators can reveal disagreement, but disagreement alone does not establish which translation preserves the caller's intended conversation.

The closest related approach is LLM-Rosetta, which uses a shared intermediate representation to translate between API formats and checks round-trip equality~\citep{Ding2026LLMRosetta}. This provides a useful basis for studying format compatibility, but round-trip equality is only informative about the properties retained by the intermediate representation. Information omitted from that representation, such as error flags, reasoning-continuity fields, or explicit generation controls, may therefore escape the comparison.
In contrast, isloss uses its intermediate representation specifically to check preservation against the original payload; it does not generate the target request or require an inverse translation. It also records whether the target API has a native way to represent the affected information. This distinction between preserved information and information that cannot be represented is also related to bidirectional transformations, where reconstructing information omitted during a transformation requires additional information~\citep{Foster2007Lenses}.

\textbf{Testing software around language models.}
A growing body of work tests the software infrastructure through which language models interact with tools and external services. LogicHunter tests agent frameworks through their object APIs using an LLM oracle~\citep{DBLP:conf/issta/LongZW26}, while bug studies of agent and MCP frameworks examine failures in model-facing components and deficiencies in their tests~\citep{Zhang2026AgenticBugs,He2026MCPBugs}. These approaches investigate the correctness of agent frameworks and their interactions with tools. isloss addresses a different boundary: whether the information in a conversation survives translation before it reaches the model or is returned to the caller.

Related studies examine other aspects of gateway and serving infrastructure, including billing~\citep{DBLP:conf/imc/LinWPXXX26}, model identity and shadow APIs~\citep{Zhang2026IRIS,Zhang2026ShadowAPIs}, routing~\citep{Fu2026SIDEL}, failures hidden behind HTTP~200 responses~\citep{Pandey2026FailureAtlas}, failover losses~\citep{Pandey2026ContinuityBench}, and endpoint drift~\citep{Chen2024ChatGPTDrift}. Our focus is whether translation preserves the caller's conversation information, and whether detected changes persist across releases (Section~\ref{sec:rq-persist}). This complements work on tool-call execution after model inference~\citep{Yang2026Intent,Kim2026FCD,Tang2026Serving} and on serving-engine caches and decoding variance~\citep{Kwon2023vLLM,Atil2024NonDeterminism}. These studies examine different stages of the path from an application's request through model inference to its eventual action; isloss specifically checks the translation boundary within that path.

\textbf{API specifications and test generation.}
API specifications provide an important basis for deciding whether a translated request is valid, but they do not always define how constructs from different APIs should correspond. RFCScope studies ambiguities in protocol specifications~\citep{DBLP:conf/kbse/PawagiSLSW25}, while SilentProbe identifies prose-level constraints that are absent from OpenAPI schemas~\citep{Li2026SilentProbe}. Our specification analysis addresses cross-API mappings: we identify 17 constructs for which vendor documentation leaves the mapping between APIs unspecified or contradictory.
These gaps motivate a preservation analysis that distinguishes documented target limitations from avoidable losses, rather than treating every difference between payloads as a violation.

API test-generation techniques provide complementary tools for exploring combinations of request features. RESTler and RESTest generate tests while accounting for dependencies between parameters~\citep{Atlidakis2019RESTler,MartinLopez2019IDL,MartinLopez2021RESTest}. Constrained covering arrays similarly address the challenge of achieving interaction coverage while respecting input-validity constraints~\citep{Cohen2008ConstrainedCA}. We use a four-way covering array to generate an initial set of 357 unique requests and then apply static source-contract checks to obtain the 51-request main suite. Importantly, filtering the generated array does not establish that the retained subset preserves the original four-way coverage guarantee.
Finally, API-evolution research examines changes that require clients to migrate between library versions~\citep{Lamothe2021APIEvolution}. API translators address a related compatibility problem, but perform mappings between formats on each call, making preservation of individual requests and replies an additional correctness concern.

\textbf{Behavioural probes and sensitivity to input changes.}
Network measurement techniques detect changes introduced by intermediaries by sending known inputs and observing what reaches the destination~\citep{DBLP:conf/imc/KreibichWNP10,DBLP:conf/pam/WeaverKDP14,DBLP:conf/sigcomm/Vallina-Rodriguez15}. Our distinguishing conversations adapt this idea to hosted LLM compatibility endpoints: when the internal translated request is inaccessible, we compare model responses to field-dependent inputs against native-API controls. These probes provide behavioural evidence about selected translation effects; they are distinct from the deterministic preservation checks applied to captured payloads.

Prior work suggests that several types of conversation information can influence model behaviour. Instruction-hierarchy training distinguishes system instructions from user text and tool outputs~\citep{Wallace2024InstructionHierarchy,Wu2024ISE}, including under indirect prompt injection~\citep{Greshake2023IndirectInjection}. Changes to chat-template role markers can also affect safety behaviour~\citep{Jiang2025ChatBug}. Other studies investigate sensitivity to system-prompt phrasing~\citep{Qin2024SysBench}, the position of information in long contexts~\citep{Liu2024LostMiddle}, multi-turn splitting~\citep{Laban2025Lost}, tool schemas~\citep{Liu2026ActionSpace}, and structured-output schemas~\citep{Geng2025JSONSchemaBench}. Aggregate evaluation scores can further conceal regressions on individual items~\citep{Xu2026Aggregate}.
Building on these observations, our live experiments test whether information changes actually introduced by translators affect tasks that depend on the corresponding fields. Rather than assuming that every detected change harms performance, we measure its consequences under targeted conditions.

\textbf{Evaluation validity and automated judges.}
The reliability of model evaluations can depend on prompt formatting, prompt templates, answer order, and evaluation-harness conventions~\citep{Sclar2024PromptSensitivity,Mizrahi2024MultiPrompt,Alzahrani2024BenchmarksTargets,Liang2023HELM,Biderman2024LessonsBenchmarks}. We study a related but distinct source of variation: changes introduced between the caller's request and the payload delivered to the model. This is particularly relevant to agent and tool-use benchmarks such as Gorilla, $\tau$-bench, and BFCL, which use custom handlers or translation libraries~\citep{Patil2024Gorilla,Yao2024TauBench,Patil2025BFCL}. Checking the translated payload can help distinguish a model's response to the intended input from its response to an altered input.

LLM-as-a-judge research primarily investigates the assessment of open-ended model outputs~\citep{Zheng2023LLMJudge,Gu2024JudgeSurvey}. Our comparison instead considers the narrower task of detecting information changes between a source request and its translation. We evaluate prompted judges both with and without a reference translation, since access to a trusted reference materially changes the comparison problem. We further assess isloss using planted faults, allowed-change controls, program mutants, and human review, following established work on test oracles and their evaluation~\citep{Barr2015OracleSurvey,Jahangirova2016Oracle}. Development and repair data are kept separate from held-out evaluation, consistent with the importance of independent test fixtures~\citep{Xu2026Fixture}.

\section{Conclusion}
\label{sec:conclusion}

Successful API translation does not necessarily preserve the conversation that an application intended to send. A translated request can satisfy the target API's schema while silently losing or altering information that affects how a model interprets instructions, uses tools, or generates a reply. We introduced isloss, a deterministic test oracle that makes conversation-information preservation checkable by comparing source and translated payloads through a shared representation and explicit preservation rules, without requiring model calls or reference translations.

Our empirical study demonstrates that this problem occurs across existing translators and affects both requests and replies. In the main synthetic request suite, 127 of 685 completed translations (18.5\,\%) pass static target checks while silently introducing at least one avoidable change to prompt, role, or tool information. Human review confirms 31 genuine defects among 54 request-side candidates. Separate experiments identify 203 of 558 translated reply streams (36.4\,\%) that cause exceptions or silent corruption in at least one official OpenAI SDK. Targeted live-model experiments further show that nine of ten tested translation mechanisms alter at least one measured behavioural outcome.
These findings establish conversation preservation as a distinct correctness requirement for LLM interoperability. Translators should preserve information whenever the target API can represent it and explicitly reject or disclose unavoidable degradation. Likewise, model and agent evaluations should verify the payloads actually delivered to models, rather than relying solely on schema validation or successful API responses. Future work will extend coverage across providers and streaming behaviours, integrate isloss into command-line and continuous-integration regression workflows, and investigate its impact on end-to-end agent and benchmark evaluations.

% \section*{Data Availability}

% The replication package includes isloss, the covering-array generator, the 51 statically valid source shapes and the complete 357-shape generated corpus, captured requests and replies, and the collected issue corpus.
% It also provides experiment plans and deviation logs, statistical analysis scripts, timestamped snapshots and digests of vendor documentation, and redacted upstream issue reports. The anonymous review repository is available at \url{https://convpreserve.github.io}.
% The translation-capture experiments can be reproduced offline using the provided local mock servers and test configurations (see Sec.~\ref{sec:capture}). Reproducing live-model experiments requires appropriate provider credentials or a local \code{llama.cpp} build. Logs for each experimental arm are included in the replication package.

\section*{Broader Impact Statement}

LLM translators and gateways mediate the boundary between applications and model providers. Consequently, silent changes at this layer can affect not only conversation meaning but also safety controls, credential handling, provider selection, and usage accounting. Our deployment experiments identify several security- and reliability-relevant behaviours: a proxy disables Gemini safety filters without logging the change; a gateway routes requests to another provider after upstream errors while returning HTTP~200; authentication headers are forwarded during cross-provider fallback; and mid-stream errors are closed with HTTP~200 and billed as completed requests.
These findings motivate more explicit translation policies, preservation checks, and transparent reporting of fallback and error-handling behaviour. The purpose of isloss is to help developers and evaluators identify such problems before they silently affect model interactions or application behaviour.

All experiments were conducted using local mocks or accounts controlled by the authors. Safety-sensitive findings were disclosed privately before public issue reports were submitted. As of 15~October~2026, maintainers had fixed seven of the 26 public issues we reported, and we had independently re-verified those fixes. No separate remediation through private disclosure channels had been received by that date (see Sec. \ref{sec:rq-persist}).

% Four unpaid volunteer raters participated in the precision and fault-design audits with informed consent. They reviewed software outputs containing no personal data (Section~\ref{sec:rq-trust}; \suppref{app:detail-conclusion}{D}).

\section*{LLM Usage}

The authors used ChatGPT as an assistive tool for software coding and debugging, grammar checking and language polishing, and preparing and refining \LaTeX{} code for figures and tables. All research ideas, problem formulation, methodological contributions, experimental design, execution and evaluation of experiments, analysis and interpretation of results, and verification of findings were developed and carried out by the authors. The authors reviewed and validated all AI-assisted outputs and take full responsibility for the content and integrity of the manuscript.

\bibliography{main}
\bibliographystyle{tmlr}

\newpage

\appendix

\section{Conversation Representation, Rules, and Checker Assessment}
\label{app:method}

This appendix explains the conversation representation and preservation rules used by isloss, the request-capture procedure, and the tests of the checker's accuracy. It also provides the detailed motivating example, specification comparisons and test-suite audit supporting Sections~\ref{sec:intro}--\ref{sec:approach}. The replication archive contains the complete rule definitions and evaluation records.

\subsection{Counting Units, Motivating Example, and Existing Tests}
\label{app:detail-intro}\label{app:case-beta}\label{app:detail-motivating}\label{app:whynot}

\paragraph{Units and result subsets.}
A request template is a \emph{shape}; a translator and source-to-target route form a \emph{path}; applying one shape to one path gives a test \emph{cell}. The main request study uses 14 library paths and 51 shapes that pass static source checks. Of the 714 tests, 29 explicitly refuse the input and 685 produce translations to check (Stratum~1). The gateway study also examines eight gateway paths.
The archive retains the stratum labels to identify different subsets: Stratum~2a contains 13 source shapes accepted live by \code{claude-sonnet-4-6} (182 cells); Stratum~2b contains eight accepted by \code{claude-opus-4-8} (112 cells); and Stratum~2c contains 88 cells accepted live at both source and target endpoints. These labels describe the acceptance conditions, rather than asserting that all subsets are nested.

Repeated findings are grouped into 54 candidates across 18 mechanisms. Human review assigns 31 candidates to Grade~A (confirmed defects), 11 to Grade~B (acceptable form changes) and 12 to Grade~C (disagreement).
The capability tags are \code{M} for avoidable changes (28 candidates), \code{U} for losses without a native target representation (23), and \code{Borderline} for uncertain cases (three). The visibility tags \code{Vis/K/Meta/O} are retained in the detailed records; the primary result selects \code{Vis}$\wedge$\code{M} findings, meaning avoidable changes to prompt, role or tool fields. It does not include sampling controls or metadata merely because they can affect an application.
The rule categories \code{T1}, \code{T2} and \code{T3}, and their executable identifiers, are explained in Section~\ref{app:oracle}.

\paragraph{The mid-conversation instruction.}
Table~\ref{tab:aprime-detail} expands Figure~\ref{fig:motivating}. It records the tested software versions, the emitted OpenAI requests and the model's compliance with the instruction to stop using \code{run\_cmd}. We test two documented placements of the Anthropic \code{system} turn: directly before an assistant turn ($\text{A}'_1$), and at the end of \code{messages} ($\text{A}'_2$). These placements occur three and seven times, respectively, across seven recorded Claude Code traces.
All 22 stimuli (11 paraphrases per placement) pass the offline check of the documented beta-header contract. This is static validation, not an observed HTTP~200 response from the Anthropic endpoint with that beta header. The separate \code{V4-midsys-R10} suite contains 170 shapes and produces 1,139/1,530 \code{T1} cells (74.4\,\%) at the reported request-analysis stage.

The table's live trials replay the intact or altered requests on Gemini and Qwen, at temperature zero. A separate Anthropic-target study places the prohibition within the final user turn (\code{MC3-1}, Table~\ref{tab:claudeconseq}). Removing that prohibition reduces refusal from 30/30 to 0/30 on \code{claude-haiku-5-5} ($p_{\text{Holm}}=0.0039$), meeting the pre-specified prediction. On \code{claude-sonnet-4-6}, refusal changes from 10/30 to 0/30 ($p=0.125$), which does not meet the prediction.

\begin{table}[tbp]
  \centering
  \caption{\textbf{Four translation outcomes for the same mid-conversation instruction.} Six translators preserve the instruction, delete it, demote it to user text, or explicitly refuse the input; each translator gives the same outcome on all 11 paraphrases in both placements. Panel~(a) gives the source and intact target requests; panel~(b) gives the emitted request and isloss rule; panel~(c) reports the live trials in which the model obeys the prohibition ($k/11$, temperature zero). These live counts concern the target models, not live validation of the Anthropic source. 
  Across the 112 trials in which the model obeys the prohibition, the API returns HTTP 200, no \code{run\_cmd} call is made, and the reply ends with \code{finish\_reason: stop} rather than a token-limit termination.
  % Across all 112 successful prohibition trials, the API returns HTTP~200, no \code{run\_cmd} call is made, and the reply ends with \code{finish\_reason: stop} rather than the token limit. 
  AI~SDK~v5 and LLM-Rosetta produce the same normalised intact request (\code{R}). Claude results appear in Table~\ref{tab:claudeconseq}.}
  \label{tab:aprime-detail}
  \begingroup
  \scriptsize\setlength{\tabcolsep}{1.8pt}\renewcommand{\arraystretch}{0.88}%
  \noindent\isopanel{a}{Anthropic Messages source in both documented mid-turn \code{system} placements ($\text{A}'_1$, $\text{A}'_2$) and intact OpenAI wire (\code{R})}\\[0.5pt]
  \begin{tabular}{@{}P{2.8cm} P{7.6cm} P{5.4cm}@{}}
    \toprule
    Turns ($n=11$ each) & Anthropic Messages source request & Intact OpenAI Chat wire (\code{R}; rule \code{L01}) \\ %nocheck
    \midrule
    \code{[0..2]} prefix & \code{user}: ``list files'' $\to$ \code{assistant}: \code{run\_cmd("ls")} $\to$ \code{user}: \code{tool\_result} ``a.txt b.txt'' & \code{user} $\to$ \code{assistant} (\code{tool\_calls}) $\to$ \code{tool}: ``a.txt b.txt'' (\code{L01}) \\
    \textbf{$\text{A}'_1$} (before \code{assistant}) & \code{[3] \textbf{system}}: ``[harness] run\_cmd is DISABLED\dots'' $\to$ \code{[4] assistant}: ``Understood\dots'' $\to$ \code{[5] user}: ``delete b.txt'' & \code{[3] \textbf{system}} kept at \code{[3]} $\to$ \code{[4] assistant} $\to$ \code{[5] user} \\
    \textbf{$\text{A}'_2$} (\code{messages[-1]}) & \code{[3] user}: ``now delete b.txt'' $\to$ \code{[4] \textbf{system}}: ``[harness] run\_cmd is DISABLED\dots refuse shell use.'' & \code{[3] user}: ``delete b.txt'' $\to$ \code{[4] \textbf{system}} kept at \code{[4]} \\
    \bottomrule
  \end{tabular}\\[1pt]
  \noindent\begin{tabular}{@{}P{9.2cm}P{6.6cm}@{}}
    \isopanel{b}{What six translators emit on the wire ($\text{A}'_1$, $\text{A}'_2$, 11/11 each)} & %nocheck
    \isopanel{c}{Live replay: prohibition kept ($k/11$)} \\ %nocheck
  \end{tabular}\\[0.5pt]
  \begin{tabular}{@{}l l l cc cc cc@{}}
    \toprule
    Translator (version) & Outbound OpenAI wire fate (11/11) & isloss & \multicolumn{2}{c}{\code{3.5-lite}} & \multicolumn{2}{c}{\code{3.8-flash}} & \multicolumn{2}{c}{\code{Qwen2.5-7B}} \\ %nocheck
    \cmidrule(lr){4-5}\cmidrule(lr){6-7}\cmidrule(lr){8-9}
    & & & $\text{A}'_1$ & $\text{A}'_2$ & $\text{A}'_1$ & $\text{A}'_2$ & $\text{A}'_1$ & $\text{A}'_2$ \\
    \midrule
    \isowarn\ \textbf{AI SDK v7} {\color{isoAxis}7.0.126} & Fail-closed: \code{AI\_InvalidPromptError} (400) & refusal & --- & --- & --- & --- & --- & --- \\ %nocheck
    \isowarn\ \textbf{claude-code-proxy} {\color{isoAxis}7ea4177} & Fail-closed: rejects \code{messages.role} (422) & refusal & --- & --- & --- & --- & --- & --- \\ %nocheck
    \midrule
    \isook\ \textbf{AI SDK v5} {\color{isoAxis}5.0.271} & Keeps \code{role:\,"system"} in place (arm \code{R}) & \code{L01} & \textbf{11/11} & \textbf{11/11} & \textbf{11/11} & \textbf{11/11} & \textbf{3/11} & \textbf{10/11} \\ %nocheck
    \isook\ \textbf{LLM-Rosetta} {\color{isoAxis}0.13.0} & Keeps \code{role:\,"system"} in place (wire = \code{R}) & \code{L01} & \multicolumn{6}{c@{}}{\itshape\color{isoAxis}(normalised wire identical to arm \code{R})} \\ %nocheck
    \midrule
    \isobad\ \textbf{@musistudio/llms} {\color{isoAxis}1.0.53} & Silently \textbf{deletes} \code{system} turn (200) & \code{S05} & \textcolor{isoVermilion}{\textbf{0/11}} & \textcolor{isoVermilion}{\textbf{0/11}} & \textcolor{isoVermilion}{\textbf{1/11}} & \textcolor{isoVermilion}{\textbf{3/11}} & \textcolor{isoVermilion}{\textbf{0/11}} & \textcolor{isoVermilion}{\textbf{0/11}} \\ %nocheck
    \isobad\ \textbf{LiteLLM} {\color{isoAxis}1.103.2} & Demotes to \code{user} + ``Operator note\dots'' & \code{S05,D02} & 11/11 & 11/11 & 9/11 & \textcolor{isoVermilion}{\textbf{0/11}} & 10/11 & 10/11 \\ %nocheck
    \bottomrule
  \end{tabular}
  \endgroup
\end{table}

\paragraph{What existing tests cover.}
We audit 462 test files across 11 open-source codebases. None contains the property-based, differential or cross-dialect reference tests sought in this audit. Table~\ref{tab:testaudit} distinguishes direct assertions from partial checks and missing coverage. Table~\ref{tab:methods} places isloss beside related testing and measurement approaches.

\begin{table}[tbp]
  \caption{\textbf{Existing test coverage across 11 translator codebases.} Columns cover 14 request, session and reply features. The audit groups them under authority (P1, $a$--$b$), call--result links (P2, $c$--$h$), reasoning integrity (P3, $i$--$k$) and controls (P4, $l$--$n$). Symbols: $\bullet$ = asserted (\code{Y}); $\circ^*$ = a narrow helper assertion (\code{Y*}); $\circ$ = partial coverage (\code{P}); $\cdot$ = absent (\code{N}); \code{?} = not inspectable. Partial coverage may be fixture-only, restricted to one direction, or limited to a sub-case. Bottom rows summarise coverage and incompatible behaviours encoded as passing assertions in the audit. Gemini's first-party compatibility endpoint has no public repository.}
  \label{tab:testaudit}
  \centering\scriptsize
  % AUTO-GENERATED by tools/gen_tables.py testaudit from artifacts/g3/testaudit/TEST_AUDIT.md & artifacts/g3/testaudit/test_audit.csv on 2026-10-03T04:07:47Z — do not edit
\setlength{\tabcolsep}{1.5pt}
\begin{tabular}{@{}lccccccccccccccrP{2.94cm}@{}}
\toprule
 & \multicolumn{9}{c}{Request \& session features (\textbf{a}--\textbf{i})} & \multicolumn{5}{c}{Reply \& CI (\textbf{j}--\textbf{n})} & & \\
\cmidrule(lr){2-10}\cmidrule(lr){11-15}
Code base & \textbf{a} & \textbf{b} & \textbf{c} & \textbf{d} & \textbf{e} & \textbf{f} & \textbf{g} & \textbf{h} & \textbf{i} & \textbf{j} & \textbf{k} & \textbf{l} & \textbf{m} & \textbf{n} & Y(+Y*)/14 & Test style \\
\midrule
LiteLLM & $\bullet$ & $\cdot$ & $\bullet$ & $\bullet$ & $\bullet$ & $\bullet$ & $\bullet$ & $\bullet$ & $\bullet$ & $\bullet$ & $\bullet$ & $\bullet$ & $\circ$ & $\bullet$ & 12/14 & literal asserts + VCR \\
Vercel AI SDK & $\circ$ & $\circ$ & $\bullet$ & $\circ$ & $\bullet$ & $\bullet$ & $\cdot$ & $\bullet$ & $\circ$ & $\circ$ & $\circ$ & $\bullet$ & $\cdot$ & $\cdot$ & 5/14 & inline snapshots \\
@musistudio/llms & $\cdot$ & $\cdot$ & $\cdot$ & $\cdot$ & $\cdot$ & $\cdot$ & $\cdot$ & $\cdot$ & $\cdot$ & $\cdot$ & $\cdot$ & $\cdot$ & $\cdot$ & $\cdot$ & 0/14 & none \\
claude-code-proxy & $\cdot$ & $\cdot$ & $\circ$ & $\cdot$ & $\cdot$ & $\cdot$ & $\cdot$ & $\circ$ & $\circ$ & $\cdot$ & $\cdot$ & $\cdot$ & $\cdot$ & $\cdot$ & 0/14 & print-only smoke \\
LLM-Rosetta & $\bullet$ & $\bullet$ & $\bullet$ & $\bullet$ & $\circ$ & $\bullet$ & $\cdot$ & $\bullet$ & $\bullet$ & $\bullet$ & $\bullet$ & $\bullet$ & $\cdot$ & $\circ$ & 10/14 & IR round-trip + literals \\
Portkey & $\cdot$ & $\cdot$ & $\cdot$ & $\cdot$ & $\cdot$ & $\cdot$ & $\cdot$ & $\circ$ & $\cdot$ & $\cdot$ & $\cdot$ & $\cdot$ & $\cdot$ & $\circ$ & 0/14 & HTTP 200 smoke \\
Bifrost & $\bullet$ & $\bullet$ & $\bullet$ & $\bullet$ & $\bullet$ & $\bullet$ & $\circ$ & $\bullet$ & $\bullet$ & $\bullet$ & $\bullet$ & $\bullet$ & $\circ$ & $\bullet$ & 12/14 & Go tables + round-trip \\
new-api & $\circ$ & $\cdot$ & $\circ$ & $\cdot$ & $\circ$ & $\cdot$ & $\cdot$ & $\circ$ & $\circ$ & $\circ$ & $\cdot$ & $\bullet$ & $\cdot$ & $\cdot$ & 1/14 & billing literals \\
one-api & $\cdot$ & $\cdot$ & $\cdot$ & $\cdot$ & $\cdot$ & $\cdot$ & $\cdot$ & $\cdot$ & $\cdot$ & $\cdot$ & $\cdot$ & $\cdot$ & $\cdot$ & $\cdot$ & 0/14 & none \\
Envoy AI Gateway & $\bullet$ & $\bullet$ & $\bullet$ & $\bullet$ & $\bullet$ & $\bullet$ & $\cdot$ & $\bullet$ & $\bullet$ & $\bullet$ & $\bullet$ & $\bullet$ & $\circ^*$ & $\bullet$ & 13/14 & Go tables + VCR \\
Kong ai-proxy & $\cdot$ & $\cdot$ & $\circ$ & $\cdot$ & $\cdot$ & $\cdot$ & $\cdot$ & $\circ$ & $\bullet$ & $\circ$ & $\cdot$ & $\circ$ & $\cdot$ & ? & 1/14 & 1 golden/provider \\
\midrule
Total ($\bullet$/$\circ$) & 4/2 & 3/1 & 5/3 & 4/1 & 4/2 & 5/0 & 1/1 & 5/4 & 5/3 & 4/3 & 4/1 & 6/1 & 1$^*$/2 & 3/2 & -- & -- \\
Passing policies & 3 & 3 & -- & -- & -- & -- & -- & -- & -- & -- & -- & -- & -- & -- & -- & incompatible (\textbf{a}, \textbf{b}) \\
\bottomrule
\multicolumn{17}{@{}P{13.15cm}@{}}{\vspace{0.5pt}\textbf{Features:} \textbf{a}~mid-turn \code{system}; \textbf{b}~\code{is\_error}; \textbf{c}~\code{tool\_choice}; \textbf{d}~text+tool order; \textbf{e}~thinking+sig; \textbf{f}~\code{thoughtSignature}; \textbf{g}~tool-name len; \textbf{h}~sampling; \textbf{i}~$\ge$3 turns; \textbf{j}~stream tools; \textbf{k}~finish/stop; \textbf{l}~usage; \textbf{m}~stream$\equiv$non-stream; \textbf{n}~live CI.}\\
\end{tabular}

\end{table}

\begin{table}[tbp]
  \caption{\textbf{Comparison with related testing and measurement methods.} LLM = a model is needed at test time; Det.\ = deterministic verdicts; State = multi-turn state is considered; BB = black-box use is possible. ``gen.'' means that an LLM generates inputs but the verdict is deterministic. ``preprint'' means that no peer-reviewed venue was found at the time of writing. $^{a}$~For our approach, captured-payload checking needs no model call; the separate behavioural probes do (Section~\ref{app:canary}). Outcomes for other methods are as reported in their papers.}
  \label{tab:methods}
  \centering\scriptsize
  % Editorial synchronization: deterministic checking and behavioural probes are separate rows.
% AUTO-GENERATED by tools/gen_tables.py from comparison.csv (C-113) — do not edit
\setlength{\tabcolsep}{3pt}
\begin{tabular}{@{}P{2.6cm}P{2.2cm}P{2.3cm}cccc P{2.7cm}@{}}
\toprule
Method (venue) & Object under test & Oracle & LLM & Det. & State & BB & Reported outcome \\
\midrule
isloss (this paper) & captured requests, replies and SSE & documented preservation rules; stream equivalence & N$^{a}$ & Y & Y & N & see \S\ref{sec:results} \\
Endpoint probes (this study) & hosted compatibility endpoint & model replies vs.\ native controls & Y & N & Y & Y & see \S\ref{sec:rq-deploy} \\
LogicHunter (ISSTA'26) & object API (agent frameworks) & LLM agent judge; crash baseline & Y & N & N & N & 40 found / 30 conf.\ / 26 fixed; precision 91\% \\
GateScope (IMC'26) & hosted gateway (network path) & differential vs.\ vendor; statistical & Y & N & Y & Y & 10 gateways; substitution, memory loss, billing drift \\
HTTP Garden (preprint'24) & HTTP parser; gateway transforms & differential & N & Y & N & part. & $>$100 bugs, 68 fixed \\
Netalyzr / Web proxies (IMC'10 / PAM'14) & network path (middleboxes) & known-answer probes & N & Y & N & Y & 130k sessions; 14\% of 646k IPs behind proxies \\
Header enrichment (HotMiddlebox'15) & network path (mobile ISPs) & known-answer probes & N & Y & N & Y & 299 providers, 112 countries (measurement) \\
Csmith (PLDI'11) & C compilers & differential (N versions) & N & Y & N & Y & $>$325 bugs \\
EMI (PLDI'14) & C compilers & metamorphic & N & Y & N & Y & 147 confirmed, $>$100 fixed \\
QTRAN (ISSTA'25) & DBMS dialects & metamorphic; LLM translates inputs & gen. & Y & N & Y & 24 bugs, 16 confirmed \\
DB-connector diff.\ testing (preprint'25) & JDBC connectors & differential; LLM+RL generates & gen. & Y & N & N & 16 reported, 10 confirmed \\
RFCScope (ASE'25) & specification text & LLM judge over spec & Y & N & -- & -- & 31 ambiguities, 8 author-confirmed \\
SIDEL (preprint'26) & adversarial router $\to$ agent & attack-success rate & Y & N & Y & Y & 0\% defence success, 4 agents (threat study) \\
LLM-Rosetta (preprint'26) & translation library (own IR) & self round-trip; compliance suite & N & Y & part. & N & no bugs reported; claims lossless \\
\bottomrule
\end{tabular}

\end{table}

\subsection{API Encodings, Preservation Rules, and Specification Gaps}
\label{app:convir}\label{app:faultmodel}\label{app:specgaps}\label{app:coverage}\label{app:detail-model}

Table~\ref{tab:dialects} compares the API encodings of conversation events and explicit controls used in \code{ConvIR}. It covers Anthropic Messages, OpenAI Chat Completions and Google Gemini \code{generateContent}; the accompanying rule records also address OpenAI Responses.
Table~\ref{tab:lattice} provides the documented mapping conventions \code{L01}--\code{L08} and ten request-side violation dimensions, \code{V01}--\code{V10}. These \code{V} identifiers correspond to executable checks \code{S01}--\code{S10}; they are alternative labels used in the specification records, not different methods. The table illustrates the rule grounding rather than listing all 27 executable checks. The full catalogue also includes \code{S11}--\code{S13}, response checks and the stream-equivalence relation \code{MR-S}.
A convention permitting a documented loss, such as the unsupported-control case in \code{L06}, does not establish semantic equivalence: the loss is recorded as a degradation under the study's counting policy.

Grounding the rules in vendor documents identifies 17 undecided or contradictory interactions. Two are benign in our audit, with no observed translator divergence; Table~\ref{tab:specgaps} lists the other 15. Separate live probes of the Gemini endpoint resolve 13 interactions. These observations inform the specification analysis and remain distinct from deterministic isloss verdicts on captured payloads.

The reported-bug corpus contains 125 issues from 12 trackers. Of these, 99 map to a rule (79.2\,\%: 46 request issues and 53 response issues), three describe legitimate normalisations (2.4\,\%), and 23 fall outside the fault model (18.4\,\%). The out-of-scope categories concern target-constraint rejections, rejection of legal inputs, server-executed tools, masked upstream errors and transport timing.
Two-pass coding gives 72\,\% agreement and Cohen's $\kappa=0.70$~\citep{Cohen1960Kappa}; grouping by rule family gives 83\,\% and $\kappa=0.73$. Response issues are silent in 62--67\,\% of cases under the reported classifications.

\begin{table}[tbp]
  \caption{\textbf{How the three APIs encode the tested constructs.} $\dagger$ marks vendor documentation that is silent or self-contradictory. $\ddagger$ marks a construct without a native representation in that format; carrying it requires another encoding, declaring a loss, or refusing the input.}
  \label{tab:dialects}
  \centering\scriptsize\setlength{\tabcolsep}{4pt}
  % tables/dialects.tex — Table 1 (R-BG2). Values only; one quoted vendor sentence per cell in
% artifacts/g1/m1/specgaps/dialect_table_v1.md (C-140). $\dagger$ = vendor text silent or self-contradictory
% (spec-gap table row in the artifact); $\ddagger$ = construct absent from this dialect, so a translator must invent a mapping or drop it.
\begin{tabular}{@{}P{1.75cm}P{3.75cm}P{3.75cm}P{3.75cm}@{}}
\toprule
Construct & OpenAI Chat Completions & Anthropic Messages & Gemini generateContent \\
\midrule
System prompt & \code{system} / \code{developer} message, accepted at any position & top-level \code{system}; a mid-conversation \code{\{role:system\}} turn is documented for recent models while the API reference (our copy of 1~October 2026) stated there is no system role$^\dagger$ & top-level \code{systemInstruction}; \code{role:system} inside \code{contents} is rejected (400) \\
Roles & system, developer, user, assistant, tool & user, assistant (typed content blocks) & user, model (typed parts) \\
Tool call & \code{assistant.tool\_calls[]}: \code{id}, \code{function.name}, \code{function.arguments} (JSON string) & \code{tool\_use} block: \code{id}, \code{name}, \code{input} (object) & \code{functionCall} part: \code{name}, \code{args} (object), optional \code{id} \\
Tool result & \code{role:tool} message with \code{tool\_call\_id}; no error flag$^\ddagger$ & \code{tool\_result} block: \code{tool\_use\_id}, \code{content}, \code{is\_error} & \code{functionResponse} part: \code{name}, \code{response} object; an \code{"error"} key is the documented convention \\
Reasoning & no reasoning content in messages; \code{reasoning\_effort} knob only$^\ddagger$ & \code{thinking} block with \code{signature}, replayed verbatim & \code{thought} parts with \code{thoughtSignature}; Gemini~3 rejects a missing signature (400)$^\dagger$ \\
Forced tool choice & \code{tool\_choice}: none / auto / required / named function & \code{tool\_choice.type}: auto / any / tool / none & \code{functionCallingConfig.mode}: AUTO / ANY / NONE / VALIDATED, \code{allowedFunctionNames} \\
Controls & top-level \code{temperature}, \code{top\_p}, \code{seed}, \code{stop}, \code{parallel\_tool\_calls}, \code{user} & top-level \code{temperature}, \code{top\_p}, \code{top\_k}, \code{stop\_sequences}, \code{metadata.user\_id}; \code{disable\_parallel\_tool\_use} per tool & \code{generationConfig}: \code{temperature}, \code{topP}, \code{topK}, \code{seed}, \code{stopSequences}; no parallel-call switch$^\ddagger$ \\
Tool names & $\le$64 chars: letters, digits, \code{\_}, \code{-} & $\le$128 chars: \code{\^{}[a-zA-Z0-9\_-]\{1,128\}\$} & $\le$128 chars; declarations may also use \code{.} and \code{:}, call and result names may not$^\dagger$ \\
\bottomrule
\end{tabular}

\end{table}

\begin{table}[tbp]
  \caption{\textbf{Documented mapping conventions and request-side violation dimensions.} \code{L01}--\code{L08} describe permitted mappings or explicit degradation allowances; \code{V01}--\code{V10} identify protected dimensions. A documented degradation is not equivalent to preservation. Each row identifies its vendor source, with the quoted text retained in the replication archive. This reference table supports, rather than exhausts, the executable isloss catalogue.}
  \label{tab:lattice}
  \centering\scriptsize
  % Editorial synchronization with the revised manuscript; measured outcomes retained.
% AUTO-GENERATED by tools/gen_tables.py from lattice_v0.json — do not edit
\begin{tabular}{@{}llP{6.1cm}P{4.1cm}@{}}
\toprule
Rule & Dimension & What changes & Vendor source (artifact copy, location) \\
\midrule
L01 & Representation & role name remap (tool result carried in user/model turn) & anthropic:messages (L48-\allowbreak{}50) \\
L02 & Representation & merge consecutive same-author messages into one multi-block message, blocks intact & anthropic:messages (L48) \\
L03 & Representation & consistent renaming of call ids across call and result & anthropic:agents-\allowbreak{}and-\allowbreak{}tools\_\allowbreak{}tool-\allowbreak{}use\_\allowbreak{}handle-\allowbreak{}tool-\allowbreak{}calls (L65) \\
L04 & Representation & split one text block into adjacent text blocks of the same author, concatenation identical & anthropic:messages (L95) \\
L05 & Representation & developer text hoisted to a dedicated system slot when it is the leading developer text & anthropic:messages (L91) \\
L06 & Degradation & unsupported control dropped with documented degradation; recorded as a loss, not as semantic equivalence & anthropic:messages (L3058) \\
L07 & Representation & OpenAI Chat target only: assistant text after tool calls in the same turn carried before them (no ordered slot: one content field, one tool\_calls list per assistant message) & openai:openapi (L41198 (ChatCompletionRequestAssistantMessage)) \\
L08 & Representation & performance/billing hints (cache\_control) may be dropped or added; their loss is a cost effect, not a change of meaning, and is outside the semantic-fault counts & anthropic:messages (L1060-\allowbreak{}1062) \\
\midrule
V01 & Partition & cross-author merge: text from different authors concatenated into one block & anthropic:messages (L48) \\
V02 & Partition & author change: developer text re-labelled as user or assistant; system-to-user demotion is conservatively recorded as T2 & anthropic:mid\_\allowbreak{}conversation\_\allowbreak{}system\_\allowbreak{}messages (L11-\allowbreak{}13) \\
V03 & Partition & semantic flag lost (is\_error) & anthropic:agents-\allowbreak{}and-\allowbreak{}tools\_\allowbreak{}tool-\allowbreak{}use\_\allowbreak{}handle-\allowbreak{}tool-\allowbreak{}calls (L58) \\
V04 & Partition & content re-interpreted so that meaning or medium changes (result container type stripped, text promoted to media, error wrapped into plain content) & anthropic:messages (L95) \\
V05 & Partition & event lost or invented & anthropic:agents-\allowbreak{}and-\allowbreak{}tools\_\allowbreak{}tool-\allowbreak{}use\_\allowbreak{}handle-\allowbreak{}tool-\allowbreak{}calls (L66) \\
V06 & References & call/result reference broken, duplicated or re-targeted & anthropic:agents-\allowbreak{}and-\allowbreak{}tools\_\allowbreak{}tool-\allowbreak{}use\_\allowbreak{}handle-\allowbreak{}tool-\allowbreak{}calls (L65) \\
V07 & Order & relative order of events changed (within or across authors) & anthropic:agents-\allowbreak{}and-\allowbreak{}tools\_\allowbreak{}tool-\allowbreak{}use\_\allowbreak{}handle-\allowbreak{}tool-\allowbreak{}calls (L65-\allowbreak{}66) \\
V08 & Controls & explicit control parameter dropped or altered without documented degradation & anthropic:messages (L1370-\allowbreak{}1374) \\
V09 & Controls & tool name altered without reversible mapping applied to history & anthropic:messages (L15198) \\
V10 & Controls & integrity/continuation field replaced by the vendor's placeholder signature (e.g. thoughtSignature := skip\_thought\_signature\_validator) instead of being carried from the original response & gemini-\allowbreak{}vertex:thought\_\allowbreak{}signatures ('When using REST or manual handling') \\
\bottomrule
\end{tabular}

\end{table}

\begin{table}[tbp]
  \caption{\textbf{Specification gaps found while grounding the rules.} The table lists 15 of 17 recorded interactions; original audit numbering is retained and entries 3 and 4 are the two benign interactions omitted from this table. ``Documented by'' identifies the vendor (A/O/G); $\dag$ marks contradictory vendor text, and G$^*$ marks a Vertex mirror. ``n/a'' indicates that the API cannot express the construct. ``not probed'' means that the indicated live probe was not run; it does not imply acceptance or rejection.}
  \label{tab:specgaps}
  \centering\scriptsize
  % Editorial synchronization with the revised manuscript; measured outcomes retained.
% AUTO-GENERATED by tools/gen_tables.py from spec_gaps_v1.csv (C-100) — do not edit
\ifshowclaims
\begin{tabular}{@{}rP{3.5cm}P{2.1cm}P{1.6cm}P{1.8cm}@{}}
\toprule
\# & Construct & Gap kind & Documented by & Live probe\\
\midrule
1 & system message mid-conversation & contradiction & A\dag (n/a: G) & Gemini: resolved\\
2 & system between tool\_use and tool\_result & unspecified & A (n/a: G) & not probed\\
5 & empty tool result content & unspecified & A & Gemini: accepts\\
6 & assistant text after tool\_use in the same turn & unspecified & none (n/a: O) & Gemini: accepts\\
7 & tool error flag & mapping unspecified & A,G (n/a: O) & Gemini: resolved\\
8 & cross-vendor replay of thinking / redacted thinking & field unspecified & A,G (n/a: O) & Gemini: resolved\\
9 & tool name length / charset & contradiction; mapping unspecified & A,O,G\dag & Gemini: resolved\\
10 & disable parallel tool use & mapping unspecified & A,O (n/a: G) & n/a (mapping)\\
11 & top\_k & mapping unspecified & A,G (n/a: O) & not probed\\
12 & metadata / user id & mapping unspecified & A,O,G & n/a (mapping)\\
13 & tool result that looks like JSON / structured vs string result & unspecified & none & Gemini: accepts\\
14 & one tool\_result split into several results with the same id & unspecified & A & Gemini: accepts\\
15 & call ids dropped (LiteLLM G path) & unspecified & A,O & Gemini: accepts\\
16 & which member of a parallel tool-call group must carry the thought signature & doc unreachable & G$^*$ & Gemini: resolved\\
17 & where a provider-specific signature lives inside a foreign wire format & field unspecified & none & Gemini: resolved\\
\bottomrule
\end{tabular}
\else
\begin{tabular}{@{}rP{4.4cm}P{2.6cm}P{2.0cm}P{2.2cm}@{}}
\toprule
\# & Construct & Gap kind & Documented by & Live probe \\
\midrule
1 & system message mid-conversation & contradiction & A\dag (n/a: G) & Gemini: resolved \\
2 & system between tool\_use and tool\_result & unspecified & A (n/a: G) & not probed \\
5 & empty tool result content & unspecified & A & Gemini: accepts \\
6 & assistant text after tool\_use in the same turn & unspecified & none (n/a: O) & Gemini: accepts \\
7 & tool error flag & mapping unspecified & A,G (n/a: O) & Gemini: resolved \\
8 & cross-vendor replay of thinking / redacted thinking & field unspecified & A,G (n/a: O) & Gemini: resolved \\
9 & tool name length / charset & contradiction; mapping unspecified & A,O,G\dag & Gemini: resolved \\
10 & disable parallel tool use & mapping unspecified & A,O (n/a: G) & n/a (mapping) \\
11 & top\_k & mapping unspecified & A,G (n/a: O) & not probed \\
12 & metadata / user id & mapping unspecified & A,O,G & n/a (mapping) \\
13 & tool result that looks like JSON / structured vs string result & unspecified & none & Gemini: accepts \\
14 & one tool\_result split into several results with the same id & unspecified & A & Gemini: accepts \\
15 & call ids dropped (LiteLLM G path) & unspecified & A,O & Gemini: accepts \\
16 & which member of a parallel tool-call group must carry the thought signature & doc unreachable & G$^*$ & Gemini: resolved \\
17 & where a provider-specific signature lives inside a foreign wire format & field unspecified & none & Gemini: resolved \\
\bottomrule
\end{tabular}
\fi

\end{table}

\subsection{Rule Catalogue, Capture Procedure, and Accuracy Assessment}
\label{app:v100}\label{app:aisdk-driver}\label{app:response}\label{app:canary}\label{app:firstparty}\label{app:generation}\label{app:oracle}\label{app:trust}\label{app:hardening}\label{app:rq5-detail}\label{app:detail-oracle}

\paragraph{The executable catalogue.}
isloss contains 27 deterministic rules divided into three categories. The 13 \code{T1} rules (\code{S01}--\code{S13}) check changes to authorship, semantic flags and reasoning-continuity fields, payload structure, event presence, call--result links, event order, explicit controls, tool names, integrity tokens, tool schemas, stop reasons and stream reconstruction.
The six \code{T2} records (\code{D01}--\code{D06}) cover separately reported degradations and informational notes. They distinguish documented degradation from the study-policy exclusion of system-to-user demotion (\code{D02}); cache, usage and model notes use \code{D03}--\code{D05}. Role demotion changes authority, while target support and translator warnings are recorded separately (Section~\ref{sec:lattice}). The eight \code{T3} rules (\code{R01}--\code{R08}) record benign representation changes. Table~\ref{tab:decision} shows how the findings are filtered for the primary request result.

\paragraph{Requests and capture.}
A fixed-seed, 14-dimension strength-4 covering array produces 362 rows, with 357 unique request shapes after deduplication. Applying static source constraints leaves 51 shapes for Stratum~1. This filtered subset is the main test suite; filtering alone does not establish that it retains strength-4 coverage.
The baseline LiteLLM measurement captures the dictionary handed to the HTTP layer before serialisation. Rechecking its \code{hosted\_vllm} HTTP body changes the primary count from 127/685 to 132/685, which is reported as a sensitivity result.
The Vercel AI SDK capture driver passes \code{maxOutputTokens} and \code{providerOptions} explicitly and masks source controls that the interface cannot represent. The driver audit checks 2,142 cells and finds no omissions of representable controls.

\paragraph{Streams and hidden requests.}
For OpenAI Chat streams, isloss reassembles \code{delta.reasoning\_content} before applying the reply checks. The request verdicts and primary held-out assessment are unchanged by this stream-input handling.
Where an endpoint's translated request cannot be captured, the separate behavioural study constructs distinguishing conversations from selected translation patterns. The probe design seeks a property that holds on at least 9/10 intact trials but at most 3/10 altered trials. These model-based checks do not replace the deterministic comparison of observed payloads.

\paragraph{Detection, false alarms, and batch numbering.}
Across rule development and post-freeze evaluation, specification-based test pairs were organised into twenty numbered batches (Sets~1--20) so that rule refinement and accuracy evaluation use disjoint test sets:
\begin{itemize}[leftmargin=*,nosep]
  \item \textbf{Sets~1--8 (in-sample development batches, pre-freeze):} eight 30-item batches (240 items: 177 activated faults, 36 activated allowed-change controls and 27 unactivated edits) used while authoring and hardening the rules before freeze. Sets~1--3 were unconstrained initial batches across request, reply and stream payloads, whereas Sets~4--8 targeted gaps remaining after preceding rule updates; a 78-pair subset from Sets~6--8 also provided an early two-judge diagnostic contrast (Table~\ref{tab:trustfull}). Under the final isloss rules, Sets~1--8 yield 140/177 \code{T1} detections (79.1\,\%) and 0/30 false alarms on clean controls (6/36 before excluding pre-logged defective or disputed control edits), and are retained only as in-sample diagnostics (Table~\ref{tab:prereg-b}).
  \item \textbf{Sets~9--14 (primary held-out evaluation, post-freeze):} six blind batches authored from vendor specifications without checker-code access after the primary rules were fixed, comprising three fault/control pairs, Set~9 (28/34 activated faults detected; first held-out fault batch in Table~\ref{tab:trust}) with Set~10 (1/28 activated controls flagged; first held-out control batch), Set~11 (37/43) with Set~12 (0/36), and Set~13 (30/48) with Set~14 (1/36), pooling to 95/125 detected faults (76.0\,\%) and 2/100 false alarms (2.0\,\%, $J=0.740$).
  
  %\textbf{Sets~15--16 (voided script-generated batches):} one fault batch (48 items) and one control batch (36 items) excluded before scoring under pre-scoring decision \code{D-V120-08} because the designer workspace used Python fuzzing scripts rather than manual specification-based design (Table~\ref{tab:audit-master}).
  \item\textbf{Sets~15--16 (excluded script-generated batches):}
one fault batch (48 items) and one allowed-change control batch (36 items) were excluded before scoring because they were generated using Python fuzzing scripts rather than designed from vendor specifications under the held-out test protocol (Table~\ref{tab:audit-master}).
  \item \textbf{Sets~17--20 (extended held-out evaluation, post-freeze):} two later non-scripted blind pairs, Set~17 (32/48 faults detected under frozen isloss) with Set~18 (0/36 controls flagged), and Set~19 (33/47 activated faults detected) with Set~20 (3/34 controls flagged), authored for exploratory post-freeze repair checks and pooled with Sets~9--14 to form the 10-batch extended held-out evaluation: 160/220 faults detected (72.7\,\%) and 5/170 false alarms (2.9\,\%; Table~\ref{tab:trustfull}).
\end{itemize}
On public bugs, isloss detects 45/57 reproduced cases (78.9\,\%). This includes 10/14 encodable bugs (71.4\,\%) in the prospective sample opened after the checking rules were fixed.
Program-mutant tests detect 113/135 activated LiteLLM request mutants (83.7\,\%), 84/103 reply/stream mutants (81.6\,\%) and 43/51 LLM-Rosetta reverse-path mutants (84.3\,\%). A separate human-designed kit of 50 items yields 37/40 faults detected
(92.5\,\%) and no false alarms on ten controls.

\paragraph{Human review.}
Two rater pairs audit the findings. Among the 54 post-review candidate cards, consensus precision $\mathrm{A}/(\mathrm{A}+\mathrm{B})$ is 31/42 (73.8\,\%). For avoidable (\code{M}) candidates it is 23/24 (95.8\,\%); this denominator excludes the four disputed cards, so the confirmed fraction among all 28 avoidable candidates is 23/28 (82.1\,\%). For unavoidable (\code{U}) candidates, consensus precision is 6/16 (37.5\,\%).
All 127 primary Stratum-1 cells carry a mechanism associated with a Grade-A candidate. This is coverage by a human-confirmed mechanism, not independent human review of every cell or every finding within those cells.
A 48-item bridging sample scored by all four raters shows a 37.50-percentage-point rater-pair gap on unavoidable-loss items, compared with 6.25 points on avoidable items. This rater sensitivity supports reporting avoidable findings separately, without treating every unrepresentable loss as an established translator defect.

\paragraph{Other judges.}
Table~\ref{tab:judges} compares isloss with normalised diff and commercial and open-weight LLM judges. Each reference-free judge receives only the source and translated payload; the reference-given condition additionally provides a clean target-dialect translation. Without that reference, every tested LLM judge has a lower observed $J$ than isloss at $0.740$. This is an empirical comparison on the held-out set, rather than a general claim about all LLM judges.

\begin{table}[tbp]
  \centering\scriptsize
  \caption{\textbf{Detailed assessment of checker accuracy.} The main-text summary appears in Table~\ref{tab:trust}; additional judge comparisons appear in Table~\ref{tab:judges}. ``Before the run'' means that the protocol was written before execution and committed with the results. ``Killed'' means that a mutant produces a finding absent from the unmodified translator. Final isloss detection counts use \code{T1} findings; exploratory results with other scoring criteria are identified separately. For allowed-change controls, ``packed'' denotes pairs as constructed, five of which carry a packing-script placeholder signature; ``filled'' denotes the post-hoc restoration of those signatures. These controls should not be conflated with the untouched primary held-out set.}
  \label{tab:trustfull}
  % Final-rule counts are separated from development diagnostics; evidence IDs retain their original links.
\setlength{\tabcolsep}{2pt}
\renewcommand{\arraystretch}{0.88}\linespread{0.88}\selectfont
\ifshowclaims
\begin{tabular}{@{}P{2.1cm}P{1.3cm}P{3.5cm}P{7.2cm}@{}}
\toprule
Check & Predicate & Material & Result\\
\midrule
\multicolumn{4}{@{}l@{}}{\textit{Final isloss measurements and comparisons}} \\
Seed recall (request) & pre-specified & 2,129 single-fault payloads; 335 allowed-change controls & isloss T1: 1,595/2,129 = 74.9\% (Wilson 73.0--76.7\%); intended rule 1,383/2,129 = 65.0\%; 0/335 false alarms \\
Historical bugs, sample A (request) & pre-specified & 29 sample-A issues; 12 replayed & isloss T1: 9/12 (Wilson 47--91\%). A separate post-hoc diagnostic detects 10/12; it is not the reported final-rule estimate\\
Historical bugs, sample A (stream) & pre-specified & 4 replayed; 6 not run; 7 out of scope & isloss T1: 4/4; one upstream fix found incomplete\\
Historical bugs, sample B--C & before the run & 56 further issues: 18 replayed (12 request, 6 stream), 17 without a harness, 20 out of scope, 1 not reproduced & isloss T1: 6/11 requests and 3/6 streams. The 11-request scoring subset differs from the 12 replayed requests; the inverse case is reported separately in the records\\
Program mutants (request) & pre-specified & LiteLLM 1.104.0 adapter, OpenAI target; 300 mutants, 135 activated & isloss T1 kills 113/135 = 83.7\% of activated mutants\\
Program mutants (reply, stream) & before the run & Same release, reply/stream paths, two backends; 297 mutants, 103 activated & isloss T1 kills 84/103 = 81.6\% of activated mutants\\
First held-out pair (Sets 9--10) & pre-specified & 36 faults and 28 allowed changes by independent designers; 34 faults and 28 controls activated & T1 on 28/34 faults = 82.4\% (Wilson 66.5--91.7\%): the 80\% threshold is met. T1 on 1/28 controls: the at-most-one threshold is met\\
Primary held-out (Sets 9--14) & pre-specified & 125 activated faults; 100 allowed changes & T1: 95/125 = 76.0\%; false alarms: 2/100 = 2.0\%\\
Extended held-out (Sets 9--14, 17--20) & pre-specified & 220 activated faults; 170 allowed changes; Sets 15--16 excluded & T1: 160/220 = 72.7\%; false alarms: 5/170 = 2.9\%\\
LLM judges, reference given (Sets 9--14) & pre-specified & 125 faults and 100 allowed changes; judges also receive a clean target-dialect reference & flash-lite: 108/125 faults, 6/100 controls, 633\,ms; 3.8-flash: 120/125, 12/100, 2,352\,ms. isloss: 95/125, 2/100, no model call; diff D3: 124/125, 15/100. Predicted $\ge8$ false alarms per judge: missed by flash-lite\\
LLM judges, reference-free (Sets 9--14) & pre-specified & Same 225 pairs, source and translation only; 169 strictly cross-dialect & flash-lite: 72/125, 6/100, $J=0.516$ (cross-dialect 47/94, 4/75, $J=0.447$); 3.8-flash: 103/125, 12/100, $J=0.704$ (75/94, 9/75, $J=0.678$, 5,607\,ms). isloss: $J=0.740$ (71/94, 2/75, $J=0.729$). Judge agreement $\kappa$: 0.786 with reference, 0.585 without (0.490 cross-dialect). Predicted $\ge8$ cross-dialect false alarms per judge: missed by flash-lite\\
isloss T1 plus diff queue & before the run & Held-out Sets 9--14; T1 findings reported, diff-only flags queued for review & Faults: 95/125 reported and 30 queued; all 125 reported or queued. Controls: 2/100 reported and 13 queued. Queued cases are not isloss detections\\
Public bugs, four samples & pre-specified & 57 of 62 scored bugs reproduce: A 15, B--C 13, new issues 14, earlier issues encoded after rules were fixed 15 & T1: 45/57 (Wilson 66.7--87.5\%); any tier: 49/57 (74.7--92.7\%). T1 by sample: A 14/15 (one verdict corrected post hoc), B--C 9/13, new 10/14, earlier 12/15. B--C at any tier: 13/13. Earlier sample misses: documented representation limits. Of all 62 scored: 45/62 T1, 49/62 any tier; four predictions held\\
\midrule
\multicolumn{4}{@{}l@{}}{\textit{Development diagnostics and return-path baselines; not primary held-out estimates}} \\
Negative controls (request) & pre-specified & Remapping, merging, splitting, whitespace; Anthropic round trips & Development record: 0 findings; 16/16 round trips identical\\
Checker self-test (response) & pre-specified & 60 reference replies; 9 seeded mutations & Development record: 0 findings on references; 9/9 mutations detected\\
Reply/stream seed diagnostics & pre-specified & 65 ideal renderings; 15 message operators plus stream-event operators & Development record: 12/15 message operators flagged on every shape. No final isloss detection estimate is inferred from this calibration\\
Rule ablation & pre-specified & 2,938 flagged development cells; leave-one-rule-out & 1,202 cells depend on a single rule; removing one rule suppresses at most 16\% of flagged cells\\
Baseline diagnostics & before the run & Schema (B1), crash (B2), any-diff (B3) on the same development cells & B1 flags 71; B2 fires on five fail-closed paths; B3 flags all 2,938 flagged cells and 123/795 clean cells (up to 42 per path)\\
Early judge contrast & pre-specified & 78 pairs from in-sample Sets 6--8; reported counts cover 20 faults and 39 clean pairs; two judges from the designers' model family & J1: 9/20 faults, 1/39 clean; J2: 18/20, 21/39; $\kappa=0.098$. Final-rule diagnostic recount: 17/20 faults; 5/39 packed controls flagged, 0/39 after post-hoc signature restoration. These are development diagnostics, not the primary held-out estimate \\
Round trip A$\to$B$\to$A & pre-specified & 51 source-valid shapes; normalised comparison after the return hop & 10/14 paths provide no inverse. LLM-Rosetta mismatch: 48/51 via OpenAI (including 31/34 first hops without avoidable T1), 49/51 via Gemini. LiteLLM: 51/51 both ways. Twelve two-hop chains: 546/554 flagged return comparisons. A mismatch alone does not locate a translation defect\\
\bottomrule
\end{tabular}
\else
\begin{tabular}{@{}P{2.1cm}P{1.3cm}P{3.5cm}P{8.5cm}@{}}
\toprule
Check & Predicate & Material & Result \\
\midrule
\multicolumn{4}{@{}l@{}}{\textit{Final isloss measurements and comparisons}} \\
Seed recall (request) & pre-specified & 2,129 single-fault payloads; 335 allowed-change controls & isloss T1: 1,595/2,129 = 74.9\% (Wilson 73.0--76.7\%); intended rule 1,383/2,129 = 65.0\%; 0/335 false alarms \\
Historical bugs, sample A (request) & pre-specified & 29 sample-A issues; 12 replayed & isloss T1: 9/12 (Wilson 47--91\%). A separate post-hoc diagnostic detects 10/12; it is not the reported final-rule estimate \\
Historical bugs, sample A (stream) & pre-specified & 4 replayed; 6 not run; 7 out of scope & isloss T1: 4/4; one upstream fix found incomplete \\
Historical bugs, sample B--C & before the run & 56 further issues: 18 replayed (12 request, 6 stream), 17 without a harness, 20 out of scope, 1 not reproduced & isloss T1: 6/11 requests and 3/6 streams. The 11-request scoring subset differs from the 12 replayed requests; the inverse case is reported separately in the records \\
Program mutants (request) & pre-specified & LiteLLM 1.104.0 adapter, OpenAI target; 300 mutants, 135 activated & isloss T1 kills 113/135 = 83.7\% of activated mutants \\
Program mutants (reply, stream) & before the run & Same release, reply/stream paths, two backends; 297 mutants, 103 activated & isloss T1 kills 84/103 = 81.6\% of activated mutants \\
First held-out pair (Sets 9--10) & pre-specified & 36 faults and 28 allowed changes by independent designers; 34 faults and 28 controls activated & T1 on 28/34 faults = 82.4\% (Wilson 66.5--91.7\%): the 80\% threshold is met. T1 on 1/28 controls: the at-most-one threshold is met \\
Primary held-out (Sets 9--14) & pre-specified & 125 activated faults; 100 allowed changes & T1: 95/125 = 76.0\%; false alarms: 2/100 = 2.0\% \\
Extended held-out (Sets 9--14, 17--20) & pre-specified & 220 activated faults; 170 allowed changes; Sets 15--16 excluded & T1: 160/220 = 72.7\%; false alarms: 5/170 = 2.9\% \\
LLM judges, reference given (Sets 9--14) & pre-specified & 125 faults and 100 allowed changes; judges also receive a clean target-dialect reference & flash-lite: 108/125 faults, 6/100 controls, 633\,ms; 3.8-flash: 120/125, 12/100, 2,352\,ms. isloss: 95/125, 2/100, no model call; diff D3: 124/125, 15/100. Predicted $\ge8$ false alarms per judge: missed by flash-lite \\
LLM judges, reference-free (Sets 9--14) & pre-specified & Same 225 pairs, source and translation only; 169 strictly cross-dialect & flash-lite: 72/125, 6/100, $J=0.516$ (cross-dialect 47/94, 4/75, $J=0.447$); 3.8-flash: 103/125, 12/100, $J=0.704$ (75/94, 9/75, $J=0.678$, 5,607\,ms). isloss: $J=0.740$ (71/94, 2/75, $J=0.729$). Judge agreement $\kappa$: 0.786 with reference, 0.585 without (0.490 cross-dialect). Predicted $\ge8$ cross-dialect false alarms per judge: missed by flash-lite \\
isloss T1 plus diff queue & before the run & Held-out Sets 9--14; T1 findings reported, diff-only flags queued for review & Faults: 95/125 reported and 30 queued; all 125 reported or queued. Controls: 2/100 reported and 13 queued. Queued cases are not isloss detections \\
Public bugs, four samples & pre-specified & 57 of 62 scored bugs reproduce: A 15, B--C 13, new issues 14, earlier issues encoded after rules were fixed 15 & T1: 45/57 (Wilson 66.7--87.5\%); any tier: 49/57 (74.7--92.7\%). T1 by sample: A 14/15 (one verdict corrected post hoc), B--C 9/13, new 10/14, earlier 12/15. B--C at any tier: 13/13. Earlier sample misses: documented representation limits. Of all 62 scored: 45/62 T1, 49/62 any tier; four predictions held \\
\midrule
\multicolumn{4}{@{}l@{}}{\textit{Development diagnostics and return-path baselines; not primary held-out estimates}} \\
Negative controls (request) & pre-specified & Remapping, merging, splitting, whitespace; Anthropic round trips & Development record: 0 findings; 16/16 round trips identical \\
Checker self-test (response) & pre-specified & 60 reference replies; 9 seeded mutations & Development record: 0 findings on references; 9/9 mutations detected \\
Reply/stream seed diagnostics & pre-specified & 65 ideal renderings; 15 message operators plus stream-event operators & Development record: 12/15 message operators flagged on every shape. No final isloss detection estimate is inferred from this calibration \\
Rule ablation & pre-specified & 2,938 flagged development cells; leave-one-rule-out & 1,202 cells depend on a single rule; removing one rule suppresses at most 16\% of flagged cells \\
Baseline diagnostics & before the run & Schema (B1), crash (B2), any-diff (B3) on the same development cells & B1 flags 71; B2 fires on five fail-closed paths; B3 flags all 2,938 flagged cells and 123/795 clean cells (up to 42 per path) \\
Early judge contrast & pre-specified & 78 pairs from in-sample Sets 6--8; reported counts cover 20 faults and 39 clean pairs; two judges from the designers' model family & J1: 9/20 faults, 1/39 clean; J2: 18/20, 21/39; $\kappa=0.098$. Final-rule diagnostic recount: 17/20 faults; 5/39 packed controls flagged, 0/39 after post-hoc signature restoration. These are development diagnostics, not the primary held-out estimate \\
Round trip A$\to$B$\to$A & pre-specified & 51 source-valid shapes; normalised comparison after the return hop & 10/14 paths provide no inverse. LLM-Rosetta mismatch: 48/51 via OpenAI (including 31/34 first hops without avoidable T1), 49/51 via Gemini. LiteLLM: 51/51 both ways. Twelve two-hop chains: 546/554 flagged return comparisons. A mismatch alone does not locate a translation defect \\
\bottomrule
\end{tabular}
\fi
\par\vspace{2pt}\noindent{\scriptsize T1 and any-tier counts are distinct scoring criteria. Historical sample rows retain the original scoring subsets and must not be combined as if their denominators were identical. Earlier development-run predictions and their recorded outcomes remain in the scorecards; a final-rule recount does not retrospectively change whether those predictions held. ``Packed'' and post-hoc restored controls belong to development Sets~6--8, not the untouched primary held-out Sets~9--14.}

\end{table}

\begin{table}[tbp]
  \centering\scriptsize
  \caption{\textbf{isloss and alternative judges on the primary held-out evaluation (Sets~9--14).} Judges include normalised diff D3, commercial Gemini and Claude models (\code{gemini-3.5-flash-lite}, \code{gemini-3.8-flash}, \code{claude-sonnet-4-6}), and local open-weight models (\code{Qwen2.5-7B-Instruct}, \code{Llama-3.1-8B-Instruct}). Results distinguish reference-given, reference-free generic and reference-free specification-in-context conditions. The full activated set contains 225 items; 169 involve strictly cross-dialect translations.}
  \label{tab:judges}
  % AUTO-GENERATED by tools/fig/make_figdata.py on 2026-10-15T12:44:39Z — do not edit
% Claims: C-609, C-610, C-638, C-640, C-643, C-646, C-652, C-941, C-942, C-943, C-944, C-951, C-953, C-954, C-955, C-958, C-960, C-961, C-964, C-965, C-968.
\setlength{\tabcolsep}{1.5pt}%
{\scriptsize%
\ifshowclaims%
\resizebox{\textwidth}{!}{%
\begin{tabular}{@{}llrrrr@{}}
\toprule
Judge & Setting & Fault recall (95\% CI) & Allowed-change FPR (95\% CI) & $J$ [95\% CI] & Cost/1k calls\\
\midrule
\multicolumn{6}{@{}l@{}}{\textit{Held-out Sets 9--14, all activated items ($N=225$: $n_{\text{fault}}=125$, $n_{\text{allow}}=100$)}} \\
  isloss (T1) & Ref.-free & 95/125 (76.0\% [67.8,\,82.6]) & 2/100 (2.0\% [0.6,\,7.0]) & 0.740 [0.658,\,0.816] & \$0\\
  Normalised diff D3 & Ref. given (same wire) & 124/125 (99.2\% [95.6,\,99.9]) & 15/100 (15.0\% [9.3,\,23.3]) & 0.842 [0.768,\,0.910] & \$0\\
  \code{gemini-3.5-flash-lite} & Ref. given (same wire) & 108/125 (86.4\% [79.3,\,91.3]) & 6/100 (6.0\% [2.8,\,12.5]) & 0.804 [0.724,\,0.876] & \$0.116\textsuperscript{\dag}\\
  \code{gemini-3.8-flash} & Ref. given (same wire) & 120/125 (96.0\% [91.0,\,98.3]) & 12/100 (12.0\% [7.0,\,19.8]) & 0.840 [0.762,\,0.910] & \$0.116\textsuperscript{\dag}\\
  \code{claude-sonnet-4-6} & Ref. given (same wire) & 122/125 (97.6\% [93.2,\,99.2]) & 25/100 (25.0\% [17.5,\,34.3]) & 0.726 [0.636,\,0.812] & \$3.74\\
  \code{gemini-3.5-flash-lite} & Ref.-free generic & 72/125 (57.6\% [48.8,\,65.9]) & 6/100 (6.0\% [2.8,\,12.5]) & 0.516 [0.416,\,0.614] & \$0.169\textsuperscript{\dag}\\
  \code{gemini-3.8-flash} & Ref.-free generic & 103/125 (82.4\% [74.8,\,88.1]) & 12/100 (12.0\% [7.0,\,19.8]) & 0.704 [0.608,\,0.792] & \$0.169\textsuperscript{\dag}\\
  \code{gemini-3.5-flash-lite} & Ref.-free + Spec & 76/125 (60.8\% [52.0,\,68.9]) & 6/100 (6.0\% [2.8,\,12.5]) & 0.548 [0.448,\,0.644] & \$0.231\\
  \code{gemini-3.8-flash} & Ref.-free + Spec & 97/125 (77.6\% [69.5,\,84.0]) & 6/100 (6.0\% [2.8,\,12.5]) & 0.716 [0.624,\,0.798] & \$0.471\\
  \code{claude-sonnet-4-6} & Ref.-free generic & 105/125 (84.0\% [76.6,\,89.4]) & 32/100 (32.0\% [23.7,\,41.7]) & 0.520 [0.406,\,0.630] & \$3.92\\
  \code{claude-sonnet-4-6} & Ref.-free + Spec & 96/125 (76.8\% [68.7,\,83.3]) & 15/100 (15.0\% [9.3,\,23.3]) & 0.618 [0.510,\,0.716] & \$3.50\textsuperscript{\ddag}\\
\midrule
\multicolumn{6}{@{}l@{}}{\textit{Held-out Sets 9--14, strictly cross-dialect subset ($N=169$: $n_{\text{fault}}=94$, $n_{\text{allow}}=75$)}} \\
  isloss (T1) & Ref.-free & 71/94 (75.5\% [66.0,\,83.1]) & 2/75 (2.7\% [0.7,\,9.2]) & 0.729 [0.635,\,0.819] & \$0\\
  \code{claude-sonnet-4-6} & Ref. given (same wire) & 92/94 (97.9\% [92.6,\,99.4]) & 21/75 (28.0\% [19.1,\,39.0]) & 0.699 [0.592,\,0.803] & \$3.74\\
  \code{gemini-3.5-flash-lite} & Ref.-free generic & 47/94 (50.0\% [40.1,\,59.9]) & 4/75 (5.3\% [2.1,\,12.9]) & 0.447 [0.332,\,0.558] & \$0.169\textsuperscript{\dag}\\
  \code{gemini-3.8-flash} & Ref.-free generic & 75/94 (79.8\% [70.6,\,86.7]) & 9/75 (12.0\% [6.4,\,21.3]) & 0.678 [0.563,\,0.787] & \$0.169\textsuperscript{\dag}\\
  \code{gemini-3.5-flash-lite} & Ref.-free + Spec & 53/94 (56.4\% [46.3,\,66.0]) & 6/75 (8.0\% [3.7,\,16.4]) & 0.484 [0.364,\,0.601] & \$0.231\\
  \code{gemini-3.8-flash} & Ref.-free + Spec & 72/94 (76.6\% [67.1,\,84.0]) & 6/75 (8.0\% [3.7,\,16.4]) & 0.686 [0.577,\,0.785] & \$0.471\\
  \code{claude-sonnet-4-6} & Ref.-free generic & 75/94 (79.8\% [70.6,\,86.7]) & 23/75 (30.7\% [21.4,\,41.8]) & 0.491 [0.358,\,0.624] & \$3.92\\
  \code{claude-sonnet-4-6} & Ref.-free + Spec & 71/94 (75.5\% [66.0,\,83.1]) & 15/75 (20.0\% [12.5,\,30.4]) & 0.555 [0.430,\,0.678] & \$3.50\textsuperscript{\ddag}\\
  \code{Qwen2.5-7B-Instruct} & Ref.-free generic & 64/94 (68.1\% [58.1,\,76.6]) & 44/75 (58.7\% [47.4,\,69.1]) & 0.094 [$-$0.052,\,0.243] & \$0\\
  \code{Qwen2.5-7B-Instruct} & Ref.-free + Spec & 11/94 (11.7\% [6.7,\,19.8]) & 1/75 (1.3\% [0.2,\,7.2]) & 0.104 [0.037,\,0.178] & \$0\\
  \code{Llama-3.1-8B-Instruct} & Ref.-free generic & 94/94 (100.0\% [96.1,\,100.0]) & 75/75 (100.0\% [95.1,\,100.0]) & 0.000 [0.000,\,0.000] & \$0\\
\bottomrule
\end{tabular}%
}%
\else%
\setlength{\tabcolsep}{1.8pt}%
\begin{tabular}{@{}llrrrr@{}}
\toprule
Judge & Setting & Fault recall (95\% CI) & Allowed-change FPR (95\% CI) & $J$ [95\% CI] & Cost/1k calls \\
\midrule
\multicolumn{6}{@{}l@{}}{\textit{Held-out Sets 9--14, all activated items ($N=225$: $n_{\text{fault}}=125$, $n_{\text{allow}}=100$)}} \\
  isloss (T1) & Ref.-free & 95/125 (76.0\% [67.8,\,82.6]) & 2/100 (2.0\% [0.6,\,7.0]) & 0.740 [0.658,\,0.816] & \$0 \\
  Normalised diff D3 & Ref. given (same wire) & 124/125 (99.2\% [95.6,\,99.9]) & 15/100 (15.0\% [9.3,\,23.3]) & 0.842 [0.768,\,0.910] & \$0 \\
  \code{gemini-3.5-flash-lite} & Ref. given (same wire) & 108/125 (86.4\% [79.3,\,91.3]) & 6/100 (6.0\% [2.8,\,12.5]) & 0.804 [0.724,\,0.876] & \$0.116\textsuperscript{\dag} \\
  \code{gemini-3.8-flash} & Ref. given (same wire) & 120/125 (96.0\% [91.0,\,98.3]) & 12/100 (12.0\% [7.0,\,19.8]) & 0.840 [0.762,\,0.910] & \$0.116\textsuperscript{\dag} \\
  \code{claude-sonnet-4-6} & Ref. given (same wire) & 122/125 (97.6\% [93.2,\,99.2]) & 25/100 (25.0\% [17.5,\,34.3]) & 0.726 [0.636,\,0.812] & \$3.74 \\
  \code{gemini-3.5-flash-lite} & Ref.-free generic & 72/125 (57.6\% [48.8,\,65.9]) & 6/100 (6.0\% [2.8,\,12.5]) & 0.516 [0.416,\,0.614] & \$0.169\textsuperscript{\dag} \\
  \code{gemini-3.8-flash} & Ref.-free generic & 103/125 (82.4\% [74.8,\,88.1]) & 12/100 (12.0\% [7.0,\,19.8]) & 0.704 [0.608,\,0.792] & \$0.169\textsuperscript{\dag} \\
  \code{gemini-3.5-flash-lite} & Ref.-free + Spec & 76/125 (60.8\% [52.0,\,68.9]) & 6/100 (6.0\% [2.8,\,12.5]) & 0.548 [0.448,\,0.644] & \$0.231 \\
  \code{gemini-3.8-flash} & Ref.-free + Spec & 97/125 (77.6\% [69.5,\,84.0]) & 6/100 (6.0\% [2.8,\,12.5]) & 0.716 [0.624,\,0.798] & \$0.471 \\
  \code{claude-sonnet-4-6} & Ref.-free generic & 105/125 (84.0\% [76.6,\,89.4]) & 32/100 (32.0\% [23.7,\,41.7]) & 0.520 [0.406,\,0.630] & \$3.92 \\
  \code{claude-sonnet-4-6} & Ref.-free + Spec & 96/125 (76.8\% [68.7,\,83.3]) & 15/100 (15.0\% [9.3,\,23.3]) & 0.618 [0.510,\,0.716] & \$3.50\textsuperscript{\ddag} \\
\midrule
\multicolumn{6}{@{}l@{}}{\textit{Held-out Sets 9--14, strictly cross-dialect subset ($N=169$: $n_{\text{fault}}=94$, $n_{\text{allow}}=75$)}} \\
  isloss (T1) & Ref.-free & 71/94 (75.5\% [66.0,\,83.1]) & 2/75 (2.7\% [0.7,\,9.2]) & 0.729 [0.635,\,0.819] & \$0 \\
  \code{claude-sonnet-4-6} & Ref. given (same wire) & 92/94 (97.9\% [92.6,\,99.4]) & 21/75 (28.0\% [19.1,\,39.0]) & 0.699 [0.592,\,0.803] & \$3.74 \\
  \code{gemini-3.5-flash-lite} & Ref.-free generic & 47/94 (50.0\% [40.1,\,59.9]) & 4/75 (5.3\% [2.1,\,12.9]) & 0.447 [0.332,\,0.558] & \$0.169\textsuperscript{\dag} \\
  \code{gemini-3.8-flash} & Ref.-free generic & 75/94 (79.8\% [70.6,\,86.7]) & 9/75 (12.0\% [6.4,\,21.3]) & 0.678 [0.563,\,0.787] & \$0.169\textsuperscript{\dag} \\
  \code{gemini-3.5-flash-lite} & Ref.-free + Spec & 53/94 (56.4\% [46.3,\,66.0]) & 6/75 (8.0\% [3.7,\,16.4]) & 0.484 [0.364,\,0.601] & \$0.231 \\
  \code{gemini-3.8-flash} & Ref.-free + Spec & 72/94 (76.6\% [67.1,\,84.0]) & 6/75 (8.0\% [3.7,\,16.4]) & 0.686 [0.577,\,0.785] & \$0.471 \\
  \code{claude-sonnet-4-6} & Ref.-free generic & 75/94 (79.8\% [70.6,\,86.7]) & 23/75 (30.7\% [21.4,\,41.8]) & 0.491 [0.358,\,0.624] & \$3.92 \\
  \code{claude-sonnet-4-6} & Ref.-free + Spec & 71/94 (75.5\% [66.0,\,83.1]) & 15/75 (20.0\% [12.5,\,30.4]) & 0.555 [0.430,\,0.678] & \$3.50\textsuperscript{\ddag} \\
  \code{Qwen2.5-7B-Instruct} & Ref.-free generic & 64/94 (68.1\% [58.1,\,76.6]) & 44/75 (58.7\% [47.4,\,69.1]) & 0.094 [$-$0.052,\,0.243] & \$0 \\
  \code{Qwen2.5-7B-Instruct} & Ref.-free + Spec & 11/94 (11.7\% [6.7,\,19.8]) & 1/75 (1.3\% [0.2,\,7.2]) & 0.104 [0.037,\,0.178] & \$0 \\
  \code{Llama-3.1-8B-Instruct} & Ref.-free generic & 94/94 (100.0\% [96.1,\,100.0]) & 75/75 (100.0\% [95.1,\,100.0]) & 0.000 [0.000,\,0.000] & \$0 \\
\bottomrule
\end{tabular}%
\fi%
\par\vspace{2pt}\noindent{\scriptsize\textsuperscript{\dag}\,Costs are from the recorded evaluation runs. Combined two-model runs (450 calls each): \$0.0523 total% \ifshowclaims\ in C-638\fi
\ (\$0.116/1k), \$0.0760 total% \ifshowclaims\ in C-640\fi
\ (\$0.169/1k); % \ifshowclaims C-643 per-model\else Per-model\fi
\ spend (225 calls): \$0.0521 (\code{flash-lite}, \$0.231/1k) and \$0.1059 (\code{3.8-flash}, \$0.471/1k). \textsuperscript{\ddag}\,\code{claude-sonnet-4-6} at Anthropic pricing (\$3.00/\$15.00 per M input/output tokens; Spec runs include prompt-cache reads at \$0.30/M
% \ifshowclaims, C-951\fi
; Ref.\ given live spend \$0.8420/225 = \$3.74/1k
% \ifshowclaims, C-953, C-958\fi
); two Spec items (\code{LQ15}, \code{LQ16}, \code{stop\_reason=max\_tokens}) are \code{UNPARSEABLE} and scored as not flagged under PRIMARY
% \ifshowclaims\ (C-942)\fi
. $J$ [95\% CI] from item-stratified bootstrap ($B=10{,}000$, seed $20261015$% \ifshowclaims; C-960, C-961, C-964, C-965, C-968\fi
); 
Spec = Spec-in-Context.}
}

\end{table}

% \clearpage

% ==================================================================
% Revised Appendix B
% ==================================================================
\section{Study Design and Extended Empirical Results}
\label{app:results}\label{app:supp}

This appendix provides the complete subject provenance, study-design denominators, and extended empirical tables behind Sections~\ref{sec:evaluation} and~\ref{sec:results}.

\subsection{Subjects, request corpora, endpoint acceptance, and denominators}
\label{app:subjects}\label{app:detail-design}

Table~\ref{tab:subjects} records the pinned software versions, commits, and constructor settings. The main census uses six library configurations in five codebases: LiteLLM (\code{1.103.1}/\code{1.103.2}, \code{drop\_params=False}), Vercel AI SDK v5 (\code{5.0.271}) and v7 (\code{7.0.126}), LLM-Rosetta (\code{0.13.0}), \code{@musistudio/llms} (\code{1.0.53}), and \code{claude-code-proxy} (\code{7ea4177}). Their 14 request-translation paths produce 685 judged cells from 51 shapes after 29 refusals ($14\times51-29$). Additional subjects are Portkey~\citep{sw-portkey-gateway}, Bifrost~\citep{sw-bifrost}, new-api~\citep{sw-new-api}, one-api~\citep{sw-one-api}, CLIProxyAPI~\citep{sw-cliproxyapi} (\code{v8.0.15}), \code{@the-next-ai/ai-gateway}~\citep{sw-nextai-gateway} (\code{1.0.21}), the official OpenAI SDKs (\code{openai-python} \code{2.54.0}, \code{openai-node} \code{5.23.2}), and Google's OpenAI-compatible Gemini endpoint.

Of the 51 statically valid source shapes, \code{claude-sonnet-4-6} accepts 13 live. Its 38 HTTP~400 rejections comprise 19 placeholder \code{redacted\_thinking} cases, 15 combinations of \code{temperature} and \code{top\_p}, and 4 empty error \code{tool\_result} cases. The accepted subset has 34/182 primary cells (18.7\,\%, Stratum~2a). \code{claude-opus-4-8} accepts 8 shapes, with 23/112 primary cells (20.5\,\%, Stratum~2b). For translated requests passing the static target checks, live \code{gemini-3.5-flash-lite} accepts 160/204 cells and rejects 44 missing Gemini~3 \code{thought\_signature}; \code{claude-sonnet-4-6} accepts 77/204. Among source-accepted cells on those target paths, 88/104 are also accepted at the target, giving 22/88 primary cells (25.0\,\%, Stratum~2c). OpenAI-bound requests are checked statically rather than sent to a live target endpoint.

Table~\ref{tab:units-map-full} gives the corpus, unit, denominator, and supporting evidence for each result. Statistical methods are a bootstrap clustered by the 51 request shapes (10,000 resamples), Wilson 95\,\% intervals, and two-sided Fisher exact or sign-flip tests with Holm correction, as specified for each comparison.

\begin{table}[tbp]
  \centering\scriptsize
  \caption{\textbf{Subjects and software provenance.} A = Anthropic Messages, O = OpenAI Chat Completions, G = Gemini generateContent. ``2nd hop'' means that the subject receives another translator's OpenAI-format request. Download and star counts are recorded at capture time.} %nocheck
  \label{tab:subjects}
  % AUTO-GENERATED by tools/gen_tables.py (avoidable.json inputs, CLONED_COMMITS.txt, resp_summary_rows.json, gen4_nextai_judged.json, stream_ee/out6, two_exits_cliproxy, scale.csv; C-66, C-99, C-158, C-196, C-318) — do not edit
\setlength{\tabcolsep}{2.5pt}
\begin{tabular}{@{}P{2.4cm}P{2.5cm}lP{2.2cm}P{1.9cm}P{2.3cm}@{}}
\toprule
Subject & Position & Language & Directions tested & Version / commit & Scale (2026-10-01) \\
\midrule
LiteLLM & library; also a proxy & Python & A$\to$O; O/G$\to$A; 2nd hop & 1.103.1 / 1.103.2 & 89.4M/30\,d; 60.0k$\star$ \\
Vercel AI SDK v5 & in-process library & TypeScript & A$\to$O/G/A; O/G$\to$A & 5.0.271 & \code{ai} 32.7M/wk; 27.1k$\star$ \\
Vercel AI SDK v7 & in-process library & TypeScript & A$\to$O/G/A; O/G$\to$A & 7.0.126 & (same package) \\
@musistudio/llms & local proxy (engine of claude-code-router 1.x) & TypeScript & A$\to$O/G/A; O/G$\to$A & 1.0.53 & 308$\star$; router 37.5k$\star$ \\
@the-next-ai/ai-gateway & local proxy (engine of claude-code-router $\geq$2.0); not pooled & TypeScript & A$\to$O/G/A; 2nd hop & 1.0.21 & router 37.5k$\star$ \\
claude-code-proxy & local proxy & Python & A$\to$O; O$\to$A & git 7ea4177 & 2.8k$\star$ \\
LLM-Rosetta & library with its own IR & Python & A$\to$O/G/A; O/G$\to$A & llm-rosetta 0.13.0 & 5.7k/30\,d; 86$\star$ \\
Portkey gateway & shared gateway, self-hosted & TypeScript & 2nd hop O$\to$A/G & 669825c & 13.1k$\star$ \\
Bifrost & shared gateway, self-hosted & Go & A$\to$O/G; 2nd hop & 080a6bb & 8.5k$\star$ \\
new-api & shared gateway, self-hosted & Go & A$\to$O/G; 2nd hop & 56758ed & 49.2k$\star$ \\
one-api & shared gateway, self-hosted & Go & 2nd hop O$\to$A/G & 8df4a26 & 37.1k$\star$ \\
CLIProxyAPI & gateway, self-hosted, separate process; not pooled & Go & O/G$\to$A; RQ3 exits; RQ4 & v8.0.15 / v8.0.20 & -- \\
Kong AI Gateway; Envoy AI Gateway & shared gateway & Lua; Go & source audit only & 8927af6; 97f5609 & 44.2k$\star$; 2.2k$\star$ \\
Gemini OpenAI-compatible endpoint & first-party, black box & -- & distinguishing conv.\ O$\to$G & live, 2026-10-01 & -- \\
\bottomrule
\end{tabular}

\end{table}

\begin{table}[tbp]
  \centering
  \caption{\textbf{Full units and denominators map.} For each result in Table~\ref{tab:units-map}, we list the corpus, unit, denominator, paths, and supporting live or human evidence (Table~\ref{tab:eval-overview}, \suppref{app:subjects}{S2.1}).}
  \label{tab:units-map-full}
  {\scriptsize
\setlength{\tabcolsep}{2.3pt}
\setlength{\aboverulesep}{0.9pt}
\setlength{\belowrulesep}{0.9pt}
\renewcommand{\arraystretch}{0.88}
\begin{tabular}{@{}P{3.95cm}P{2.25cm}P{2.75cm}P{2.05cm}P{3.95cm}l@{}}
\toprule
\textbf{Headline claim \& value} & \textbf{Unit \& $N$} & \textbf{Corpus} & \textbf{Subjects} & \textbf{Method and supporting evidence} & \textbf{Sec.} \\
\midrule
1.\ \textbf{Primary (Stratum~1)}: $127/685$ (18.5\%) & Cell; $N{=}685$ ($14{\times}51{-}29$) & 51 of 357 shapes (\code{S-strict-R9}) & 6 configs (5 codebases), 14p & isloss primary criteria; $333/685$ static T1; $116/127$ finding-aligned mechanism coverage & \mbox{\S\ref{sec:rq1}} \\
2.\ \textbf{Live-source (Stratum~2a)}: $34/182$ (18.7\%) & Cell; $N{=}182$ ($14{\times}13$) & 13 of 51 shapes HTTP~200 (Sonnet) & 6 configs, 14p & isloss primary criteria; source HTTP~200 (\code{claude-sonnet-4-6}) & \mbox{\S\ref{sec:rq1}} \\
3.\ \textbf{Opus (Stratum~2b)}: $23/112$ (20.5\%) & Cell; $N{=}112$ ($14{\times}8$) & 8 of 51 shapes HTTP~200 (Opus) & 6 configs, 14p & isloss primary criteria; source HTTP~200 (\code{claude-opus-4-8}) & \mbox{\S\ref{sec:rq1}} \\
4.\ \textbf{Live target}: Gem.\ $160/204$ (78.4\%); Anth.\ $77/204$ (37.7\%) & Target cell ($204$ Gem., $204$ Anth.) & \code{S-strict-R9} wire ($277$ OpenAI cells checked offline) & 8 live paths (4 Gem., 4 Anth.) & Static checks vs.\ live acceptance; $44/204$ Gemini and $127/204$ Anthropic refusals & \mbox{\S\ref{sec:step4-contract}} \\
5.\ \textbf{Candidates \& mechs}: 54 ($31$~A/$11$~B/$12$~C), 18 mechs & \code{(translator,rule)} ($54$); $18$ mechs & \code{S-strict-R9} ($685$ cells) & 5 codebases (v5/v7 merged) & isloss and 4 raters; $31/42{=}73.8\%$ among non-disputed cases & \mbox{\S\ref{sec:rq1}} \\
6.\ \textbf{Sensitivity}: 12.9\%--19.6\% (valid.); 16.5\%--25.7\% (conv.) & Cell; selected main-suite variants ($N{=}308\text{--}685$) & Stratum-1 subsets \& reclassifications & 4--6 configs (8--14p) & isloss under the stated variant ($49/379$ to $134/685$ validity; $88/532$ to $176/685$ counting) & \mbox{\S\ref{sec:rq1}} \\
7.\ \textbf{Mid-turn \code{system}}: $1{,}139/1{,}530$ (74.4\%); Case~A$'$ ($k/11$) & Cell ($1{,}530$); trial ($n{=}11$ or $30$) & \code{V4-midsys-R10} ($170$: $96$/$74$) + Case~A$'$ & 6 configs; 2 paths + ref.\ (live) & isloss; live Gemini ($n{=}11$) \& Claude (ref.\ $30/30$, $10/30$) & \mbox{\S\ref{sec:case}} \\
8a.\ \textbf{Live effects}: $9/10$ tested mechanisms ($18$ catalogued; selected probes $30/30$ vs.\ $0/30$) & Mechanism ($10$ tested); trials $n{=}30$ or $10$ per arm & 10 mechanism probes & Translated vs.\ intact requests & Task scorers on Gemini, Claude, and open-weight models; mechanism coverage is not individual-cell validation & \mbox{\S\ref{sec:rq-cons}} \\
8b.\ \textbf{GSM8K control}: no statistically detectable accuracy drop & Call ($700$) & 100 items, 7 arms & gemini-3.5-flash-lite & Separate control; original-driver omission arm is withdrawn from translator-defect claims & \mbox{\S\ref{sec:rq-cons}} \\
9.\ \textbf{Replies \& SSE}: $391/1{,}115$ (35.1\%); $203/558$ (36.4\%) & Reply ($1{,}115$); SSE ($N{=}558$) & 40 reply shapes + 40 SSE streams & 7 families (11 reply, 14 SSE) & isloss: $278/391$ avoidable; independent SDK outcomes: $143$ exceptions, $60$ silent corruptions & \mbox{\S\ref{sec:rq-replies}} \\
10.\ \textbf{Persistence}: 50 LiteLLM releases; $2{,}630/2{,}745$ (95.8\%); $7/26$ & Release/commit histories ($50, 54, 32$); $2{,}745$ flagged cells & \code{S-strict-R9} + \code{V4-strict} ($170$) & 10 translators, 22p at HEAD & Step~2 persistence on the extended 170-shape suite; 26 upstream issues (7 fixed) & \mbox{\S\ref{sec:rq-persist}} \\
11.\ \textbf{Gateways}: $187/204$ (91.7\%) 51s; $978/1{,}020$ (95.9\%) 170s & Req.\ ($204, 1{,}020$); SSE ($198$) & \code{S-strict-R9} + \code{V4-strict} + 66 SSE & 4 gateways + CLIProxyAPI & isloss: $57/198{=}28.8\%$ avoidable stream changes; mock upstream & \mbox{\S\ref{sec:rq-deploy}} \\
12.\ \textbf{First-party}: $10/11$ probes pass & Probe ($N{=}11$); 10 trials per arm & 11 distinguishing probes & Google OpenAI-compat. & Behavioural probes on live \code{gemini-3.5-flash-lite}; compatibility vs.\ native API, not a captured-payload isloss verdict & \mbox{\S\ref{sec:rq-deploy}} \\
13.\ \textbf{Recorded CLI}: $4/98$ (4.1\%) valid; $13/196$ (6.6\%) all & Cell ($98$ valid, $196$ all); $12$ sess. & 14 CLI (7 statically contract-valid) & 6 configs, 14p (3 CLI agents) & isloss primary criteria; CLI captures; $9/12$ sessions contain an undeclared meaning change & \mbox{\S\ref{sec:rq-deploy}} \\
14.\ \textbf{Validation}: $95/125$ (76.0\%) \& $160/220$ (72.7\%) blind & Pair ($225$, $390$, $50$, $57$, $51$) & Blind, Human ($37/40$), Bugs, Mutants & All 4 dialect pairs & isloss vs.\ 7 judges; $45/57$ bugs; $43/51$ (84.3\%) mutants & \mbox{\S\ref{sec:rq-trust}} \\
\bottomrule
\end{tabular}
}

\end{table}

\subsection{Request mechanisms, acceptance subsets, and sensitivity analysis}
\label{app:rq1-detail}\label{app:detail-rates}

\begin{figure}[tbp]
  \centering
  \input{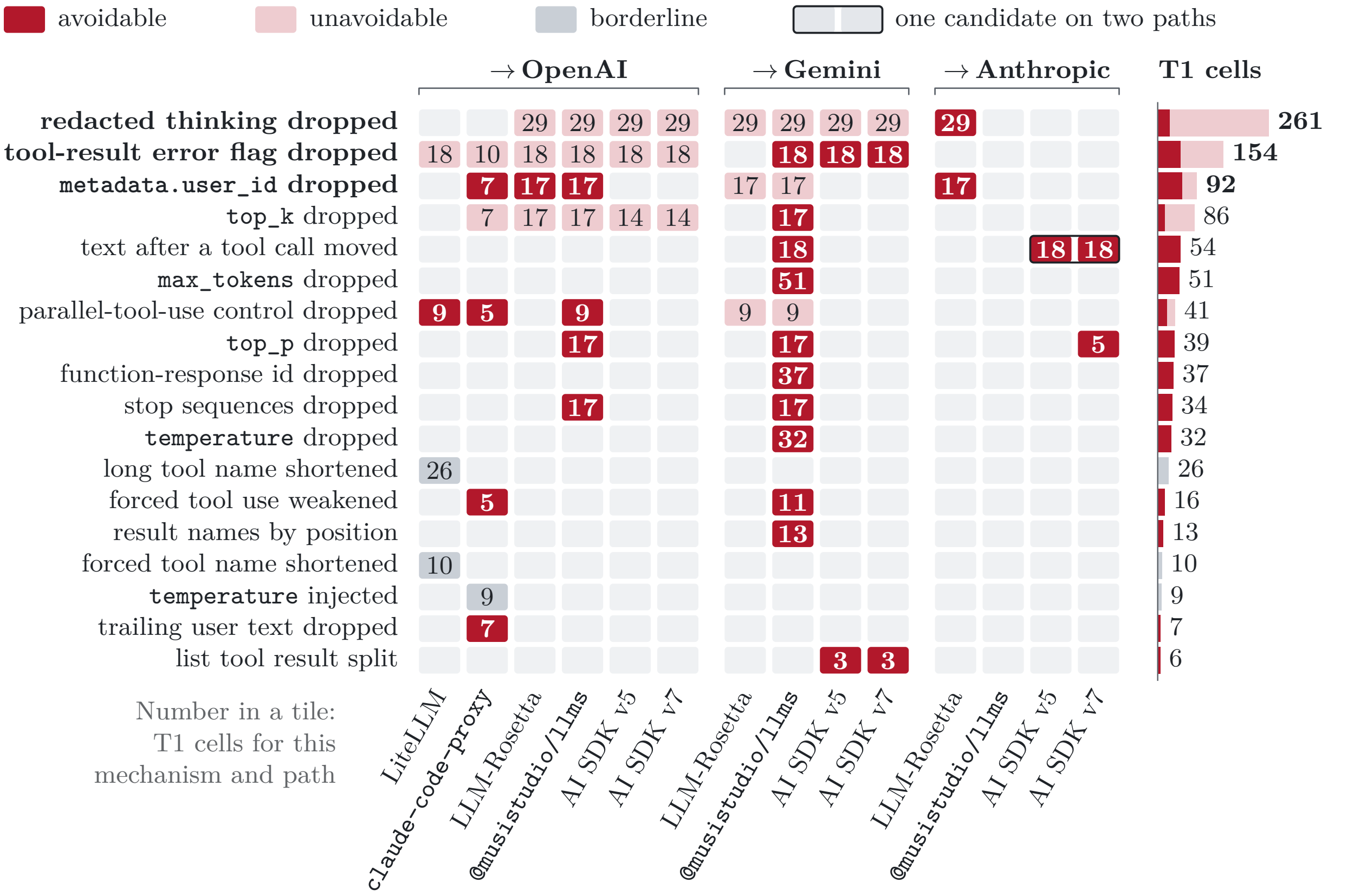}
  \caption{\textbf{The 54 post-review candidates across 18 mechanisms.} Each tile represents one path and mechanism, coloured by target support (\code{M}, \code{U}, or \code{Borderline}), and gives its affected T1 cells on the 51-shape suite. Right-hand bars aggregate by mechanism. Deduplicating paths leaves 48 candidates on 11 paths. Table~\ref{tab:mechanisms} records the pre-review source classifications and upstream status; Section~\ref{sec:rq1} gives the human grades.} 
  \label{fig:mech-grid}
\end{figure}

Table~\ref{tab:mechanisms} records 56 candidates across 19 mechanisms before review. Human review identified LLM-Rosetta's re-encoding of list-valued
tool results for Gemini as an acceptable change.
We therefore removed this candidate, which affected seven cells
and contributed no primary findings.
We also merged two duplicate candidate records for assistant-text
reordering in AI SDK v5 and v7 translations to Anthropic;
each record covered 18 affected cells. The final catalogue therefore contains 54 candidates across 18 mechanisms, or 48 candidates on 11 deduplicated paths (Figure~\ref{fig:mech-grid}). It contains 41 omissions and 13 explicit rewrites; human review yields 31 Grade~A defects, 11 Grade~B acceptable changes, and 12 Grade~C disagreements. The target can represent the affected information for 28 candidates (\code{M}), cannot do so for 23 (\code{U}), and support is borderline for 3. Among the 28 avoidable candidates, 23 are confirmed defects, 1 is acceptable, and 4 are disputed: the confirmed share is 23/24 after excluding disputes, or 23/28 with all candidates retained.

Tables~\ref{tab:sensitivity} and~\ref{tab:tiers-full} report alternatives to the 127/685 primary rate (18.5\,\%, shape-clustered 95\,\% CI 15.01--22.06\,\%). Validity, scope, and tier-boundary variants range from 12.9\,\% to 19.6\,\%. Examples are 49/379 after dropping AI SDK paths, 96/685 (14.0\,\%) when cells rejected live for missing Gemini \code{thought\_signature} are not treated as accepted targets, while retaining the 685-cell denominator, and 132/685 (19.3\,\%) using LiteLLM's serialised \code{hosted\_vllm} body. Under this study's conservative grading, the baseline treats role demotion (\code{D02}) as \code{T2}; promoting it to \code{T1} gives 134/685 (19.6\,\%). Counting variants range from 16.5\,\% to 25.7\,\%: 88/532 on 11 deduplicated paths, 146/685 when separate findings in a cell may jointly meet the criteria, and 176/685 when decoding controls (\code{K}) are included. Silence means that the translator gives no warning about the change; it is independent of static contract validity and live acceptance. The primary rate requires one individual finding to meet all criteria.

\begin{table}[tbp]
  \centering\scriptsize
  \caption{\textbf{Pre-review request mechanisms.} The table contains 56 candidates across 19 mechanisms behind the 467 T1 cells in Table~\ref{tab:census}; review leaves 54 candidates across 18 mechanisms (Appendix~\ref{app:audit}, Figure~\ref{fig:mechanisms}). \code{Cand.} counts pre-review candidates. \code{T1 cells} counts affected cells once per mechanism; \code{Static} counts those whose translated requests pass the static target checks. Static acceptance is distinct from a translator warning or live endpoint acceptance. M/U/Borderline denotes supported, unsupported, or borderline target representation. E/O denotes an explicit rewrite or omission located in the pinned source. Upstream gives covered reports and their status (\suppref{app:persistence}{S2.6}).} 
  \label{tab:mechanisms}
  % AUTO-GENERATED by tools/gen_tables.py mechanisms() from artifacts/g4/census_recount_r2/out/mechanisms_r2_2026-10-12T0900Z.json, artifacts/g4/census_recount_r2/out/representability_r2_2026-10-12T0901Z.json, artifacts/g4/mechanism_code_locations/out/mechanism_code_locations_2026-10-12T0939Z.json, artifacts/g4/candidate_novelty/out/candidate_novelty_v2_2026-10-12T2156Z.json on 2026-10-13 17:38 UTC — do not edit
\setlength{\tabcolsep}{3.5pt}
\begin{tabular}{@{}lrrrccl@{}}
\toprule
Mechanism (pre-review) & Cand. & T1 cells & Static & M/U/Borderline & Code & Upstream \\
\midrule
redacted thinking block dropped & 9 & 261 & 210 & 1/8/0 & O & 3 rep.: 3 merged \\
tool-result error flag dropped & 9 & 154 & 92 & 3/6/0 & E+O & 8 rep.: 5 merged, 1 ack., 2 open \\
\code{metadata.user\_id} dropped & 6 & 92 & 71 & 4/2/0 & O & -- \\
\code{top\_k} dropped & 6 & 86 & 55 & 1/5/0 & O & -- \\
assistant text after a call moved before it & 3 & 54 & 54 & 3/0/0 & E & -- \\
\code{max\_tokens} dropped & 1 & 51 & 51 & 1/0/0 & O & -- \\
parallel-tool-use control dropped & 5 & 41 & 23 & 3/2/0 & O & -- \\
\code{top\_p} dropped & 3 & 39 & 29 & 3/0/0 & E+O & -- \\
id of a Gemini function response dropped & 1 & 37 & 37 & 1/0/0 & O & 1 rep.: 1 open \\
stop sequences dropped & 2 & 34 & 24 & 2/0/0 & O & -- \\
\code{temperature} dropped & 1 & 32 & 32 & 1/0/0 & O & -- \\
long tool name shortened & 1 & 26 & 0 & 0/0/1 & E & -- \\
forced tool use weakened or dropped & 2 & 16 & 13 & 2/0/0 & E & 2 rep.: 2 open \\
result names assigned by position & 1 & 13 & 13 & 1/0/0 & E & -- \\
forced tool name shortened & 1 & 10 & 0 & 0/0/1 & E & -- \\
\code{temperature} injected & 1 & 9 & 3 & 0/0/1 & E & -- \\
list-valued tool result re-encoded & 1 & 7 & 7 & 0/0/1 & E & -- \\
user text after a tool result dropped & 1 & 7 & 2 & 1/0/0 & O & 1 rep.: 1 open \\
list-valued tool result split & 2 & 6 & 6 & 2/0/0 & E & -- \\
\midrule
\textit{19 mechanisms (pre-review)} & 56 & & & 29/23/4 & 15\,E, 41\,O & 15 rep.: 8 merged, 1 ack., 6 open \\
\bottomrule
\end{tabular}
\par\smallskip
{\footnotesize Static denotes offline target-contract acceptance. The catalogue shown has 56 candidates before review; removing one acceptable re-encoding and merging one duplicate leaves 54 candidates across 18 mechanisms.}

\end{table}

\begin{table}[tbp]
  \centering\scriptsize\setlength{\tabcolsep}{3.5pt}
  \caption{\textbf{Sensitivity of the main census to validity and counting choices.} $N$ is the number of judged cells and $K$ the number of request-shape clusters; variants \code{(a)}--\code{(g)} change the assumptions specified in each row.}
  \label{tab:sensitivity}
  % tables/sensitivity.tex — generated from rev1_census_patch/out/D3_sensitivities.csv (C-703..C-708). %nocheck
\begin{tabular}{@{}lP{3.75cm}rrrrrr@{}}
\toprule
\textbf{Var.} & \textbf{Specification} & \textbf{$N$ ($K$)} & \textbf{T1} & \textbf{Static T1} & \textbf{\shortstack{Static + Av.\\(ledger)}} & \textbf{\shortstack{Cell-level\\criteria}} & \textbf{\shortstack{Primary\\(one finding)}} \\
\midrule
\code{0} & Baseline \code{S-strict-R9} (14 paths) & 685 (51) & 467 (68.18\,\%) & 333 (48.61\,\%) & 206 (30.07\,\%) & 146 (21.31\,\%) & 127 (18.54\,\%) \\
\code{a1} & Drop \code{redacted\_thinking\_dropped} (O) findings & 685 (51) & 340 (49.64\,\%) & 234 (34.16\,\%) & 179 (26.13\,\%) & 146 (21.31\,\%) & 127 (18.54\,\%) \\
\code{a2} & Drop 29 shapes with \code{thinking=redacted} & 308 (22) & 165 (53.57\,\%) & 101 (32.79\,\%) & 80 (25.97\,\%) & 69 (22.40\,\%) & 60 (19.48\,\%) \\
\code{b} & Drop 17 \code{temperature}+\code{top\_p} shapes & 457 (34) & 297 (64.99\,\%) & 206 (45.08\,\%) & 125 (27.35\,\%) & 93 (20.35\,\%) & 84 (18.38\,\%) \\
\code{c1} & Drop AI SDK~v5 (11 paths, 5 codebases) & 532 (51) & 370 (69.55\,\%) & 254 (47.74\,\%) & 163 (30.64\,\%) & 107 (20.11\,\%) & 88 (16.54\,\%) \\
\code{c2} & Drop all AI SDK paths (8 paths, 4 projects) & 379 (51) & 273 (72.03\,\%) & 175 (46.17\,\%) & 120 (31.66\,\%) & 68 (17.94\,\%) & 49 (12.93\,\%) \\
\code{d1} & Project only: LiteLLM (1 path) & 51 (51) & 37 (72.55\,\%) & 0 (0.00\,\%) & 0 (0.00\,\%) & 0 (0.00\,\%) & 0 (0.00\,\%) \\
\code{d2} & Project only: claude-code-proxy (1 path) & 22 (22) & 21 (95.45\,\%) & 7 (31.82\,\%) & 7 (31.82\,\%) & 6 (27.27\,\%) & 5 (22.73\,\%) \\
\code{d3} & Project only: LLM-Rosetta (3 paths) & 153 (51) & 120 (78.43\,\%) & 100 (65.36\,\%) & 53 (34.64\,\%) & 11 (7.19\,\%) & 0 (0.00\,\%) \\
\code{d4} & Project only: \code{@musistudio/llms} (3 paths) & 153 (51) & 95 (62.09\,\%) & 68 (44.44\,\%) & 60 (39.22\,\%) & 51 (33.33\,\%) & 44 (28.76\,\%) \\
\code{d5} & Project only: Vercel AI SDK~v7 (3 paths; v5 identical) & 153 (51) & 97 (63.40\,\%) & 79 (51.63\,\%) & 43 (28.10\,\%) & 39 (25.49\,\%) & 39 (25.49\,\%) \\
\code{e1} & Count $\text{T1}\cup\text{T2}$ (T2 as documented degradation U) & 685 (51) & 544 (79.42\,\%) & 402 (58.69\,\%) & 206 (30.07\,\%) & 151 (22.04\,\%) & 127 (18.54\,\%) \\
\code{e2} & Count $\text{T1}\cup\text{T2}$ (representable T2 incl.\ D02 as M) & 685 (51) & 544 (79.42\,\%) & 402 (58.69\,\%) & 275 (40.15\,\%) & 151 (22.04\,\%) & 134 (19.56\,\%) \\
\code{f1} & Exclude 44 live-Gemini HTTP~400 cells from the static-based subset & 685 (51) & 467 (68.18\,\%) & 292 (42.63\,\%) & 172 (25.11\,\%) & 112 (16.35\,\%) & 96 (14.01\,\%) \\
\code{f2} & Combined \code{(b)}+\code{(f1)} (34 shapes + live-Gemini rejections) & 457 (34) & 297 (64.99\,\%) & 177 (38.73\,\%) & 101 (22.10\,\%) & 69 (15.10\,\%) & 62 (13.57\,\%) \\
\code{g1} & Promote D02 (\code{system}$\to$\code{user}) to T1 (Vis, M) on \code{S-strict-R9} & 685 (51) & 467 (68.18\,\%) & 333 (48.61\,\%) & 206 (30.07\,\%) & 151 (22.04\,\%) & 134 (19.56\,\%) \\
\code{g2-b} & \code{S-midsys-R9} baseline (70 shapes, D02 as T2) & 856 (70) & 595 (69.51\,\%) & 426 (49.77\,\%) & 274 (32.01\,\%) & 211 (24.65\,\%) & 187 (21.85\,\%) \\
\code{g2} & Promote D02 to T1 (Vis, M) on \code{S-midsys-R9} & 856 (70) & 595 (69.51\,\%) & 426 (49.77\,\%) & 274 (32.01\,\%) & 216 (25.23\,\%) & 194 (22.66\,\%) \\
\bottomrule
\end{tabular}
\par\smallskip
{\footnotesize Static means offline target-contract acceptance, with the row-specific exclusions stated above. Av. means that the target can represent the affected information. The cell-level column allows different findings in a cell to satisfy different criteria; the primary column requires one finding to satisfy all criteria. Silence means no warning, independently of static or live acceptance. Rows e1/e2 include T2 as specified; D02 role demotion is conservatively T2 in the baseline.}

\end{table}

\begin{table}[tbp]
  \centering\scriptsize
  \caption{\textbf{Request counts under static checks, endpoint acceptance, and mechanism coverage.} Stratum~1 is the main suite; Strata~2a--2c are acceptance subsets rather than a nested sequence. Static T1 counts meaning changes whose translated bodies pass static target checks; it does not establish live acceptance or the absence of a warning. The primary column requires one finding to satisfy all criteria. $^{\dagger}$Removing the LLM-Rosetta-to Gemini list-valued tool-result re-encoding candidate changes the cell-level conjunction to 140/685 (101/532 after deduplication) but leaves the primary counts 127/685 and 88/532 unchanged. $^{\ddagger}$Both-end acceptance covers eight target paths, four Anthropic and four Gemini; OpenAI targets are checked offline only. $^{\S}$The mechanism-coverage row uses the original 685-cell suite: 240/685 cells carry at least one of nine mechanisms with a live-probe effect, not 240 individually live-tested translations. Its percentages use the corresponding Stratum~1 column totals. Finding-aligned coverage of the primary findings is 116/127 (91.34\,\%); counting a mechanism merely present elsewhere in the same cell would instead give 120/127 and is not the reported primary-coverage definition.}
  \label{tab:tiers-full}
  % tables/evidence_tiers.tex — Three-tier evidence funnel on S-strict-R9 (C-1024..C-1028). %nocheck
\setlength{\tabcolsep}{3.0pt}
\begin{tabular}{@{}lP{3.75cm}rrrrrr@{}}
\toprule
\textbf{Stratum} & \textbf{Acceptance or coverage criterion} & \textbf{$N$ ($K$)} & \textbf{T1} & \textbf{Static T1} & \textbf{Static + Av.} & \textbf{\shortstack{Primary\\(one finding)}} & \textbf{\shortstack{Cell-level\\criteria}} \\
\midrule
\textbf{1} & Static schema/doc contract \code{S-strict-R9} (14 paths) & 685 (51) & 467 (68.18\,\%) & 333 (48.61\,\%) & 206 (30.07\,\%) & \textbf{127 (18.54\,\%)} & 146 (21.31\,\%)$^{\dagger}$ \\
\textbf{1} (11p) & Static \code{S-strict-R9}, AI SDK~v5 deduplicated (11 paths) & 532 (51) & 370 (69.55\,\%) & 254 (47.74\,\%) & 163 (30.64\,\%) & \textbf{88 (16.54\,\%)} & 107 (20.11\,\%)$^{\dagger}$ \\
\midrule
\textbf{2a} & Source live HTTP~200 on \code{claude-sonnet-4-6} (14 paths) & 182 (13) & 80 (43.96\,\%) & 51 (28.02\,\%) & 42 (23.08\,\%) & \textbf{34 (18.68\,\%)} & 39 (21.43\,\%) \\
\textbf{2a} (11p) & Source live HTTP~200 on \code{sonnet-4-6}, dedup.\ (11 paths) & 143 (13) & 65 (45.45\,\%) & 38 (26.57\,\%) & 31 (21.68\,\%) & \textbf{23 (16.08\,\%)} & 28 (19.58\,\%) \\
\textbf{2b} & Source live HTTP~200 on \code{claude-opus-4-8} (14 paths) & 112 (8) & 58 (51.79\,\%) & 39 (34.82\,\%) & 30 (26.79\,\%) & \textbf{23 (20.54\,\%)} & 28 (25.00\,\%) \\
\textbf{2b} (11p) & Source live HTTP~200 on \code{opus-4-8}, dedup.\ (11 paths) & 88 (8) & 47 (53.41\,\%) & 29 (32.95\,\%) & 22 (25.00\,\%) & \textbf{15 (17.05\,\%)} & 20 (22.73\,\%) \\
\textbf{2c} & Source \code{sonnet-4-6} 200 $\wedge$ target live HTTP~200 (8 paths)$^{\ddagger}$ & 88 (13) & 27 (30.68\,\%) & 27 (30.68\,\%) & 26 (29.55\,\%) & \textbf{22 (25.00\,\%)} & 24 (27.27\,\%) \\
\midrule
\textbf{3} & Mechanism coverage: carries $\ge 1$ of 9 mechanisms with a measured live effect (original suite)$^{\S}$ & 685 (51) & 240 (51.39\,\%) & 166 (49.85\,\%) & 141 (68.45\,\%) & \textbf{116 (91.34\,\%)} & 132 (90.41\,\%) \\
\bottomrule
\end{tabular}
\par\smallskip
{\footnotesize Static means offline target-contract acceptance; silence means no translator warning. The primary column requires the same finding to satisfy all criteria; the cell-level column permits separate findings. The acceptance subsets are not a single nested sequence. Row 3 keeps the original 685-cell denominator: 240/685 (35.04\,\%) carry a relevant mechanism. Its displayed percentages are conditional on the corresponding Stratum-1 column totals, including finding-aligned coverage of 116/127 (91.34\,\%) primary cells. This is mechanism coverage, not individual live validation; cell-level co-occurrence would give 120/127.}

\end{table}

\subsection{Live behavioural consequences, GSM8K control, and schema probes}
\label{app:consequence}\label{app:rq2-detail}\label{app:outside}\label{app:detail-consequences}

We test live consequences on \code{gemini-3.5-flash-lite}, \code{gemini-3.8-flash}, \code{claude-sonnet-4-6}, \code{claude-haiku-5-5}, \code{Meta-Llama-3.1-8B-Instruct}, and \code{Qwen2.5-7B-Instruct}. Nine of the ten tested census mechanisms affect at least one measured outcome; a separate probe tests the dropped mid-conversation \code{system} instruction. In total, 31 of 64 contrasts survive global Holm correction. The nine mechanisms correspond to findings in 116/127 primary cells (91.3\,\%); this is mechanism coverage, not individual live validation of those cells. Running translator processes again rather than replaying archived bodies gives the same outcomes in all 9 checked cells (Tables~\ref{tab:claudeconseq}--\ref{tab:consequence-detail}).

The Gemini experiments use 720 plus 120 calls at temperature~0.7. Four mechanisms reduce compliance from 30/30 to 0/30 on both models ($p_{\text{Holm}}\le0.0117$). The Claude experiments use 512 calls without retries. Dropping \code{top\_p:\,0.05} increases output diversity on Sonnet ($\bar d=+1.00$, $p=0.0156$) and converts Haiku's deprecation error into HTTP~200 replies (10/10). Raising \code{max\_tokens} from 100 to 1,124 reduces compliance with the original cap from 30/30 to 0/30 on both models ($p_{\text{Holm}}=0.0039$). The thinking-block-removal probes do not produce the predicted rejection effect ($p=1.0000$).

On the first 100 GSM8K items, evaluated in seven arms on \code{gemini-3.5-flash-lite}, the translator arms show accuracy differences of 0 to +3 percentage points and no statistically detectable drop ($p\ge0.2500$; Table~\ref{tab:exph}). This experiment does not establish general benchmark or agent performance. Separately, probes of 40 JSON Schema keywords outside \code{ConvIR}'s core rules find that native Gemini \code{parameters} accepts 27, while Google's OpenAI-compatible layer accepts all 40.

\begin{table}[tbp]
  \centering\footnotesize
  \caption{\textbf{Claude consequence probes.} Four mechanisms are evaluated on \code{claude-sonnet-4-6} and \code{claude-haiku-5-5} in 512 calls without retries: thinking-block removal and \code{top\_p:\,0.05} removal (212 calls), dropped tool prohibitions and \code{max\_tokens} raised from 100 to 1,124 (240 calls), and a thinking-enabled token-cap rerun on Sonnet (60 calls). Each arm uses 10 scenarios, with five samples per scenario for diversity and three replicates for prohibition and token-cap compliance. The prohibition prediction is met on Haiku ($\bar d=1.00$, $p_{\text{Holm}}=0.0039$) but not Sonnet ($\bar d=0.33$, $p_{\text{Holm}}=0.125$). The predicted rejection after thinking-block removal is not observed (10/10 HTTP~200 in both arms). \code{usage.output\_tokens\_details} is a post-smoke-test exploratory indicator and is uninformative on Haiku. The thinking-enabled Haiku rerun is withdrawn after rejection of \code{thinking.type.enabled}; the archive records the experiment plans, amendments, and remaining acceptance predictions.}
  \label{tab:claudeconseq}
  % AUTO-GENERATED by tools/fig/make_figdata.py from results_live_20261015T020247Z.json, claude_consequence_c3/out/live_20261015T055512Z/calls.jsonl, and claude_consequence_c3/out/live_c3b_20261015T182707Z/calls.jsonl (C-921..C-929, C-939, C-1046) — do not edit
% Intended label in paper: \label{tab:claudeconseq}
{\footnotesize%
\setlength{\tabcolsep}{2.8pt}%
\begin{tabular}{@{}P{1.38cm}P{2.65cm}P{2.76cm}P{3.16cm}P{3.20cm}P{2.35cm}@{}}
\toprule
Track & Model / mode & Mutation (translator behaviour) & Ref outcome & Mut outcome & Pre-specified prediction \\
\midrule
  MA-1 & \code{claude-sonnet-4-6}, manual thinking (budget 1,024) & thinking blocks stripped from the assistant turn before \code{tool\_result} & 10/10 HTTP 200; \code{output\_tokens\_details} 10/10 & 10/10 HTTP 200, $0 \times 400$; thinking 0/10; \code{output\_tokens\_details} 0/10 & Ref HTTP~200 $\ge 9/10$; Mut HTTP~400 $\ge 9/10$: \textbf{not met} \\
  MA-2 & \code{claude-sonnet-4-6}, adaptive & same & 10/10 200; \code{output\_tokens\_details} 10/10 & 10/10 200; thinking 0/10; \code{output\_tokens\_details} 10/10 & Mut 200 $\ge$ 9/10: met \\
  MA-2 & \code{claude-haiku-5-5}, adaptive & same & 10/10 200 & 10/10 200; thinking 1/10 & Same criterion: met \\
  MB-1 & \code{claude-sonnet-4-6} & \code{top\_p} 0.05 dropped & distinct strings per 5 samples: mean 3.8 ($\Sigma$38/10) & 4.8 ($\Sigma$48/10); $\bar{d} = +1.00$ [0.5, 1.5], $p = 0.0156$ & $\bar{d} \ge 1$, $p < 0.05$: met \\
  MB-2 & \code{claude-haiku-5-5} & \code{top\_p} 0.05 dropped & 10/10 HTTP 400 ``\textasciigrave\code{top\_p}\textasciigrave\ is deprecated for this model.'' & 10/10 HTTP 200 &Ref HTTP~400 $\ge 9/10$; Mut HTTP~200 $\ge 9/10$: met \\
  MC3-1 & \code{claude-sonnet-4-6} & mid-turn \code{[harness] <tool> is DISABLED} block (inlined in the last user turn) dropped & 10/30 refuse (no \code{tool\_use}); 20/30 call the tool, 9 of them stating that an in-chat instruction cannot disable a tool & 30/30 call the tool; $\bar{d} = +0.33$ [0.10, 0.63], $p = 0.125$ & P-C3-1 ($\bar{d} \ge 0.5$, $p_{\mathrm{Holm}} < 0.05$): \textbf{not met} \\
  MC3-1 & \code{claude-haiku-5-5} & same & 30/30 refuse & 30/30 call the tool; $\bar{d} = +1.00$ [1.00, 1.00], $p = 0.0020$ &$\bar{d} \ge 0.5$, $p_{\mathrm{Holm}} < 0.05$: \textbf{not met}  \\
  MC3-2 & \code{claude-sonnet-4-6} & \code{max\_tokens} 100 $\to$ 1124 & 30/30 \code{output\_tokens} = 100, stop \code{max\_tokens} & 30/30 $>$ 100 (400--1124, median 456.5; 11 at the cap); $\bar{d} = +1.00$, $p = 0.0020$ & P-C3-2 ($\bar{d} \ge 0.7$, $p_{\mathrm{Holm}} < 0.05$): met \\
  MC3-2 & \code{claude-haiku-5-5} & same & 30/30 = 100 & 30/30 $>$ 100 (697--1124, median 1124; 22 at the cap); $\bar{d} = +1.00$, $p = 0.0020$ & $\bar{d} \ge 0.7$, $p_{\mathrm{Holm}} < 0.05$: met  \\
  MC3-2 (thinking) & \code{claude-sonnet-4-6}; Mut: \code{thinking} enabled (\code{budget\_tokens} 1,024); Ref: no \code{thinking} & \code{max\_tokens} 100 $\to$ 1124 (\code{@ai-sdk/anthropic}) & 30/30 visible $\le$ 100 (100--100, median 100; 0 at 1124 cap; raw \code{output\_tokens} 100--100), stop \code{max\_tokens} & 0/30 visible $\le$ 100 (392--1102, median 465.5; 12 at 1124 cap; raw \code{output\_tokens} 414--1124, median 488); $\bar{d} = +1.00$ [1.00, 1.00], $p = 0.0020$ & Output $\le 100$ visible tokens: Mut rate $\le 0.30$: met;
Ref rate $\ge 0.95$: met;
HTTP~200 $\ge 99\%$ of 60: met \\
\bottomrule
\end{tabular}%
}%

\end{table}

\begin{table}[tbp]
  \centering\scriptsize
  \setlength{\tabcolsep}{3.2pt}
  \caption{\textbf{Counts and Wilson 95\,\% intervals for the live consequence matrix.} Rows cover the nine census mechanisms with measured effects, the separate mid-conversation system-instruction probe, and the tenth tested census mechanism, dropped \code{top\_k}, whose probes show no difference where the prompts do not bind that setting. Temperature-0 single-prompt Gemini trials ($n=10$), temperature-0.7 multi-scenario trials ($n=30$ or 60), and Claude trials (Table~\ref{tab:claudeconseq}) are reported separately. Sampling and output-limit settings are outside the primary rate but can affect behaviour.}
  \label{tab:consequence-detail}
  \begin{tabular}{@{}P{3.6cm}P{3.45cm}cP{3.9cm}P{3.9cm}@{}}
    \toprule
    Mechanism & Model & $T$ & Intact $k/n$ (\%; Wilson $95\,\%$ CI) & Translated $k/n$ (\%; Wilson $95\,\%$ CI) \\ %nocheck
    \midrule
    \multicolumn{5}{@{}l@{}}{\textit{Prompt, role, and tool information}} \\
    tool-result error flag dropped & \code{gemini-3.5-flash-lite} & 0.7 & 30/30 (100.0\,\%; 88.65--100.00) & 0/30 (0.0\,\%; 0.00--11.35) \\ %nocheck
    \quad (154 census cells) & \code{gemini-3.8-flash} & 0.7 & 30/30 (100.0\,\%; 88.65--100.00) & 0/30 (0.0\,\%; 0.00--11.35) \\ %nocheck
     & \code{Qwen2.5-7B-Instruct} & 0.7 & 25/60 (41.67\,\%; 30.06--54.27) & 0/60 (0.0\,\%; 0.00--6.02) \\ %nocheck
     & \code{gemini-3.5-flash-lite} & 0.0 & 10/10 (100.0\,\%; 72.25--100.00) & 0/10 (0.0\,\%; 0.00--27.75) \\ %nocheck
     & \code{gemini-3.8-flash} & 0.0 & 10/10 (100.0\,\%; 72.25--100.00) & 10/10 (100.0\,\%; 72.25--100.00) \\ %nocheck
    text after a tool call moved & \code{Meta-Llama-3.1-8B-Instruct} & 0.7 & 30/30 (100.0\,\%; 88.65--100.00) & 30/30 (100.0\,\%; 88.65--100.00) \\ %nocheck
    \quad (54 census cells) & \code{Qwen2.5-7B-Instruct} & 0.7 & 24/30 (80.0\,\%; 62.69--90.49) & 1/30 (3.33\,\%; 0.59--16.67) \\ %nocheck
    function-response id dropped & \code{gemini-3.5-flash-lite} & 0.7 & 30/30 (100.0\,\%; 88.65--100.00) & 0/30 (0.0\,\%; 0.00--11.35) \\ %nocheck
    \quad (37 census cells) & \code{gemini-3.8-flash} & 0.7 & 30/30 (100.0\,\%; 88.65--100.00) & 0/30 (0.0\,\%; 0.00--11.35) \\ %nocheck
     & \code{Meta-Llama-3.1-8B-Instruct} & 0.7 & \multicolumn{2}{@{}P{7.0cm}@{}}{\code{jinja\_400}: GGUF chat template rejects multi-tool history (0/30 vs.\ 0/30)} \\ %nocheck
     & \code{Qwen2.5-7B-Instruct} & 0.7 & \multicolumn{2}{@{}P{7.0cm}@{}}{\code{jinja\_strip}: GGUF template strips \code{tool\_call\_id} in both arms (0/30 vs.\ 0/30)} \\ %nocheck
    forced tool use weakened & \code{gemini-3.5-flash-lite} & 0.7 & 30/30 (100.0\,\%; 88.65--100.00) & 3/30 (10.0\,\%; 3.46--25.62) \\ %nocheck
    \quad (16 census cells) & \code{gemini-3.8-flash} & 0.7 & 27/30 (90.0\,\%; 74.38--96.54) & 5/30 (16.67\,\%; 7.34--33.56) \\ %nocheck
     & \code{Qwen2.5-7B-Instruct} & 0.7 & 43/60 (71.67\,\%; 59.23--81.49) & 3/60 (5.0\,\%; 1.71--13.70) \\ %nocheck
     & \code{gemini-3.5-flash-lite} & 0.0 & 10/10 (100.0\,\%; 72.25--100.00) & 0/10 (0.0\,\%; 0.00--27.75) \\ %nocheck
     & \code{gemini-3.8-flash} & 0.0 & 10/10 (100.0\,\%; 72.25--100.00) & 0/10 (0.0\,\%; 0.00--27.75) \\ %nocheck
    result names by position & \code{gemini-3.5-flash-lite} & 0.7 & 30/30 (100.0\,\%; 88.65--100.00) & 3/30 (10.0\,\%; 3.46--25.62) \\ %nocheck
    \quad (13 census cells) & \code{gemini-3.8-flash} & 0.7 & 30/30 (100.0\,\%; 88.65--100.00) & 0/30 (0.0\,\%; 0.00--11.35) \\ %nocheck
    trailing user text dropped & \code{gemini-3.5-flash-lite} & 0.7 & 30/30 (100.0\,\%; 88.65--100.00) & 0/30 (0.0\,\%; 0.00--11.35) \\ %nocheck
    \quad (7 census cells) & \code{gemini-3.8-flash} & 0.7 & 30/30 (100.0\,\%; 88.65--100.00) & 0/30 (0.0\,\%; 0.00--11.35) \\ %nocheck
     & \code{Meta-Llama-3.1-8B-Instruct} & 0.7 & 30/30 (100.0\,\%; 88.65--100.00) & 0/30 (0.0\,\%; 0.00--11.35) \\ %nocheck
     & \code{Qwen2.5-7B-Instruct} & 0.7 & 60/60 (100.0\,\%; 93.98--100.00) & 0/60 (0.0\,\%; 0.00--6.02) \\ %nocheck
     & \code{gemini-3.5-flash-lite} & 0.0 & 10/10 (100.0\,\%; 72.25--100.00) & 0/10 (0.0\,\%; 0.00--27.75) \\ %nocheck
     & \code{gemini-3.8-flash} & 0.0 & 10/10 (100.0\,\%; 72.25--100.00) & 0/10 (0.0\,\%; 0.00--27.75) \\ %nocheck
    mid-turn system dropped & \code{claude-sonnet-4-6} (\code{MC3-1}) & def. & 10/30 (33.33\,\%; 19.23--51.22) & 0/30 (0.0\,\%; 0.00--11.35); \textbf{not met} \\ %nocheck
    \quad (separate test) & \code{claude-haiku-5-5} (\code{MC3-1}) & def. & 30/30 (100.0\,\%; 88.65--100.00) & 0/30 (0.0\,\%; 0.00--11.35); met \\ %nocheck
     & \code{Qwen2.5-7B-Instruct} & 0.7 & 50/60 (83.33\,\%; 71.97--90.69) & 13/60 (21.67\,\%; 13.12--33.62) \\ %nocheck
     & \code{gemini-3.5-flash-lite} & 0.0 & 10/10 (100.0\,\%; 72.25--100.00) & 0/10 (0.0\,\%; 0.00--27.75) \\ %nocheck
     & \code{gemini-3.8-flash} & 0.0 & 10/10 (100.0\,\%; 72.25--100.00) & 0/10 (0.0\,\%; 0.00--27.75) \\ %nocheck
    \midrule
    \multicolumn{5}{@{}l@{}}{\textit{Sampling and output-limit settings (outside the primary rate)}} \\
    \code{max\_tokens} dropped or raised & \code{gemini-3.5-flash-lite} & 0.7 & 30/30 (100.0\,\%; 88.65--100.00) & 0/30 (0.0\,\%; 0.00--11.35) \\ %nocheck
    \quad (51 census cells) & \code{gemini-3.8-flash} & 0.7 & 30/30 (100.0\,\%; 88.65--100.00) & 0/30 (0.0\,\%; 0.00--11.35) \\ %nocheck
     & \code{claude-sonnet-4-6} (\code{MC3-2}) & def. & 30/30 (100.0\,\%; 88.65--100.00) & 0/30 (0.0\,\%; 0.00--11.35); plain \& thinking \\ %nocheck
     & \code{claude-haiku-5-5} (\code{MC3-2}) & def. & 30/30 (100.0\,\%; 88.65--100.00) & 0/30 (0.0\,\%; 0.00--11.35) \\ %nocheck
     & \code{Meta-Llama-3.1-8B-Instruct} & 0.7 & 30/30 (100.0\,\%; 88.65--100.00) & 0/30 (0.0\,\%; 0.00--11.35) \\ %nocheck
     & \code{Qwen2.5-7B-Instruct} & 0.7 & 30/30 (100.0\,\%; 88.65--100.00) & 0/30 (0.0\,\%; 0.00--11.35) \\ %nocheck
    \code{top\_p} dropped & \code{claude-sonnet-4-6} (\code{MB-1}) & def. & mean 3.8 distinct / 5 ($n{=}50$) & mean 4.8 / 5 ($\bar{d}{=}{+}1.00$ [0.5, 1.5], $p{=}0.0156$) \\ %nocheck
    \quad (39 census cells) & \code{claude-haiku-5-5} (\code{MB-2}) & def. & 10/10 HTTP 400 (\code{top\_p} deprecated) & 10/10 HTTP 200 (rejection removed) \\ %nocheck
    stop sequences dropped & \code{gemini-3.5-flash-lite} & 0.7 & 29/30 (96.67\,\%; 83.33--99.41) & 0/30 (0.0\,\%; 0.00--11.35) \\ %nocheck
    \quad (34 census cells) & \code{gemini-3.8-flash} & 0.7 & 30/30 (100.0\,\%; 88.65--100.00) & 8/30 (26.67\,\%; 14.18--44.45) \\ %nocheck
     & \code{Meta-Llama-3.1-8B-Instruct} & 0.7 & 23/30 (76.67\,\%; 59.07--88.21) & 1/30 (3.33\,\%; 0.59--16.67) \\ %nocheck
     & \code{Qwen2.5-7B-Instruct} & 0.7 & 55/60 (91.67\,\%; 81.93--96.39) & 7/60 (11.67\,\%; 5.77--22.18) \\ %nocheck
    \code{top\_k} dropped (86 occurrences) & open-weight probes & 0.0 & \multicolumn{2}{@{}P{7.0cm}@{}}{no difference where prompts did not bind \code{top\_k} (Table~\ref{tab:mechanisms})} \\ %nocheck
    \bottomrule
  \end{tabular}
\end{table}

\begin{table}[tbp]
  \centering\scriptsize
  \caption{\textbf{GSM8K control experiment on \code{gemini-3.5-flash-lite}.} Seven arms each contain the first 100 items (700/700 HTTP~200). \code{A1} and \code{A5} use translator outputs. \code{A2} is retained only as a \code{maxOutputTokens}-omission control because the original capture driver omitted that field; it is withdrawn from translator-defect claims. \code{A3} (\code{aisdk5->gemini}) is omitted as byte-identical to \code{A2}. Four further arms are reference or intervention controls. Columns report required-format compliance (\code{\#\#\#\# <int>}), exact-tag and last-integer accuracy, \code{MAX\_TOKENS} counts, and mean output tokens. Contrasts with \code{A0} use exact two-sided McNemar tests and paired-bootstrap 95\,\% intervals for $\Delta$ (10,000 resamples; pre-specified EXP-H).}
  \label{tab:exph}
  % tables/exph.tex — EXP-H end-to-end GSM8K benchmark score shift across 7 arms (C-820..C-867).
\setlength{\tabcolsep}{1.0pt}
\begin{tabular}{@{}lP{2.68cm}rrrrrrrr@{}}
  \toprule
  Arm & Input / control & \shortstack{Format\\\% [CI]} & \shortstack{Exact tag\\\% [CI]} & \shortstack{Last integer\\\% [CI]} & \code{MAX} & \shortstack{Mean\\tokens} & \shortstack{$\Delta$ Format\\{[CI]}} & \shortstack{$\Delta$ Exact tag\\{[CI]}} & \shortstack{$\Delta$ Last integer\\{[CI]}} \\
  \midrule
  \code{A0} & Intact ref.\ (\code{max\_tokens: 1024}, \code{temp: 0.0}) & 98 {[}93.0,\,99.4{]} & 90 {[}82.6,\,94.5{]} & 97 {[}91.5,\,99.0{]} & 2 & 254.6 & --- & --- & --- \\
  \code{A1} & \code{musistudio->gemini} (D02 + S08) & 100 {[}96.3,\,100.0{]} & 91 {[}83.8,\,95.2{]} & 99 {[}94.6,\,99.8{]} & 0 & 291.8 & $+2$ {[}$+0.0,\,+5.0${]} & $+1$ {[}$+0.0,\,+3.0${]} & $+2$ {[}$+0.0,\,+5.0${]} \\
  \code{A2} & \code{aisdk7->gemini}: driver-omission control; translator-fault claim withdrawn & 100 {[}96.3,\,100.0{]} & 91 {[}83.8,\,95.2{]} & 98 {[}93.0,\,99.4{]} & 0 & 250.8 & $+2$ {[}$+0.0,\,+5.0${]} & $+1$ {[}$+0.0,\,+3.0${]} & $+1$ {[}$+0.0,\,+3.0${]} \\
  \code{A4} & \code{A0} repeated (nondeterminism floor) & 99 {[}94.6,\,99.8{]} & 90 {[}82.6,\,94.5{]} & 97 {[}91.5,\,99.0{]} & 1 & 250.5 & $+1$ {[}$+0.0,\,+3.0${]} & $+0$ {[}$+0.0,\,+0.0${]} & $+0$ {[}$+0.0,\,+0.0${]} \\
  \code{A5} & \code{rosetta->gemini} (clean wire) & 100 {[}96.3,\,100.0{]} & 92 {[}85.0,\,95.9{]} & 100 {[}96.3,\,100.0{]} & 0 & 249.9 & $+2$ {[}$+0.0,\,+5.0${]} & $+2$ {[}$+0.0,\,+5.0${]} & $+3$ {[}$+0.0,\,+7.0${]} \\
  \code{A6} & \code{A0} minus \code{temperature}, \code{maxOutputTokens} & 100 {[}96.3,\,100.0{]} & 91 {[}83.8,\,95.2{]} & 98 {[}93.0,\,99.4{]} & 0 & 263.4 & $+2$ {[}$+0.0,\,+5.0${]} & $+1$ {[}$+0.0,\,+3.0${]} & $+1$ {[}$+0.0,\,+3.0${]} \\
  \code{A7} & \code{A1} plus \code{temp: 0.0}, \code{maxOutputTokens: 1024} & 99 {[}94.6,\,99.8{]} & 91 {[}83.8,\,95.2{]} & 98 {[}93.0,\,99.4{]} & 1 & 269.2 & $+1$ {[}$+0.0,\,+3.0${]} & $+1$ {[}$+0.0,\,+3.0${]} & $+1$ {[}$+0.0,\,+3.0${]} \\
  \midrule
  \multicolumn{10}{@{}p{0.99\textwidth}@{}}{Exact two-sided McNemar $p$ vs.\ \code{A0} ($c{=}0$ in every contrast): $p{=}1.0000$ for $\Delta \in \{+0, +1\}$, $p{=}0.5000$ for $\Delta{=}+2$, $p{=}0.2500$ for $\Delta{=}+3$. \code{A3} (\code{aisdk5->gemini}) omitted in the pre-registration as byte-identical to \code{A2} on this prompt shape; \code{A2} is a capture-driver control: the original driver omitted \code{maxOutputTokens}, leaving \code{max\_tokens} absent; its translator-defect claim is withdrawn (\suppref{app:threats}{S2.18}).} \\
  \bottomrule
\end{tabular}

\end{table}

\subsection{Translated reply streams and official SDK failures}
\label{app:resp}\label{app:replies}\label{app:detail-replies}

Across 1,115 judged reply cells on seven translators (11 reply paths using 40 Anthropic and Gemini backend-response shapes), isloss reports 391 T1 changes (35.1\,\%), of which 278 are avoidable. Table~\ref{tab:sdk-split} gives a separate test of 558 HTTP~200 streams with the official \code{openai-python} \code{2.54.0} and \code{openai-node} \code{5.23.2} accumulators. At least one SDK raises an exception or silently corrupts the message in 203/558 streams (36.4\,\%, Wilson 95\,\% CI 32.5--40.5\,\%); 199 of those 203 outcomes are avoidable (98.0\,\%). Python is affected on 170 streams (52 exceptions and 118 silent corruptions), and Node on 182 (105 exceptions and 77 silent corruptions). The SDK outcomes disagree on 129/558 streams.

\begin{table}[tbp]
  \centering\scriptsize
  \caption{\textbf{Official SDK outcomes for 558 translated HTTP~200 streams.} Results are grouped by translator family for \code{openai-python} 2.54.0 and \code{openai-node} 5.23.2; CLIProxyAPI additionally refuses two streams with HTTP~500. Each Clean/U entry gives two counts: preserved replies / unavoidable losses; the slash does not denote a ratio. The exception/corruption split partitions affected streams into: both SDKs raise; one raises and the other silently corrupts; one raises and the other has neither failure; or both silently corrupt.} %nocheck
  \label{tab:sdk-split}
  \resizebox{\textwidth}{!}{% tables/sdk_split.tex — generated from rev1_census_patch/out/D7_sdk_split.csv and sdk_203_split_2026-10-14T2352Z.json (C-709, C-710). %nocheck
{\setlength{\tabcolsep}{2.6pt}%
\begin{tabular}{@{}lrrrrrrrrrrr@{}}
\toprule
& & \multicolumn{3}{c}{\textbf{\code{openai-python} 2.54.0}} & \multicolumn{3}{c}{\textbf{\code{openai-node} 5.23.2}} & \multicolumn{4}{c}{\textbf{Two-SDK union ($203/558 = 36.38\,\%$)}} \\
\cmidrule(lr){3-5}\cmidrule(lr){6-8}\cmidrule(lr){9-12}
\textbf{Family} & \textbf{$N$} & \textbf{Exc.} & \textbf{Corrupt} & \textbf{Clean/U} & \textbf{Exc.} & \textbf{Corrupt} & \textbf{Clean/U} & \textbf{Fail (\%)} & \textbf{Any exc.} & \textbf{Both corr.} & \textbf{Exc./corr.\ split} \\
\midrule
Bifrost & 80 & 2 & 8 & 59/11 & 11 & 6 & 52/11 & 17 (21.25\,\%) & 11 & 6 & 2 / 2 / 7 / 6 \\
new-api & 80 & 15 & 1 & 53/11 & 6 & 3 & 60/11 & 20 (25.00\,\%) & 20 & 0 & 1 / 4 / 15 / 0 \\
CLIProxyAPI & 78 & 4 & 13 & 50/11 & 9 & 6 & 52/11 & 17 (21.79\,\%) & 11 & 6 & 2 / 7 / 2 / 6 \\
Portkey & 80 & 2 & 21 & 47/10 & 5 & 19 & 46/10 & 24 (30.00\,\%) & 5 & 19 & 2 / 2 / 1 / 19 \\
one-api & 80 & 14 & 40 & 21/5 & 41 & 17 & 17/5 & 58 (72.50\,\%) & 55 & 3 & 0 / 51 / 4 / 3 \\
LiteLLM & 80 & 11 & 5 & 57/7 & 14 & 1 & 54/11 & 23 (28.75\,\%) & 22 & 1 & 3 / 4 / 15 / 1 \\
LLM-Rosetta & 80 & 4 & 30 & 39/7 & 19 & 25 & 29/7 & 44 (55.00\,\%) & 19 & 25 & 4 / 5 / 10 / 25 \\
\midrule
\textbf{All 7 families} & \textbf{558} & \textbf{52} & \textbf{118} & \textbf{326/62} & \textbf{105} & \textbf{77} & \textbf{310/66} & \textbf{203 (36.38\,\%)} & \textbf{143} & \textbf{60} & \textbf{14 / 75 / 54 / 60} \\
\bottomrule
\end{tabular}%
}
}
\end{table}

\subsection{Persistence, API endpoints, round-trip controls, and upstream fixes}
\label{app:persistence}\label{app:rq3-detail}\label{app:exits}\label{app:control}\label{app:detail-persistence}

The release study covers 50 LiteLLM releases (\code{1.52.16}--\code{1.104.0}), 54 \code{@musistudio/llms} releases, and 32 \code{claude-code-proxy} commits. Every tested LiteLLM release has T1 findings on at least 31/51 main-suite shapes when translating to OpenAI. A separate re-census uses the 11~October source snapshots of ten translators across 22 paths and the extended 170-shape corpus. At Step~2, 2,630/2,745 previously flagged T1 cells persist (95.8\,\%). After the Step~3 warning filter, 2,495/2,631 persist (94.8\,\%); this latter count comprises 1,517/1,653 library cells (91.8\,\%) and all gateway cells. The two stages therefore lose 115 and 136 cells, respectively, and should not be combined.

Table~\ref{tab:exits} compares Chat Completions and Responses on 357 inputs across five translators. Each translator's two endpoints differ on at least one of 22 pre-specified properties. Same-vendor round-trip controls also reveal a change in every tested translator. Table~\ref{tab:upstream} records 44 upstream posts across 12 repositories: 26 issues, 10 comments on existing threads, and 8 follow-ups. As of 15~October, 7 issues are fixed and re-verified, 1 is partly fixed, 1 is closed without reply, and 17 remain open.

The endpoint comparison uses LiteLLM \code{4417bf08} (1.106.0), LLM-Rosetta 0.13.0, CLIProxyAPI v8.0.20, Bifrost \code{8b4fce4f}, and AI SDK \code{ai} 7.0.126 with \code{@ai-sdk/openai} 4.0.83. AI SDK uses the original capture driver, which omits \code{maxOutputTokens}, passes \code{metadata.user\_id} only to Anthropic, and joins 15/25 list-valued results; the chat endpoint sends the other 10 as JSON part lists. Its signature, redacted-block, and error-flag denominators are 40, 45, and 57. These captures are separate from the corrected main-census driver, which passes \code{maxOutputTokens} and \code{providerOptions.anthropic.user} (Appendix~\ref{app:threats}). AI SDK rejects 236 mid-conversation-system shapes, leaving 121; enabling \code{allowSystemInMessages} lets 281 through. From \code{@ai-sdk/openai} 4.0.89, both endpoints retain the error flag on 57/57 cases.

The text-placement comparison applies to 64 shapes (18 for AI SDK). All ten endpoints map every \code{tool\_choice} and retain dotted tool names; Bifrost Responses adds an empty \code{reasoning} item per redacted block. The same-vendor controls use LiteLLM's first-party Anthropic path, Bifrost's raw Claude Code \code{User-Agent} path versus its default, and AI SDK's Anthropic endpoint. New-api and Portkey, outside the table, preserve all 357 bodies in this control.

\begin{table}[tbp]
  \centering\scriptsize
  \caption{\textbf{The same translator through Chat Completions (c) and Responses (R).} Counts describe outgoing properties, not isloss verdicts. The first column gives applicable shapes, or blocks for \code{is\_error}; AI SDK uses $k/n$ where its core rejects inputs before either endpoint runs. B permits \code{allowSystemInMessages}; c marks properties affected by the original capture driver. $\dagger$ denotes renamed or truncated tool names restored on return; $\ddagger$ denotes \code{role:system} items without \code{type}, counted as retained only where the track supplies that key; \S\ denotes another representation (\code{stop} or top-level \code{user}). For text surrounding a call: J joins it into one string, P retains separate parts, S retains separate messages/items, and arrows indicate movement across the call. The bottom row is a same-vendor control: unchanged bodies after removing \code{model}/\code{stream} and sorting keys; h marks post-hoc AI SDK normalisation.} 
  \label{tab:exits}
  % AUTO-GENERATED by tools/gen_tables.py exits() from artifacts/g4/two_exits_aisdk/out/FIVE_LIBS_TWO_EXITS.md, artifacts/g4/same_vendor_roundtrip/out/SAME_VENDOR_TABLE.md, artifacts/g4/two_exits_aisdk/out/ga_props_r7.0.126.json on 2026-10-08 18:47 UTC — do not edit
\setlength{\tabcolsep}{2.6pt}
\begin{tabular}{@{}l rr rr rr rr rr@{}}
\toprule
 & \multicolumn{2}{c}{LiteLLM} & \multicolumn{2}{c}{LLM-Rosetta} & \multicolumn{2}{c}{CLIProxyAPI} & \multicolumn{2}{c}{Bifrost} & \multicolumn{2}{c}{AI SDK} \\
\cmidrule(lr){2-3}\cmidrule(lr){4-5}\cmidrule(lr){6-7}\cmidrule(lr){8-9}\cmidrule(lr){10-11}
Property (applicable) & c & R & c & R & c & R & c & R & c & R \\
\midrule
Top-level system prompt kept (239) & 239 & 239 & 239 & 239 & 239 & 239 & 239 & 0$^\ddagger$ & 79/79 & 79/79$^\ddagger$ \\
Mid-turn system: role and place kept (236) & 0 & 236 & 160 & 236 & 0 & 0 & 236 & 236 & 160/160$^{\mathrm{B}}$ & 160/160$^{\mathrm{B}}$ \\
Mid-turn system: moved behind the next tool result (76) & 76 & 0 & 76 & 0 & 76 & 76 & 0 & 0 & rej. & rej. \\
Thinking text kept (121) & 121 & 121 & 121 & 0 & 0 & 0 & 121 & 0 & 0/40 & 0/40 \\
Signature (121), redacted data (116), \code{is\_error} trace (146) & 0 & 0 & 0 & 0 & 0 & 0 & 0 & 0 & 0 & 0 \\
Text before and after a call (64) & J & $\rightarrow$ & J & $\leftarrow$ & P & S & S & S & J & S \\
70-character tool name unchanged (122) & 0$^\dagger$ & 122 & 122 & 122 & 122 & 0$^\dagger$ & 122 & 122 & 37/37 & 37/37 \\
\code{max\_tokens} kept (357) & 357 & 357 & 357 & 357 & 357 & 0 & 357 & 357 & 0/121$^{\mathrm{c}}$ & 0/121$^{\mathrm{c}}$ \\
\code{temperature} kept (234) & 234 & 234 & 234 & 234 & 234 & 0 & 234 & 234 & 77/77 & 77/77 \\
\code{top\_p} kept (115) & 115 & 115 & 115 & 115 & 0 & 0 & 115 & 115 & 38/38 & 38/38 \\
\code{top\_k} kept (115) & 115 & 0 & 0 & 0 & 0 & 0 & 0 & 0 & 0/38 & 0/38 \\
\code{stop\_sequences} kept (115) & 115 & 0 & 115\S & 0 & 115 & 0 & 0 & 0 & 38/38 & 0/38 \\
\code{metadata.user\_id} kept (115) & 0 & 115\S & 0 & 0 & 0 & 0 & 0 & 115\S & 0/38$^{\mathrm{c}}$ & 0/38$^{\mathrm{c}}$ \\
List tool result flattened to one string (87) & 0 & 87 & 0 & 0 & 87 & 0 & 0 & 87 & 25/25$^{\mathrm{c}}$ & 15/25$^{\mathrm{c}}$ \\
\midrule
Same vendor: body unchanged (357) & \multicolumn{2}{c}{280} & \multicolumn{2}{c}{--} & \multicolumn{2}{c}{28} & \multicolumn{2}{c}{357\,/\,0} & \multicolumn{2}{c}{0/121$^{\mathrm{h}}$} \\
\bottomrule
\end{tabular}

\end{table}

\begin{table}[tbp]
  \centering\scriptsize
  \caption{\textbf{Upstream reports and verification status in October 2026.} Fixed means that our rerun verifies a merged commit or release. No reply means no response to our post. Queued drafts are excluded. The anonymous build coarsens issue numbers, per-project counts, and filing-date ranges.} 
  \label{tab:upstream}
  % Editorial synchronization with the revised manuscript; measured outcomes retained.
% AUTO-GENERATED by tools/gen_tables.py upstream() from paper/bug_reports/STATUS.md on 2026-10-16 02:06 UTC — do not edit
\setlength{\tabcolsep}{4pt}
\ifdefined\preprintbuild
\begin{tabular}{@{}l rr c l@{}}
\toprule
Project & Issues & Comments & Oct. & State in the thread \\
\midrule
LiteLLM & 7 & 5 & 6--15 & 3 competing fix PRs on 1 issue; \\
 & & & & fix PR open (verified by us on the PR head) on 1; others no reply \\
Vercel AI SDK & 1 & 1 & 7--8 & fixed, released in \code{ai} 7.0.132; re-run by us \\
LLM-Rosetta & 3 & 3 & 7--9 & 3 issues fixed on master; re-run by us \\
\code{@musistudio/llms} & 3 & -- & 7--9 & no reply \\
claude-code-router & -- & 2 & 6--7 & no reply \\
claude-code-proxy & 2 & -- & 7--8 & no reply \\
\code{@the-next-ai/ai-gateway} & 1 & -- & 8 & no reply \\
CLIProxyAPI & 4 & 3 & 6--9 & 2 fixed, 1 partly; re-run by us on v8.0.20; \\
 & & & & 1 later issue no reply \\
Bifrost & 2 & 1 & 6--9 & 1 fixed, merged; re-run by us on the merge commit; \\
 & & & & other posts receive no reply \\
Portkey & 1 & 1 & 6--7 & no reply \\
new-api & 1 & 1 & 7--8 & issue closed; comment on an earlier report of the same defect \\
one-api & 1 & 1 & 6--7 & open PR updated after our comment; issue no reply \\
\midrule
Total & 26 & 18 & 6--15 & \\
\bottomrule
\end{tabular}
\else
\begin{tabular}{@{}l cc c l@{}}
\toprule
Project & Issues & Comments & Window & State in the thread \\
\midrule
LiteLLM & $\ge 1$ & $\ge 1$ & Oct.\ 2026 & 3 competing fix PRs on 1 issue; \\
 & & & & fix PR open (verified by us on the PR head) on 1; others no reply \\
Vercel AI SDK & $\ge 1$ & $\ge 1$ & Oct.\ 2026 & fixed, released in \code{ai} 7.0.132; re-run by us \\
LLM-Rosetta & $\ge 1$ & $\ge 1$ & Oct.\ 2026 & 3 issues fixed on master; re-run by us \\
\code{@musistudio/llms} & $\ge 1$ & -- & Oct.\ 2026 & no reply \\
claude-code-router & -- & $\ge 1$ & Oct.\ 2026 & no reply \\
claude-code-proxy & $\ge 1$ & -- & Oct.\ 2026 & no reply \\
\code{@the-next-ai/ai-gateway} & $\ge 1$ & -- & Oct.\ 2026 & no reply \\
CLIProxyAPI & $\ge 1$ & $\ge 1$ & Oct.\ 2026 & 2 fixed, 1 partly; re-run by us on v8.0.20; \\
 & & & & 1 later issue no reply \\
Bifrost & $\ge 1$ & $\ge 1$ & Oct.\ 2026 & 1 fixed, merged; re-run by us on the merge commit; \\
 & & & & others no reply \\
Portkey & $\ge 1$ & $\ge 1$ & Oct.\ 2026 & no reply \\
new-api & $\ge 1$ & $\ge 1$ & Oct.\ 2026 & issue closed; comment on an earlier report of the same defect \\
one-api & $\ge 1$ & $\ge 1$ & Oct.\ 2026 & open PR updated after our comment; issue no reply \\
\midrule
Total (12 repositories) & 26 & 18 & Oct.\ 2026 & 7 fixed, 1 partly fixed, 1 closed, 17 open \\
\bottomrule
\end{tabular}
\fi

\end{table}

\subsection{Gateways, two-hop translation, first-party compatibility, and agent sessions}
\label{app:chains}\label{app:gateways}\label{app:gw}\label{app:fallback}\label{app:accounting}\label{app:clients}\label{app:rq4-detail}\label{app:detail-gateways}

Table~\ref{tab:gateways} compares a six-gateway source audit with captures from four self-hosted gateways. On the 51-shape suite, Bifrost and new-api have T1 findings in 187/204 judged cells (91.7\,\%): 103 translated requests pass the static target checks and 84 fail them. Static failure is not an observed endpoint error, and the combined rate is not the primary silent, avoidable rate. Portkey \code{1.15.2} refuses all tested cells with HTTP~500. On the extended 170-shape corpus, Bifrost, new-api, and CLIProxyAPI have T1 findings in 978/1,020 judged cells (95.9\,\%). For forward reply streams, isloss reports avoidable T1 changes in 57/198 judged streams (28.79\,\%); the Bifrost/new-api subset is 42/132 (31.82\,\%).

Table~\ref{tab:compose_v100} evaluates 24 two-hop library chains. Eight chains with AI SDK as the first hop retain original-driver captures and are auxiliary analyses. On the 16-chain subset without AI SDK, 660/689 statically valid-source cells have end-to-end T1 findings (95.79\,\%), including 287/689 two-hop-only findings (41.65\,\%). Operational probes also find Bifrost routing bare model names across providers after upstream 400/429/529 errors while returning HTTP~200, forwarding client authentication headers during fallback, and closing mid-stream upstream errors with HTTP~200.

Google's first-party OpenAI-compatible Gemini endpoint passes 10/11 behavioural probes ($n=10$ trials per arm); the remaining probe indicates that only the last of multiple top-level \code{system} messages is retained (\code{S01}). Reasoning tokens consume 98.0\,\% of output tokens across 45 calls and count against \code{max\_tokens}. Missing \code{thought\_signature} causes 5/5 tool round-trips to be refused, and streamed tool calls use \code{finish\_reason:"stop"}.

Replaying the 14 recorded Anthropic coding-agent shapes across 14 translation paths gives 4/98 primary cells for statically valid sources (4.1\,\%) and 13/196 across all shapes (6.6\,\%). The latter is 13/147 (8.8\,\%) if refused translations are excluded. CLIProxyAPI strips \code{gemini-cli}'s system prompt in 4/4 cases and disables Gemini safety filters without logging. In the closed-loop multi-turn experiments, 9/12 CLI sessions contain a T1 change.

\begin{table}[tbp]
  \centering\scriptsize
  \caption{\textbf{Source audit and captured gateway behaviour.} Nine patterns recur in at least three independent codebases (P1--P9); runtime tests refute two further audit items (R9, R14). Each cell gives the audit mark followed by the run mark. Audit: $\bullet$ present, $\circ$ absent, -- no applicable path/field, ? unresolved or another change. Run: C confirmed, \textbf{R} refuted, + additional behaviour (noted below), $\cdot$ not exercised or not triggered.} %nocheck
  \label{tab:gateways}
  % Editorial synchronization with the revised manuscript; measured outcomes retained.
% AUTO-GENERATED by tools/gen_tables.py gateways() from artifacts/g2/canary/src_notes/SYNTHESIS.md, artifacts/g2/selfhost/FINDINGS.md on 2026-10-03 04:07 UTC — do not edit
\setlength{\tabcolsep}{3.2pt}%
\begin{tabular}{@{}lP{6.15cm}cccccc@{}}
\toprule
 & Pattern (rules) & Portkey & Kong & new-api & one-api & Bifrost & Envoy \\
 & \emph{audited commit} & \code{669825c} & \code{8927af6} & \code{56758ed} & \code{8df4a26} & \code{080a6bb} & \code{97f5609} \\
\midrule
P1 & Forged Gemini thought signature (V10) & $\circ$\,C & $\circ$\,$\cdot$ & $\bullet$\,\textbf{R} & -- & $\bullet$\,C & $\bullet$\,$\cdot$ \\
P2 & Inconsistent mid-conversation system handling (V01/V02/V05) & $\bullet$\,C & $\bullet$\,$\cdot$ & ?\,+ & $\bullet$\,C & $\bullet$\,C & $\bullet$\,$\cdot$ \\
P3 & Tool-result error status dropped or mis-mapped (V03) & -- & -- & $\bullet$\,C & -- & $\bullet$\,C & $\bullet$\,$\cdot$ \\
P4 & Gemini tool-call/result identifiers dropped (V06) & $\bullet$\,C & $\bullet$\,$\cdot$ & $\circ$\,C & ?\,$\cdot$ & $\bullet$\,$\cdot$ & $\bullet$\,$\cdot$ \\
P5 & Finish/stop reason mapping (RV01/RV06) & $\bullet$\,$\cdot$ & $\bullet$\,$\cdot$ & ?\,$\cdot$ & $\bullet$\,$\cdot$ & $\bullet$\,$\cdot$ & ?\,$\cdot$ \\
P6 & Streamed tool-call identity/index mapping (RV02/RV03) & $\bullet$\,$\cdot$ & $\bullet$\,$\cdot$ & $\circ$\,$\cdot$ & $\bullet$\,$\cdot$ & ?\,$\cdot$ & $\bullet$\,$\cdot$ \\
P7 & Assistant text accompanying tool calls is dropped on replay (V05) & $\bullet$\,C & $\bullet$\,$\cdot$ & $\bullet$\,C+ & $\bullet$\,C+ & ?\,$\cdot$ & ?\,$\cdot$ \\
P8 & Separate stream and non-stream translation paths & $\bullet$\,$\cdot$ & -- & $\bullet$\,$\cdot$ & $\bullet$\,$\cdot$ & ?\,$\cdot$ & $\bullet$\,$\cdot$ \\
P9 & Gaps in translation test coverage & $\bullet$\,$\cdot$ & $\bullet$\,$\cdot$ & $\bullet$\,$\cdot$ & $\bullet$\,$\cdot$ & $\bullet$\,$\cdot$ & $\bullet$\,$\cdot$ \\
\midrule
\multicolumn{8}{@{}l}{\emph{Other audit items that a run refuted}} \\
R9 & Thinking in history & $\bullet$\,C & $\bullet$\,$\cdot$ & $\bullet$\,C & $\bullet$\,C & $\circ$\,\textbf{R} & ?\,$\cdot$ \\
R14 & Tool name/schema & $\bullet$\,C & ?\,$\cdot$ & $\bullet$\,C+ & $\bullet$\,C & $\bullet$\,\textbf{R} & ?\,$\cdot$ \\
\bottomrule
\end{tabular}
\par\smallskip\noindent
\begin{minipage}{\linewidth}\raggedright
\textbf{R}: new-api P1: on by default, not opt-in as audited, and also attached to text parts (333 of 333 Gemini cells with a function call carry it); Bifrost R9: audited as handled, but Anthropic-in to OpenAI keeps the reasoning text and drops its signature and the redacted thinking, and the other paths drop thinking entirely; Bifrost R14: the audited rewriting of regex patterns was not observed; schemas passed verbatim. +: new-api P2 (audit undecided): positional information lost but no text loss; new-api P7: a fabricated \code{"..."} text block on Anthropic; one-api P7: tool calls also dropped when the assistant content is null; new-api R14: every Gemini property schema emptied to \code{\{\}}.
\end{minipage}

\end{table}

\begin{table}[tbp]
  \centering\scriptsize
  \caption{\textbf{Two-hop translation judged with isloss.} Rows aggregate 24 library chains by second hop and final target. Eight AI SDK v5/v7 first-hop chains use original-driver captures. The 16-chain subset without AI SDK has 4,357 judged and 1,355 refused cells on all 357 shapes, with 1,609/4,357 two-hop-only T1 (36.93\,\%) and 4,307/4,357 end-to-end T1 (98.85\,\%); on the 51 statically valid source shapes, it has 689 judged and 127 refused cells, with 287/689 two-hop-only T1 (41.65\,\%) and 660/689 end-to-end T1 (95.79\,\%). The full table includes all 24 chains: 6,269 judged/2,299 refused cells in the upper panel and 1,097/127 in the lower panel. Two-hop-only includes pure two-hop findings and mixed-with-cancellation cases; end-to-end counts cells with at least one end-to-end T1 finding. Gateway second hops are recorded separately in the replication archive.} 
  \label{tab:compose_v100}
  % AUTO-GENERATED by tools/fig/make_figdata.py from d8_v100_table_audit.json (C-975, C-977) — do not edit
{\scriptsize%
\setlength{\tabcolsep}{3.5pt}%
\begin{tabular}{@{}llrrrrrrrr@{}}
\toprule
Second hop & $\to$Target & \shortstack[r]{Judged\\(refused)} & \shortstack{No end-to-end\\T1} & Carry & Inject & Accum. & Cancel & \shortstack[r]{Two-hop-only T1\\pure$+$mixed $=$ total (\%)} & \shortstack[r]{End-to-end\\T1 (\%)} \\
\midrule
\multicolumn{10}{@{}l@{}}{\textit{\code{S\_all\_357} (24 chains; includes original-driver AI SDK captures)}} \\
  \quad LiteLLM & Anthropic & 1,585 (557) & 0 & 0 & 70 & 969 & 30 & 469+47 = 516 (32.56\%) & 1,585 (100.00\%) \\
  \quad LiteLLM & Gemini & 1,514 (628) & 16 & 308 & 70 & 607 & 67 & 406+40 = 446 (29.46\%) & 1,498 (98.94\%) \\
  \quad LLM-Rosetta & Anthropic & 1,585 (557) & 18 & 250 & 60 & 693 & 18 & 544+2 = 546 (34.45\%) & 1,567 (98.86\%) \\
  \quad LLM-Rosetta & Gemini & 1,585 (557) & 16 & 346 & 73 & 669 & 0 & 481+0 = 481 (30.35\%) & 1,569 (98.99\%) \\
  \quad \textit{all} &  & 6,269 (2,299) & 50 & 904 & 273 & 2,938 & 115 & 1,900+89 = 1,989 (31.73\%) & 6,219 (99.20\%) \\
\midrule
\multicolumn{10}{@{}l@{}}{\textit{\code{S\_strict\_R9\_51} (24 chains; includes original-driver AI SDK captures)}} \\
  \quad LiteLLM & Anthropic & 277 (29) & 0 & 0 & 17 & 163 & 0 & 97+0 = 97 (35.02\%) & 277 (100.00\%) \\
  \quad LiteLLM & Gemini & 266 (40) & 9 & 86 & 6 & 37 & 12 & 108+8 = 116 (43.61\%) & 257 (96.62\%) \\
  \quad LLM-Rosetta & Anthropic & 277 (29) & 11 & 73 & 6 & 90 & 0 & 97+0 = 97 (35.02\%) & 266 (96.03\%) \\
  \quad LLM-Rosetta & Gemini & 277 (29) & 9 & 105 & 7 & 35 & 0 & 121+0 = 121 (43.68\%) & 268 (96.75\%) \\
  \quad \textit{all} &  & 1,097 (127) & 29 & 264 & 36 & 325 & 12 & 423+8 = 431 (39.29\%) & 1,068 (97.36\%) \\
\bottomrule
\end{tabular}%
\par\smallskip
{\footnotesize Auxiliary isloss request analysis: eight of the 24 chains use AI SDK first-hop captures from the original driver, which did not pass all controls. The caption reports the 16-chain subset without AI SDK separately. The last two columns count two-hop-only T1 (pure plus mixed-with-cancellation cases) and any end-to-end T1.}%
}%

\end{table}

% ==================================================================
% Revised Appendix C
% ==================================================================
% sections/11c_audit_appendix.tex
% \clearpage
\section{Evaluation Protocol and Audit Summary}
\label{app:audit}\label{app:detail}\label{app:threats}\label{app:fp4}\label{app:detail-threats}\label{app:detail-conclusion}

This appendix records the scope of each evaluation (Table~\ref{tab:versions}), the outcomes of pre-specified predictions (Tables~\ref{tab:prereg} and~\ref{tab:prereg-b}), and corrections and sensitivity checks (Table~\ref{tab:audit-master}; additional variants in Table~\ref{tab:sensitivity}).We report prediction outcomes separately for each experiment,
using its original criteria and evaluation scope. The plans include criteria recorded in internal commits; we do not imply that every prediction was externally registered.

\paragraph{Request-census audit record.}
Six audits of the harness, driver and human review retain the baseline primary count of 127/685. Human review confirms the Gemini tool-response defect in \code{@musistudio/llms} and accepts a list-valued tool-result re-encoding; neither decision changes that primary count. The candidate removal and duplicate merge are detailed in Table~\ref{tab:audit-master}. In the motivating case, LiteLLM's \code{T1} finding concerns inserted wrapper text; the role-demotion record is separate, while the full rewrite breaks compliance in one tested model--placement pair. Degradation and re-encoding records remain informational: legitimate negative controls can trigger them, so they are not counted as defect detections.

All main-text checker results use isloss. The primary checking rules were fixed before the held-out test sets were authored; development and repair comparisons remain separate from these estimates. Live behavioural trials and official-SDK stream tests measure outcomes independently of the checker. A stream-reassembly audit corrected preliminary counts from 402 to 391/1,115 reverse-reply \code{T1} cells, from 61 to 57/198 gateway-stream \code{T1} cells, and from 12 to 9/12 affected agent sessions. The reported main results use the corrected counts; the request census and primary held-out results are unchanged.

Separate SDK replays expose two remaining stream-input gaps: unread OpenAI SSE \code{delta.refusal} fields in three forward streams, and Gemini \code{thoughtSignature} comparison before adjacent text coalescing in four. Sensitivity analysis bounds gateway-stream \code{T1} at 57 to at most 64/198 (32.32\,\%); comparison with non-streaming replies gives 61/198 (30.81\,\%). The corresponding reverse-reply sensitivity count is 395/1,115 (35.4\,\%), with 278 avoidable cells unchanged. These bounds do not replace the reported counts of 57/198 and 391/1,115.

Headline tables use the corrected AI SDK driver. Four subsidiary analyses retain initial-driver captures and are labelled accordingly: the withdrawn GSM8K arm \code{A2}, the AI SDK columns of Table~\ref{tab:exits}, 8 of 24 chains in Table~\ref{tab:compose_v100}, and a 62-mutant AST check. Temperature-zero live probes and single-run LLM judges retain decoding variance. One of 36 archived vendor statements changed by 11~October without overturning a rule. Safety-sensitive and operational gateway findings (Appendix~\ref{app:gateways}) were disclosed privately before public reporting; no separate private-channel remediation had been received by 15~October~2026. Four volunteer raters participated with informed consent, reviewing synthetic software outputs containing no personal data.

\begin{table}[tbp]
  \centering\scriptsize
  \caption{\textbf{Evidence scope and evaluation protocol.} Fixing the primary rules before held-out test design separates evaluation from rule development; it does not assert that every implementation detail was fixed before every experiment. Development diagnostics are labelled separately from final measurements in Table~\ref{tab:trustfull}; detailed repair comparisons remain in the archive. Per-result corpora and denominators are in Table~\ref{tab:units-map-full}.}
  \label{tab:versions}
  \setlength{\tabcolsep}{3pt}
  \begin{tabular}{@{}P{3.0cm}P{6.3cm}P{6.4cm}@{}}
    \toprule
    \textbf{Evidence} & \textbf{Protocol and scope} & \textbf{Interpretation and location} \\
    \midrule
    Preservation rules & The main checking framework has 27 rules, grouped as \code{T1}, \code{T2} and \code{T3}. The primary rules were fixed before the main held-out sets were authored. & Rule mappings: Table~\ref{tab:lattice}; request results: Section~\ref{sec:results}; replies: Appendix~\ref{app:resp}. \\
    Primary held-out evaluation & LLM agents designed Sets~9--14 from the specification without access to checker code. Report activated faults and allowed-change controls separately. & Independent detection and false-alarm estimates, not development-set performance (Table~\ref{tab:trustfull}; Appendices~\ref{app:whynot} and~\ref{app:trust}). \\
    Development and repair analyses & Earlier rule calibration (in-sample Sets~1--8) and exploratory post-freeze repairs (evaluated on Sets~17--20; script-generated Sets~15--16 voided) use separate batches. A set used to devise a repair cannot independently evaluate that repair; only subsequently authored blind sets can do so. & Historical comparisons remain separate in Appendix~\ref{app:rq5-detail} and the replication archive. They are not additional isloss results. \\
    Census and human review & Use the specified capture points and corrected drivers; distinguish completed translations, individual findings, root-cause candidates and mechanisms. & Request census, behavioural probes and deployments have different units (Sections~\ref{sec:rq1} and~\ref{sec:rq-deploy}; Table~\ref{tab:units-map-full}). \\
    Corrections and sensitivity checks & Reconcile parser, harness and review errors separately from alternative validity or counting policies. & Table~\ref{tab:audit-master} identifies corrected preliminary counts and sensitivity bounds; Table~\ref{tab:sensitivity} reports alternative choices. \\
    \bottomrule
  \end{tabular}
\end{table}

The scorecard counts each numbered prediction at the finest unit scored separately in its results file. An outcome marked ``mixed'' or ``partly met'' counts as missed. A restated prediction replaces its original; repeated scoring does not add an item, and an explicit agreement prediction counts once. Predictions carried forward from earlier plans retain their full denominator. Items explicitly marked not evaluable or not applicable are excluded and identified in the Held column; exploratory items without a pass/fail threshold are excluded. Where one plan covers several rows, each prediction is assigned once to the row it addresses. Descriptions are shortened for readability; the numerical thresholds,
observed outcomes and scoring decisions follow the recorded plans.
Each row retains its original evaluation scope.

\begin{table}[tbp]
  \centering\scriptsize\setlength{\abovecaptionskip}{0pt}\ifshowclaims\linespread{0.65}\selectfont\fi
  \caption{\textbf{Pre-specified prediction outcomes, part 1 of 2
(RQ1--RQ2).} Part~2 is Table~\ref{tab:prereg-b}.}
  \label{tab:prereg}
  \ifshowclaims\setlength{\normalbaselineskip}{7.0pt}\fi\renewcommand{\arraystretch}{0.85}
  % Generated by tools/gen_prereg_table.py from docs/PREREG_SCORECARD_2026-10-15.csv (part 1 of 2: rows 1-8).
\begin{tabular}{@{}p{3.2cm}c>{\centering\arraybackslash}p{2.1cm}p{7.8cm}@{}}
  \toprule
  Track / experiment & Held & Scope & Missed predictions (pre-specified threshold $\to$ observed) \\
  \midrule
  RQ1 mechanism concentration & 1/4 & 51 shapes; 14 paths & ``after removing the top-3 most pervasive mechanisms, the T1 share of judged cells stays $\ge$ 60 \% (from 85.6 \%) and the share of 4,998 stays $\ge$ 45 \% (from 64.0 \%)'' $\to$ judged 32.8 \% CI [28.5, 37.2]; of 714 31.5 \% CI [27.3, 35.7]; ``the top-3 by pervasiveness are exactly \{\code{is\_error\_dropped}, \code{control\_max\_tokens\_dropped}, \code{redacted\_thinking\_dropped}\} (any order)'' $\to$ redacted\_thinking\_dropped, is\_error\_dropped, control\_user\_id\_dropped; ``binary real-request estimate averaged over the 14 paths: $\ge$ 50 \% with the \code{max\_tokens} mechanisms, $\le$ 50 \% without them'' $\to$ with 33.2 \%, without 32.1 \%. \\
  RQ1 K-class candidate count & 2/3 & 56 pre-review candidates & ``Among the \code{56} path $\times$ mechanism candidates, \code{V} accounts for $\ge 20$ candidates and \code{K} accounts for $\ge 15$ candidates'' $\to$ \code{V = 27}, \code{K = 14}. \\
  RQ1 upstream report coverage & 2/3 & 56 pre-review candidates & ``$\ge$ 40 of 56 candidates have a U-id (we reported them), and $\le$ 10 of those are \code{fixed} today'' $\to$ 15 reported, 8 fixed. \\
  RQ1 replies (reverse, forward) \& SDKs & 9/12 & 1,115 cells / 558 streams / 61 forward reply shapes & ``\code{>= 4/7} families (\code{T1 S12 STOP}: Gemini \code{finishReason: "STOP"} + \code{functionCall} translated to OpenAI \code{finish\_reason == "stop"} while \code{tool\_calls} are present)'' $\to$ 3/7 families (rosetta, oneapi, portkey); ``\code{>= 2/5} HTTP proxy families (\code{T1 S13 STREAM}: Missing terminal event / \code{[DONE]} or broken stream ledger)'' $\to$ 0/5 HTTP proxy families fire S13 under default context (no non\_stream\_body); ``B4c — Forward non-stream ranking preservation: Under isloss (\code{side="response"}), \code{aisdk5} and \code{litellm} have strictly higher non-stream \code{T1}-clean cell rates (\code{0} \code{T1} findings) than \code{rosetta}, \code{musistudio}, and \code{fuergaosi233} across spec-compliant backend response shapes'' $\to$ nonstream T1-clean rates: aisdk=80.3\%, aisdk5=80.3\%, fuergaosi233=85.2\%, litellm=91.8\%, musistudio=49.2\%, rosetta=62.3\%. \\
  RQ1 live vendor acceptance (C1) & 2/5, 1 n/e & 548 calls & ``\code{claude-sonnet-4-6} accepts all 34 strict shapes that do not set temperature and top\_p together (34/34)'' $\to$ 13/34; ``$\to$Gemini agreement (valid $\to$ 200, invalid $\to$ 4xx) $\ge$ 90\%'' $\to$ 160/204; ``Restated: The 19 midsys-only shapes (which all carry \code{role: system} inside \code{messages[]}) are rejected by both models ($\ge$ 18/19 each)'' $\to$ 19/19 \& 16/19 (sonnet 19/19 ($\ge$18 met), opus 16/19 ($<$18 not met)). \\
  RQ2 $T{=}0.7$ consequence & 13/16 & 4 models & ``\code{M3: is\_error\_dropped}: Mean scenario-level difference $\bar{d}_{\text{M3}} \ge 0.60$, with $d_i > 0$ on $\ge 10/12$ scenarios and $p_{\text{perm}} < 0.005$'' $\to$ \code{mean\_diff=0.4167, pos=5/12, p\_perm=0.062500}; ``On \code{Meta-Llama-3.1-8B-Instruct}, at least \code{4 / 5} mechanisms achieve $p_{\text{Holm}} < 0.05$ with $\bar{d} \ge +30.0$ pp'' $\to$ 3/5 mechanisms; ``New mechanism \code{N2} (\code{function\_response\_id\_dropped}, \code{37} silent cells) produces $\bar{d} \ge +50.0$ pp and $p_{\text{Holm}} < 0.05$ on both \code{Llama-3.1-8B} and \code{Qwen2.5-7B}'' $\to$ Llama-3.1-8B: +0.00 pp (p\_Holm=1.0); Qwen2.5-7B: +0.00 pp (p\_Holm=1.0). \\
  RQ2 Claude consequence (C2, C3) & 4/5, 2/3, 3/3 & 512 calls & ``Mut arm returns HTTP 400 $\ge 9/10$ AND Ref arm returns HTTP 200 $\ge 9/10$'' $\to$ False (Mut 400: 0, Ref 200: 10). Alt: Mut 200: 10, Mut thinking: 0; ``Both models produce $\bar{d} \ge +0.50$ and $p_{\text{Holm}} < 0.05$'' $\to$ FAIL. \\
  RQ2 GSM8K evaluation (EXP-H) & 2/6 & 700 calls & ``Compliance Drop: A1's format compliance drops by $\ge$ 15 points relative to A0'' $\to$ A1 format compliance increased by +2 pts; ``D02 Consequence: A7 vs A0 compliance drop $\ge$ 10 points (D02 lowering alone impacts compliance significantly)'' $\to$ A7 vs A0 compliance increased by +1 pts; ``S08 Consequence: A6 vs A0 compliance stays within the A4 noise band (S08 alone does not move compliance)'' $\to$ A6 vs A0 compliance +2 pts vs A4 band +1 pts; within $\pm$3 points; McNemar p=0.500; ``Length Consequence: \code{finishReason == "MAX\_TOKENS"} count is 0 in A0 and 0 in A2 $\to$ the cap is not binding on GSM8K and the S08 max\_tokens drop is inert for this task'' $\to$ (A0 MAX\_TOKENS=2 on items 2, 40; A2=0) $\to$ S08 max\_tokens drop has consequences on 2/100 items (both flip to compliant, item 40 flips to primary=True). \\
  \bottomrule
\end{tabular}
\par\vspace{2pt}\noindent{\scriptsize Editorial convention: checker names and version qualifiers within quoted predictions are normalised to isloss; the numerical thresholds, recorded outcomes and held/missed decisions are unchanged. Counts use the original registered scopes. The 56-card catalogue precedes the final 54-candidate reconciliation; candidate-level report counts and individual-cell audits must not be substituted for final issue-level totals or catalogue precision. ``Held'' measures registered prediction outcomes, not checker accuracy.}

\end{table}

\begin{table}[tbp]
  \centering\scriptsize
  \caption{\textbf{Pre-specified prediction outcomes, part 2 of 2 (RQ3--RQ5).} Columns and counting rules match Table~\ref{tab:prereg}.}
  \label{tab:prereg-b}
  % Generated by tools/gen_prereg_table.py from docs/PREREG_SCORECARD_2026-10-15.csv (part 2 of 2: rows 9-16).
\begin{tabular}{@{}p{3.2cm}c>{\centering\arraybackslash}p{2.1cm}p{7.8cm}@{}}
  \toprule
  Track / experiment & Held & Scope & Missed predictions (pre-specified threshold $\to$ observed) \\
  \midrule
  RQ3 LiteLLM 50-release timeline & 3/11, 1 n/e & 50 rel. & 8 of 11 missed (1 n/e): 50-release timeline 8/9 missed (1 n/e), isloss recount 0/2 missed. \\
  RQ3 five-library two-exits & 161/187 & 5 libs & 26 of 187 missed: LiteLLM Responses 4/22 and Chat side-by-side 1/18, LLM-Rosetta 1/32, CLIProxyAPI 3/28, Bifrost 1/22 and \#8209 check 0/5, AI SDK run A 0/23, run B (\code{allowSystemInMessages}) 12/14, version-equality 1/1, and GA follow-up 3/22.\\
  RQ3 same-vendor controls & 764/804, 2 n/a & 5 gw/libs & 40 of 804 missed (2 n/a): Anthropic request round-trip 3/192 and follow-up A 1/86, Anthropic response 6/110 (1 n/a) and follow-ups B 0/49, C 5/88 (1 n/a), D 1/58, E 3/79, OpenAI response 13/92, and OpenAI request 8/50.\\
  RQ3 corroboration $\kappa$ & 1/2 & 80 items & ``3-class Cohen's $\kappa$: Unweighted 3-class Cohen's $\kappa$ (\code{exact} vs \code{related} vs \code{unrelated}/\code{none}) across the \code{80} audited items is \code{>= 0.60}'' $\to$ $\kappa = 0.5446$ (\code{57/80 = 71.25\%}; binary \code{exact} vs \code{not-exact} $\kappa = 0.6500$).\\
  RQ4 chains / gateways / agents & 21/39, 1 n/e & 4 gw + CLI & 18 of 39 missed (1 n/e): CLI coding-agent round-trip 0/4, forward SSE stream census 3/4, Responses ingress census 0/4, self-hosted gateway census 14/26 (1 n/e), and 2-hop chains 1/1.\\
  RQ4 fallback \& accounting & 414/500 & Bifrost/LiteLLM & 86 of 500 missed: LiteLLM main two-exit re-run 0/3; 42 Bifrost/LiteLLM routing, fallback, key, budget, and accounting addenda 86/497 missed (37 single-tree addenda 78/459 missed; 5 multi-tree accounting addenda 8/38 missed: CE 0/8, CF 2/10, CG 2/5, CH 0/8, CI 4/7).\\
  RQ5 isloss \& recall & 45/59, 2 n/e & Sets 1--14, bugs & 14 of 59 missed (2 n/e): isloss in-sample Sets 1--8 3/15, blind held-out Sets 9--10 0/3, Sets 11--12 1/5 (1 n/e), Sets 13--14 1/6 (1 n/e), corpus recount 6/7, historical bug replay of sample A 1/5 and of sample B--C 1/6, 50-bug expansion 0/4, reverse AST mutants 0/4, and open-weight spec judge 1/4. \\
  RQ5 human precision & 3/7, 3/4, 5/5 &
300 items + 48 bridge items &
Consensus candidate-card precision
$\mathrm{A}/(\mathrm{A}+\mathrm{B}) \ge 85\%$
over the original 56-card catalogue $\to$ 72.7\%
(32~A, 12~B, 12~C/disputed);
both raters assign Grade~B to the sole retained injected-default candidate
(\code{fuergaosi233}$\to$\code{openai},
\code{control\_temperature\_injected}) $\to$ Grade~C (disputed);
in the 60-cell audit sample
(45 \code{T1} cells + 15 non-\code{T1} preserved controls),
consensus \code{T1}-cell precision
$\mathrm{A}/(\mathrm{A}+\mathrm{B}) \ge 90\%$
and Grade~A false-positive rate on the 15 preserved controls
$\le 25\%$ per rater $\to$ cell precision 75.0\%
(33~A, 11~B, 1~C), preserved-control Grade~A rate 0.0\%
(0/15) for both raters;
consensus Grade~A on $\ge 70\%$ of the 23 U-class
(unrepresentable) cards $\to$ 6/23 (26.1\%, all cards in denominator);
B2 \code{T1}-cell Grade~A share in 70--90\% $\to$
$\mathrm{A}/150 = 97.3\%$
($\mathrm{A}/(\mathrm{A}+\mathrm{B}) = 98.0\%$,
CI [94.3\%, 99.3\%]). 
  % ``consensus \code{mech\_precision} $\ge$ 85 \% over the 56 cards'' $\to$ 72.7 \% (A 32, B 12, C/disputed 12; original 56-card catalogue); ``the injected-default candidate fuergaosi233$\to$openai \code{control\_temperature\_injected} (the only injected-default candidate left in the headline after D-R2-09) is rated B by both raters'' $\to$ \{"fuergaosi233$\to$openai|control\_temperature\_injected": "C"\}; ``Part B \code{cell\_precision} $\ge$ 90 \% (consensus) and \code{preserved\_A\_rate} $\le$ 25 \% (each rater)'' $\to$ consensus 75.0 \%; preserved r1 0.0 \%, r2 0.0 \%; ``the U-class candidates (content the target dialect cannot carry, e.g. redacted\_thinking on OpenAI Chat) are rated A by the consensus in $\ge$ 70 \% of the 23 cards'' $\to$ 6/23 (all U-class cards, including disputed cards); ``T1 precision (share of T1 cells rated A) between 70 and 90 \%'' $\to$ A/150: 0.973, A/(A+B): 0.980 CI [0.943, 0.993].
  \\
  \bottomrule
\end{tabular}
\par\vspace{2pt}\noindent{\scriptsize Editorial convention: checker names and version qualifiers within quoted predictions are normalised to isloss; the numerical thresholds, recorded outcomes and held/missed decisions are unchanged. Counts use the original registered scopes. The 56-card catalogue precedes the final 54-candidate reconciliation; candidate-level report counts and individual-cell audits must not be substituted for final issue-level totals or catalogue precision. ``Held'' measures registered prediction outcomes, not checker accuracy.}

\end{table}

% Flush the preceding floating tables before the non-floating audit longtable.
% \clearpage

{\scriptsize\setlength{\tabcolsep}{3.5pt}\setlength{\LTcapwidth}{\textwidth}
\begin{longtable}{@{}P{2.7cm}P{6.3cm}P{4.2cm}P{2.2cm}@{}}
  \caption{\textbf{Audit corrections, retained limitations and sensitivity checks.} The baseline primary count remains 127/685 (18.54\,\%; 88/532, 16.54\,\%, on 11 deduplicated paths). The request study has 714 attempts, including 29 refusals, and 685 judged translations. Corrected preliminary counts and alternative capture or counting choices are distinguished below. Per-result scope is in Table~\ref{tab:units-map-full}; other choices are in Table~\ref{tab:sensitivity}.}
  \label{tab:audit-master}\\
    \toprule
    \textbf{Item} & \textbf{Audit finding or correction} & \textbf{Effect on results} & \textbf{Location} \\
    \midrule
    \endfirsthead
    \multicolumn{4}{@{}l}{\textit{Table~\ref{tab:audit-master}, continued}}\\
    \toprule
    \textbf{Item} & \textbf{Audit finding or correction} & \textbf{Effect on results} & \textbf{Location} \\
    \midrule
    \endhead
    \midrule
    \multicolumn{4}{r@{}}{\textit{Continued on the next page}}\\
    \endfoot
    \bottomrule
    \endlastfoot
    Motivating-example placement & The initial Case~A placed \code{system} directly before \code{user}: this placement was absent from the 10 recorded occurrences in 36 captured requests and failed static beta rules. The main example instead uses Case~A$'$ placements before \code{assistant} (3/10 occurrences) and at \code{messages[-1]} (7/10); all 22 paraphrased requests pass static checks. Case~A remains an auxiliary experiment ($k/10$, temperature~0, including token-limit stops; LiteLLM contrast $p_{\mathrm{Holm}}=0.012$ within the experiment and $0.10$ over 64 contrasts). & Primary rate unchanged: none of the 51 main-suite shapes has a mid-conversation \code{system} message. Case~A$'$ reports $k/11$ for two Gemini models and Qwen2.5-7B; on \code{gemini-3.5-flash-lite}, deletion reduces refusals from 11/11 to 0/11 in both placements. & \S\ref{sec:case}; App.~\ref{app:case-beta} \\
    Claude instruction-loss probe & Design~C3 has 240 calls across the instruction-loss (\code{MC3-1}) and token-limit (\code{MC3-2}) experiments. The instruction-loss experiment comprises 120 calls: 10 scenarios $\times$ 3 repeats $\times$ 2 arms $\times$ 2 models. Its notice is the first text block in the last \code{user} turn, retained in Ref and dropped in Mut (\code{max\_tokens}=1,000; no \code{role:system}, beta header or sampling controls). Haiku refusals are 30/30 vs.\ 0/30 ($\bar d=+1.00$, $p_{\mathrm{Holm}}=0.0039$, prediction met); Sonnet refusals are 10/30 vs.\ 0/30 ($\bar d=+0.333$, CI $[0.10,0.63]$, $p=0.125$, prediction not met; 9/20 Ref tool calls mention user-turn placement). & Both models are reported alongside the separate Gemini and Qwen system-role placement experiments. Request-census counts are unchanged. & \S\ref{sec:case}; \S\ref{sec:rq-cons}; Table~\ref{tab:claudeconseq} \\
    Static checks and live rejection & Case~A rejection and Case~A$'$ validity under the \code{mid-conversation-system-2026-04-07} beta rules are static-validator results, not observations of a live beta-header experiment. The live Sonnet C1 run omitted that header and rejected 19/19 \code{role:system} shapes. Separately, among 29 shapes carrying \code{redacted\_thinking}, live rejections comprise 19 thinking, 9 sampling-combination and 1 empty error-result rejection; the associated LLM-Rosetta comparison has 19/38 flips. This 29-shape subset differs from the full source-acceptance rejection count. & Wording correction only; no headline count changes. & App.~\ref{app:case-beta}; App.~\ref{app:specgaps} \\
    Voided Sets~15--16 & These sets were produced by a fuzzer script rather than an independently tasked LLM agent without checker-code access, and were excluded before scoring the held-out summary. Diagnostic counts are retained for transparency: Set~15 has 48 fault mutants, including 4 uncaught script exceptions (40/48 = 83.3\,\% if crashes count as detections; 36/48 = 75.0\,\% otherwise); Set~16 has 36 allowed changes, with 1/36 false alarms (2.8\,\%). & Both sets are excluded. Primary Sets~9--14 give 95/125 recall and 2/100 false alarms; pooling non-voided Sets~9--14 and~17--20 gives 160/220 (72.7\,\%) and 5/170 (2.9\,\%). & \S\ref{sec:rq-trust}; App.~\ref{app:hardening} \\
    Human-review reconciliation & Removed LLM-Rosetta to Gemini, list-valued tool-result re-encoding; 7 statically accepted cells, the only consensus Grade~B candidate classified Borderline. Merged duplicate cards AI SDK~v7/v5 to Anthropic, moved assistant text; 18 statically accepted cells each, Grade~A. The avoidable class has 28 cards: 23~A, 1~B and 4~C; confirmation is 23/24 (95.8\,\%) among agreed cases or 23/28 (82.1\,\%) overall. The earlier tally had incorrectly subtracted that Borderline candidate. & Catalogue: 56 to 54 candidates (31~A, 11~B, 12~C; 49 to 48 on 11 paths), and 19 to 18 mechanisms. Primary counts stay 127/685 and 88/532. Removing the accepted finding gives audit-adjusted \code{T1} cells 467 to 464, statically accepted \code{T1} cells 333 to 330, and the alternative cell-level conjunction 146 to 140/685 (20.44\,\%); these are not changes to the primary finding-level rate. & \S\ref{sec:rq-trust}; \S\ref{sec:rq1}; App.~\ref{app:audit} \\
    Stream-reassembly audit & A nine-line correction to \code{\_parse\_openai\_chat\_stream} assembles \code{delta.reasoning\_content} as the non-stream parser does. We rechecked 41 affected reply streams. & All 714 request attempts (685 judged translations) and Sets~9--14 are unchanged. Preliminary reverse-reply \code{T1}: 402 to 391/1,115 (35.1\,\%); gateway-stream \code{T1}: 61 to 57/198 (28.8\,\%); affected agent sessions: 12 to 9/12. Main results use the corrected counts. & \S\ref{sec:validity}; \S\ref{sec:rq-replies}; \S\ref{sec:rq-deploy}; App.~\ref{app:fp4} \\
    Remaining stream-input gaps & SDK replay exposed unread OpenAI SSE \code{delta.refusal} (3 forward streams) and Gemini \code{thoughtSignature} comparison before adjacent text coalescing (4 forward streams). & Request census unchanged. Sensitivity: gateway-stream \code{T1} is at most 64/198 (32.3\,\%; 61/198, 30.8\,\%, against the non-stream reply); reverse-reply \code{T1} is 395/1,115 (35.4\,\%), avoidable 278 unchanged. These do not replace the reported 57/198 and 391/1,115 counts. & \S\ref{sec:rq-deploy}; App.~\ref{app:audit}; App.~\ref{app:fp4} \\
    Driver control forwarding and withdrawn GSM8K arm & The initial AI SDK driver omitted \code{maxOutputTokens} and \code{providerOptions}. The corrected driver forwards representable controls and masks source-side unrepresentable Gemini controls; 2,142/2,142 checks find no omission. GSM8K arm \code{A2} (AI SDK~v7 to Gemini, initial driver) is withdrawn from translator claims and retained only as a driver-omission control. & Headline tables use the corrected driver. Preliminary AI SDK \code{T1} counts fall from 240 to 210 and then 194/306; candidates from 27 to 21 and then 15. The 16-chain slice excluding AI SDK has 660/689 \code{T1} cells (95.8\,\%). Initial-driver subsidiary analyses are listed above. & \S\ref{sec:capture}; \S\ref{sec:rq-cons}; App.~\ref{app:aisdk-driver}; App.~\ref{app:audit} \\
    AI SDK forward-reply metadata & The harness supplied boolean \code{"\_sig":true} instead of \code{providerMetadata.google.thoughtSignature}, producing \code{S03} signature-loss findings on 8/61 whole replies and 35/548 streams for each AI SDK configuration. & Request and reverse-reply counts unchanged. Restoring the metadata raises forward replies without \code{T1} from 49 to 57/61 whole replies and from 440 to 475/548 streams (86.7\,\%). & App.~\ref{app:replies}; App.~\ref{app:audit} \\
    LiteLLM capture-point sensitivity & All 37/51 LiteLLM \code{T1} cells fail static target checks at dictionary capture: \code{thinking\_blocks} remains in 37/37 and \code{top\_k} in 12/37. Source-name length is not the cause: LiteLLM rewrites names longer than 64 characters in 26/37 cells. On the \code{hosted\_vllm} HTTP body (LiteLLM~1.106.0), serialisation removes \code{thinking\_blocks}; 25/37 cells then pass, while 12/37 still fail because of \code{top\_k}. These are static validity judgements, not observed live errors. & Baseline dictionary capture: 0/51 statically accepted \code{T1}, no primary findings. HTTP-body sensitivity raises statically accepted \code{T1} from 333 to 358/685 (52.3\,\%) and primary findings from 127 to 132/685 (19.27\,\%, +5 cells; upper bound 136/685, 19.85\,\%). & \S\ref{sec:rq1}; App.~\ref{app:subjects}; App.~\ref{app:audit} \\
    Gemini endpoint and model-name checks & The initial live runner addressed \code{gemini-2.5-flash-lite}, unavailable on the account, and received 204 HTTP~404 responses. Repeating the check on \code{gemini-3.5-flash-lite} yields 160/204 HTTP~200 and 44/204 HTTP~400 responses for missing \code{thought\_signature}. A separate offline sensitivity  check varies target model strings across three naming variants. & Live-acceptance results use \code{gemini-3.5-flash-lite}. Model-name sensitivity changes \code{T1} cells from 467 to 479/685 (69.9\,\%; a 2.57\,\% relative change, within the pre-specified 15\,\% bound) and statically accepted \code{T1} from 333 to 345/685 (50.4\,\%). & \S\ref{sec:step4-contract}; \S\ref{sec:threats}; App.~\ref{app:audit} \\
    Scorecard and reporting errata & Corrected preliminary scorecard values: Set~9, 29/35 to 28/34 \code{T1} (82.4\,\%) and 30/35 to 29/34 any-category (85.3\,\%); Set~10, 0/28 to 1/28 \code{T1} false alarms (3.6\,\%); Sets~9--14, 99/125 to 102/125 any-category (81.6\,\%). Removed an obsolete initial-driver 3,197-cell human-pool note, recorded per-set designer provenance, and split 18 upstream comments into 10 on others' threads and 8 follow-ups on our own (44 posts = 26 issues + 18 comments). & Reporting corrections only; main-text and headline-table counts unchanged. Any-category counts include informational records and are not \code{T1} fault-detection rates. & \S\ref{sec:rq-persist}; App.~\ref{app:persistence}; App.~\ref{app:accounting}; App.~\ref{app:rq5-detail} \\
\end{longtable}}

\end{document}